\documentclass[aps,prx,longbibliography,superscriptaddress,twocolumn,showpacs]{revtex4-1}

\usepackage[utf8]{inputenc}
\usepackage[american,british]{babel}
\usepackage[T1]{fontenc}

\usepackage{amssymb}

\usepackage{graphicx}% Include figure files
\usepackage{xcolor}
\usepackage{dcolumn}% Align table columns on decimal point
\usepackage{bm}% bold math
\usepackage[normalem]{ulem}

\usepackage{amsmath,amsthm,amssymb,mathrsfs,mathtools,amsfonts,braket}
\usepackage[caption=false]{subfig}

\usepackage{braket}
\usepackage{todonotes}
\newcounter{todocounter}
\usepackage[Option]{overpic} 
\usepackage{mathdots}
\usepackage{dsfont}
\usepackage{mathrsfs}%serif font for propagator matrix
\DeclareMathAlphabet{\mathscrbf}{OMS}{mdugm}{b}{n}

 \newcommand{\RN}[1]{%Roman numbers
  \textup{\expandafter{(\romannumeral#1)}}%
}
\usepackage{comment}

\usepackage{bigints}
\usepackage{float}
\usepackage{scalerel,stackengine,amsmath}

\DeclareMathOperator\diag{diag}

\begin{document}

\title{Non-equilibrium quantum impurity problems  \\ via matrix-product states in the temporal domain}

\author{Julian Thoenniss}
\affiliation{Department of Theoretical Physics,
University of Geneva, Quai Ernest-Ansermet 30,
1205 Geneva, Switzerland}

\author{Alessio Lerose}
\affiliation{Department of Theoretical Physics,
University of Geneva, Quai Ernest-Ansermet 30,
1205 Geneva, Switzerland}

\author{Dmitry A. Abanin}
\affiliation{Department of Theoretical Physics,
University of Geneva, Quai Ernest-Ansermet 30,
1205 Geneva, Switzerland}

\date{\today}

\begin{abstract}
Describing a quantum impurity coupled to one or more non-interacting fermionic reservoirs is a paradigmatic problem in quantum many-body physics. While historically the focus has been on the equilibrium properties of the impurity-reservoir system, recent experiments with mesoscopic and cold-atomic systems enabled studies of highly non-equilibrium impurity models, which require novel theoretical techniques. We propose an approach to analyze impurity dynamics based on the matrix-product state (MPS) representation of the Feynman-Vernon influence functional (IF). The efficiency of such a MPS representation rests on the moderate value of the temporal entanglement (TE) entropy of the IF, viewed as a fictitious ``wave function'' in the time domain. 
We obtain explicit expressions of this wave function for a family of one-dimensional reservoirs, and analyze the scaling of TE with the evolution time for different reservoir's initial states. While for initial states with short-range correlations we find temporal area-law scaling, Fermi-sea-type initial states yield logarithmic scaling with time, %reminiscent of 
closely related to the real-space entanglement scaling in critical $1d$ systems. Furthermore, we describe an efficient algorithm for converting the explicit form of the reservoirs' IF to MPS form. Once the IF is encoded by a MPS, arbitrary temporal correlation functions of the interacting impurity can be efficiently computed, irrespective of its internal structure. The approach introduced here can be applied to a number of experimental setups, including highly non-equilibrium transport via quantum dots and real-time formation of impurity-reservoir correlations.

\end{abstract}

\maketitle

%\tableofcontents

\section{Introduction}
\label{Sec:Introduction}

\begin{figure}[t]
    \centering
     \begin{overpic}[width=0.47\textwidth]{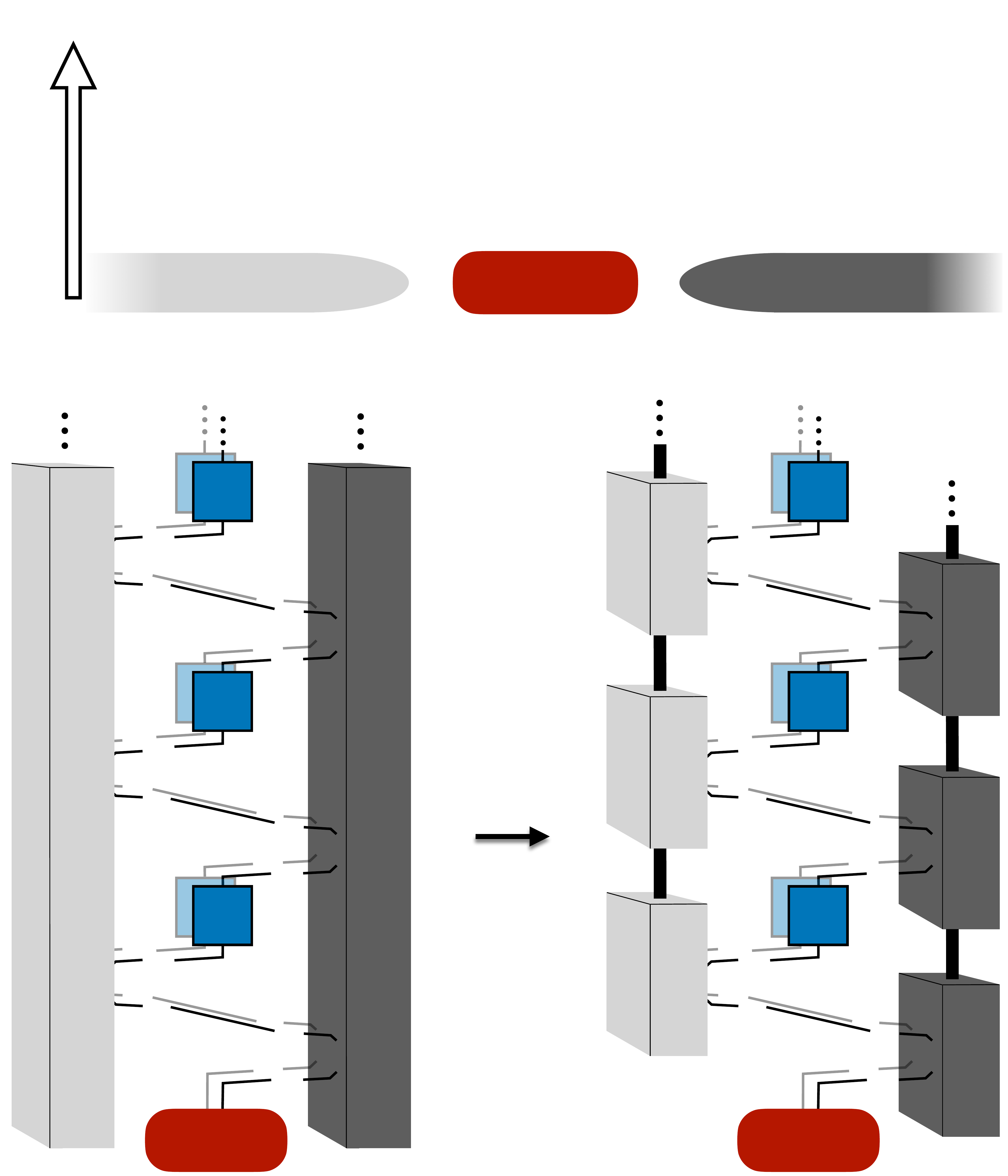}
      \put(0,100){(a)}
      \put(0,68){(b)}
     \put(9,93){Time evolution $e^{-iHt}$}
     \put(16,74.8){$\mu_L,T_L$}
       \put(65,74.8){\color{white}{$\mu_R,T_R$}}
     \put(13.7,2){\color{white}{ $\rho_\mathcal{S}(0)$}}
    \put(64.3,2){\color{white}{ $\rho_\mathcal{S}(0)$}}
     \put(40.3,74.8){\color{white}{Impurity}}
     \put(35.4,74.8){$\otimes$}
     \put(54.8,74.8){$\otimes$}
     \put(30,48){\color{white}{IM}}
      \put(4.7,48){IM}
     \put(13.7,66.5){ $\rho_\mathcal{S}(t)$}
     \put(64.3,66.5){$\rho_\mathcal{S}(t)$}
     \end{overpic}
    \caption{a) Time evolution of an interacting impurity coupled to  non-interacting fermionic reservoirs. b) The dynamical influence of a reservoir on the impurity can be encoded in a tensor acting on the impurity trajectories on the Keldysh contour --- the {\it influence matrix} (IM) --- which can be evaluated analytically.
    This multi-time tensor can be viewed as a many-body state in the temporal domain, and its ``entanglement'' can be analyzed. When this {\it temporal entanglement} is low, the IM ``wave function'' can be represented as MPS, which allows to efficiently compute the real-time dynamics of the impurity.
    }
    \label{fig:intro}
\end{figure}
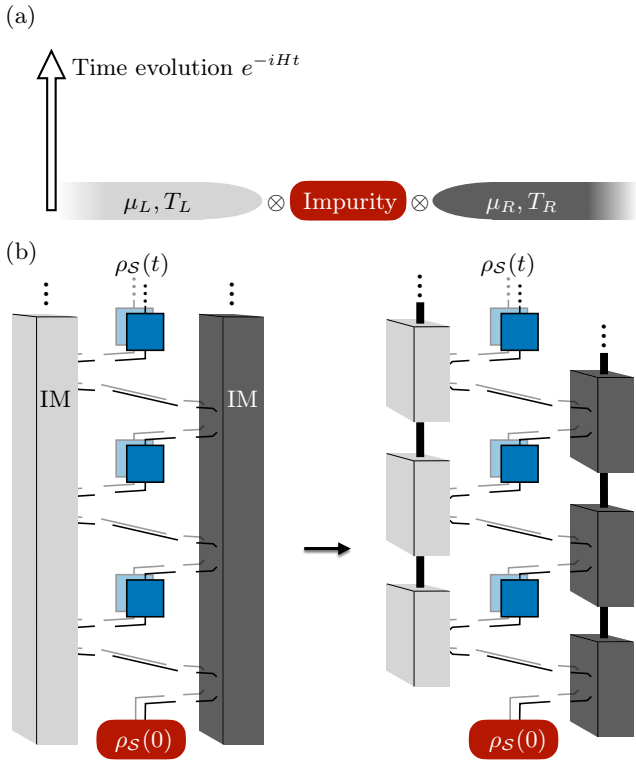

Quantum impurity models (QIMs) have long played a key role in many-body physics. QIMs typically describe a small interacting system, such as a single lattice site or a localized spin, which is embedded in a large non-interacting environment of fermions or bosons. A paradigmatic example of a QIM is the Kondo model~\cite{hewson_1993} -- a localized spin immersed in a Fermi sea.  
The intricate correlations between the impurity spin and itinerant electrons underlie the behavior of low-temperature transport in various materials~\cite{hewson_1993} and mesoscopic devices such as quantum dots~\cite{GlazmanKondo2004}, and furthermore control the properties of heavy-fermion materials. 
QIMs, including the Kondo model, also served as a testbed for theoretical techniques, such as perturbative renormalization group~\cite{Anderson1970} and Wilson's numerical renormalization group (NRG)~\cite{WilsonRMP75}. Furthermore, QIMs are a key ingredient in dynamical mean-field theory (DMFT) methods for strongly correlated materials~\cite{GeorgesRMP}. 

While ground-state and finite-temperature properties of various QIMs are well-understood, thanks to the multitude of powerful methods including NRG~\cite{WilsonRMP75,BullaNRGReview2008}, density-matrix renormalization group (DMRG)~\cite{White1992,Schollwock2005}, exact diagonalization~\cite{Caffarel94ed,capone07solving,lu2017exact}, continuous-time Monte Carlo methods~\cite{MonteCarloReview} and, in some cases exact solutions~\cite{TsvelikWiegmann83,AndreiRMP83,Sorensen_1993}, recently the focus shifted to {\it non-equilibrium} properties and {\it real-time} dynamics of QIMs~\cite{Nordlander99HowLong,Anders05Realtime,Tu08nonmarkovian,IFnanodevices,Medvedyeva13SpatiotemporalBuildup,Nuss15SpatiotemporalFormation,Schmidt16MesoscopicRydberg,nghiem17timeevolution,Schmidt_2018,skou2021non}, thanks to new experimental capabilities~\cite{DeFranceschi02outofequilibriumkondo,Tureci11_quenchKondo,latta2011quantum,fukuhara2013quantum,bauer13realizing,cetina2016ultrafast,Krinner_2017,desjardins2017observation,Riegger18LocalizedMagneticMoments,KanaszNagy18ExploringKondo,ji2021coupling,koepsell2021microscopic}. One typical experimental setup of interest in mesoscopic systems and ultracold atomic gases is that of a quantum quench, where the Hamiltonian of the system is changed suddenly (e.g., reservoirs are connected to the impurity). Another challenge arising in the context of experiments across different platforms is to analyze transport characteristics of QIMs away from the linear-response regime, e.g. at high bias voltage between the reservoirs, when the carriers in the leads have a non-equilibrium distribution, or when the system  parameters are time-dependent.  

Notable advances towards describing non-equilibrium properties of QIMs include iterative path-integral schemes~\cite{MakriMakarov94,EggerIterative2008,MillisImp2010}, diagrammatic Monte Carlo approaches~\cite{Muhlbacher08Realtime,Schiro09realtime,Werner09Diagrammatic,gull10boldline,gull11NumericallyExact,cohen11memory,cohen13neqkondo,cohen15taming,aoki14neqdmftreview}, non-Markovian master equations~\cite{Tu08nonmarkovian,IFnanodevices,dorda14auxiliary},
tensor-network~\cite{FeiguinWhite04,Vidal04Efficient,Banuls09,huang14longtime,PAECKEL2019167998,prior10efficient,WolfPRB14,TEMPO,Nusseler20Efficient,Wojtowicz20OpensystemTN}, non-equilibrium NRG~\cite{Anders05Realtime,nghiem17timeevolution,DelftNonEquil18}, and variational~\cite{Ashida2018,SHI2018245} techniques. However, understanding the power of such methods and the quality of the approximations involved, especially for QIMs involving multi-orbital impurities and several reservoirs, remains an outstanding challenge. 
Thus, other versatile approaches with theoretical efficiency guarantees, which may be able to address regimes that are difficult for other methods, are highly desirable.

In this paper, we introduce a method for non-equilibrium QIMs with non-interacting fermionic reservoirs based on a tensor-network representation of the reservoirs' influence functionals (IF)~\cite{FeynmanVernon}, illustrated in Fig.~\ref{fig:intro}. For simplicity, we focus on a quantum quench setup (Fig.~\ref{fig:intro}a): a possibly multi-orbital impurity is connected to one or more reservoirs at time $t=0$. To analyze time-dependent impurity observables, we describe the system's evolution via discretized real-time Keldysh path integral, as in previous iterative path-integral~\cite{EggerIterative2008,MillisImp2010} approaches to fermionic QIMs. Integrating out a reservoir's degrees of freedom yields an influence matrix (IM) acting on the system's trajectories, which fully encodes the dynamical influence of that reservoir on the impurity.
In the limits of vanishing or extremely strong interactions in the impurity region, knowledge of the IM allows to derive an exact non-Markovian master equation for the impurity dynamics~\cite{Tu08nonmarkovian,IFnanodevices,Mitchison_2018}.
For general interactions, however, performing summation over the impurity trajectories is a formidable problem.
The central question we address concerns the possibility of representing the IM by matrix-product states (MPS). We argue that for a large family of initial states of the reservoirs, this is indeed possible, and provide a practical algorithm for converting reservoirs' IMs to MPSs. Once a MPS form of the IMs of all reservoirs is obtained, time-dependent observables of any interacting impurity can be computed efficiently.

% oscillator bath: previous analytical approaches
Previously, various analytical and numerical IF approaches to the dynamics of a small open quantum system interacting with a thermal reservoir have been developed. Much of this work concentrated on the paradigmatic spin-boson model -- a case where the IF can be computed analytically~\cite{FeynmanVernon}.   Even though the IF is known, computing dynamics and correlations of the impurity  is, in general, a highly non-trivial task that requires a variety of approximations depending on the impurity Hamiltonian and spectral properties of the bath~\cite{LeggettRMP}.  Early numerical approaches to the problem~\cite{MakriMakarov94} are based on truncating the IF memory range, such that the temporal correlations of the bath up to a certain cutoff time are captured. Similar schemes have been developed for fermionic QIMs out-of-equilibrium~\cite{EggerIterative2008,MillisImp2010}. The main limitation of such approaches lies in the complexity of summing over the trajectories, which grows exponentially with the cutoff time. Furthermore, sampling the real-time trajectories or the interaction diagrams is generally exponentially hard as a consequence of sign problems.

% numerical approaches to IF: efficiency, challenge, bosons/fermions

Recently, Refs.~\cite{TEMPO,Jorgensen19Exploiting,Luchnikov2019,Chan21,bose2021tensor} proposed to perform a version of this IF-based propagation for the spin-boson model using a tensor-network scheme, which leads to a significantly improved efficiency when the bath's spectral density is sufficiently well-behaved. 
From a different perspective, Refs.~\cite{Banuls09,lerose2020} as well as Ref.~\cite{Chan21}  developed an IM approach to describing non-equilibrium dynamics of homogeneous {\it interacting} spin chains, and used {\it temporal entanglement} (TE) of the IM to characterize the efficiency of this approach.

Building on these recent developments, here we aim to find efficient representations of the IM of fermionic reservoirs, initially prepared in Gaussian states, as MPS. The MPS bonds encode non-local temporal correlations arising from memory effects of the reservoirs, and may be interpreted as time-local propagation of a fictitious compressed environment [see Fig.~\ref{fig:intro}c] that provides a faithful representation of the original environment's influence on the impurity. Provided the IM of each reservoir coupled to the impurity can be efficiently compressed to a MPS form with a moderate bond dimension $\chi$, time evolution of {\it any} impurity with Hilbert space dimension $q$ can then be efficiently computed by contracting the $q^2$-dimensional tensors representing the evolution of the impurity [blue squares in Fig.~\ref{fig:intro}c] with the $(q^2 \times \chi)$-dimensional tensors of the IMs' MPSs [light and dark gray boxes in Fig.~\ref{fig:intro}c], sequentially in time. Thus, non-equilibrium initial states which allow a compact MPS representation of the IM correspond to non-equilibrium QIMs that only require polynomial computational resources. The central question is therefore to understand when such a compression is efficient, i.e., under which conditions the environment can be encoded by a MPS with~$\chi$ finite or at least scaling polynomially with evolution time.

Below we will assume that reservoirs are initially in Gaussian states -- these in particular include stationary states of reservoirs, e.g. thermal equilibrium states as well as states with an arbitrary non-equilibrium distributions of quasiparticles. We view the resulting IM as a fictitious Gaussian wave function on the Keldysh temporal contour, and find that it has Bardeen-Cooper-Schrieffer(BCS)-type form. For a simple family of one-dimensional reservoir models we express the IM via the reservoir's initial state and spectral properties of quasiparticles. 

To assess the possibility of approximating such wave functions by a MPS, we will analyze their temporal entanglement properties. 
In particular, we will consider the von Neumann entanglement entropy $S(\tau,t)$ associated with a bipartition $[0,\tau]$, $[\tau,t]$ of the degrees of freedom in time, and the corresponding temporal entanglement spectrum. We find that for a wide family of short-range correlated equilibrium and non-equilibrium initial states (including finite temperature ensembles or zero-temperature states away from critical points), temporal entanglement follows an area-law: 
\begin{equation}
\max_{0\le\tau\le t} S(\tau,t)\leq C \, , \quad  t\to\infty \, ,
\end{equation}
where $C$ is a constant that depends on the reservoir properties. 
Further, for critical Fermi-sea-type initial states, we find a logarithmic violation of the temporal entanglement area-law, which is closely related to the celebrated scaling of spatial entanglement in $(1+1)$-dimensional conformal field theories~\cite{CalabreseCardyReview}: 
\begin{equation}
\max_{0\le\tau\le t} S(\tau,t)\sim \frac c 6 \log t  \, ,
\end{equation}
where $c$ is the central charge. These results, combined with the existence of sufficiently localized natural orbitals in the IM (see below) point to the theoretical feasibility of an efficient description of the IM of fermionic reservoirs as MPS.

Leveraging these results, we then present and illustrate an algorithm for converting our BCS-like IM wave function into a MPS form, by extending the Fishman-White algorithm~\cite{Fishman2015MPS} for Gaussian states with a fixed fermion number. Here we mostly concentrate on TE properties of reservoirs IM and the issue of computational efficiency, leaving physical applications to non-equilibrium QIMs for future work. 

The rest of the paper is organized as follows. 
In Sec.~\ref{sec_setup} we introduce the IM approach to QIMs and derive the explicit form of the IM for a family of models of non-interacting fermionic environment.
In Sec.~\ref{Sec:temporal_entanglement} we view the IM as a ``wave function'' in the temporal domain and  analyze the scaling of its temporal entanglement, focusing on the role of quantum criticality of the environment's initial state.
Finally, motivated by these results, in Sec.~\ref{sec:mps_conversion}  we present and discuss an explicit efficient algorithm to convert the IM of an arbitrary non-interacting fermionic environment to a MPS form.
Appendices~\ref{app:path_integral}--\ref{Sec:block_size_XY} report the technical details of the derivations encountered in the main text.

\section{Influence Matrix of non-interacting fermionic environments}

\label{sec_setup}

We start this Section by formulating the setting and reviewing the IM approach, with a focus on QIMs (Sec.~\ref{subsec_if}). Then, we introduce the environment models analyzed in this paper (Sec.~\ref{subsec_models}). Finally, using Grassman path integral, we obtain explicit expressions for their influence matrices (Sec.~\ref{subsec_method}), and discuss the relation of IM to the standard Keldysh correlation functions.

\subsection{IM approach to quantum impurity dynamics}
\label{subsec_if}

We consider the problem of a quantum quench: at time $t=0$ the impurity (with degrees of freedom denoted by $\mathcal{S}$) is suddenly coupled to an environment $\mathcal{E}$ constituted by one or more non-interacting fermionic reservoirs [see Fig.~\ref{fig:intro}a]. We will assume that the initial state $\rho_{\mathcal{E}}$ of the environment is Gaussian, and in particular we will be interested in the case when this state is stationary with respect to the internal dynamics of $\mathcal{E}$.

To describe dynamics of impurity observables, we will view it as an open quantum system. % with reservoirs playing the role of environment. 
Following Feynman and Vernon~\cite{FeynmanVernon}, we integrate out the environment's dynamical degrees of freedom, resulting in an effective action for~$\mathcal{S}$ only. The influence of $\mathcal{E}$ on $\mathcal{S}$ is expressed by an IF defined on the trajectories of $\mathcal{S}$, which is generally nonlocal in time reflecting the non-Markovian nature of the environment. In many cases of interest, this non-Markovianity is essential as no sharp separation of timescales between system and environment exists. Theoretical description of non-Markovian dynamics is known to be challenging, even for the case when environment is Gaussian~\cite{LeggettRMP}.

\begin{figure*}
    \centering
       \begin{overpic}[width=0.95\textwidth]{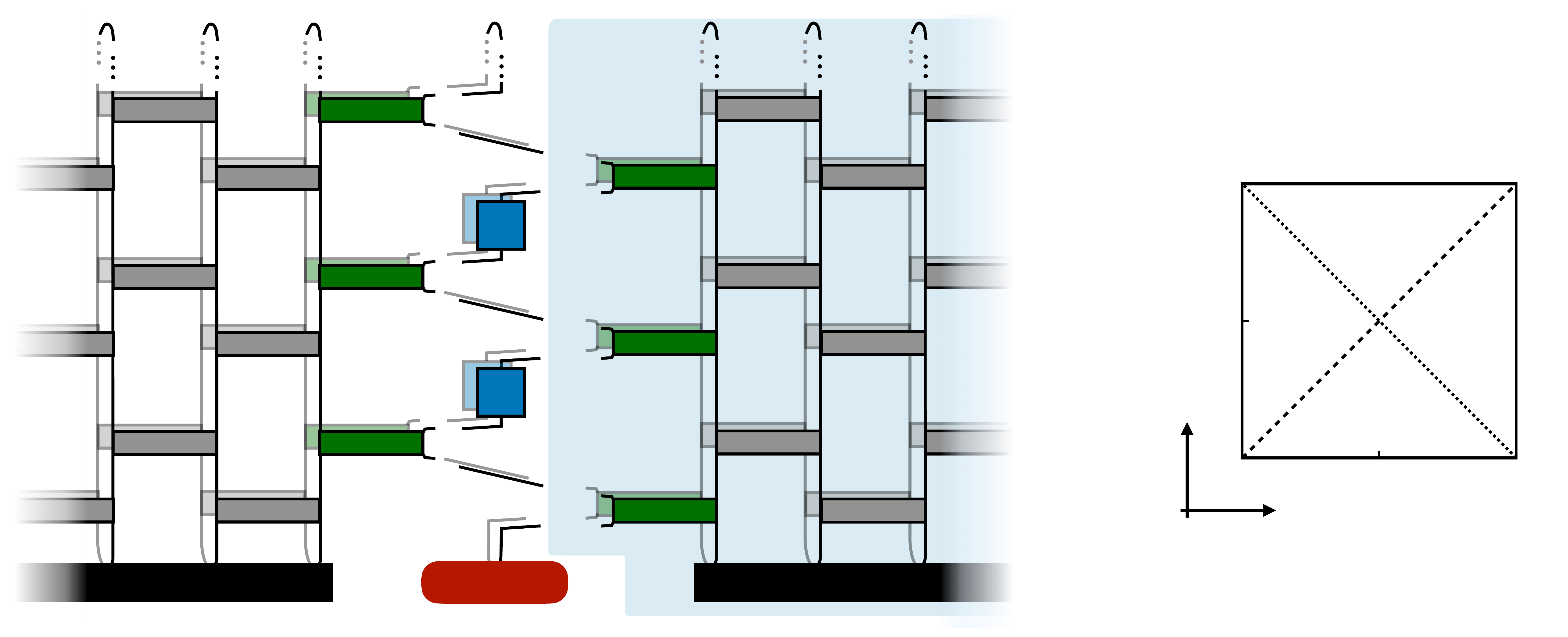}
       \put(0,41){\scriptsize (a)}
           \put(73,41){\scriptsize (b)}
  \put(30,3){\color{white}{\scriptsize$\rho_{\mathcal{S}}(0)$}}
     \put(52,3){\color{white}{\scriptsize $\rho_{\mathcal{E}_R}$}}
       \put(15,3){\color{white}{\scriptsize$\rho_{\mathcal{E}_L}$}}

        \put(50,-0.3){\scriptsize $=\mathcal{I}_R[\{\sigma^\pm,s^\pm\}]$}
        
        \put(80,6.5){\scriptsize $\mathcal{J}_x$}
          \put(73,13){\scriptsize $\mathcal{J}_y$}
           \put(86.5,9.7){\scriptsize $\pi/4$}
           \put(95,9.7){\scriptsize $\pi/2$}
           \put(76,20){\scriptsize $\pi/4$}
           \put(76,28.5){\scriptsize $\pi/2$}
           
            \put(82,20){$0$}
           \put(90.5,20){$0,\pi,\pi$}
           \put(87.5,14){$0$}
           \put(86,26){$0,\pi,\pi$}
           
           \put(38,5.8){\scriptsize $\sigma_{0}^+$}
           \put(35.5,6.3){\scriptsize \color{gray}{$\sigma_{0}^-$}}
           \put(38,9.7){\scriptsize $s_{0}^+$}
           \put(35.5,10.3){\scriptsize \color{gray}{$s_0^-$}}

           \put(38,16.6){\scriptsize $\sigma_{1}^+$}
           \put(35.5,17.1){\scriptsize \color{gray}{$\sigma_{1}^-$}}
           \put(38,20.4){\scriptsize $s_{1}^+$}
           \put(35.5,20.9){\scriptsize \color{gray}{$s_1^-$}}

           \put(38,27.1){\scriptsize $\sigma_{2}^+$}
           \put(35.5,27.6){\scriptsize \color{gray}{$\sigma_{2}^-$}}
           \put(38,31.1){\scriptsize $s_{2}^+$}
           \put(35.5,31.7){\scriptsize \color{gray}{$s_2^-$}}
       \end{overpic}
    \caption{
    a) The impurity is coupled to two independent reservoirs to its left and right.
    The two tensors arising from the summation over all trajectories of the reservoirs (i.e., from the contractions of the two tensor networks) are the two IMs.
    Gray gates acting from bottom to top represent nearest-neighbor interactions $e^{-i\delta t H_{j,j+1}}$ (foreground) and $e^{+i\delta t H_{j,j+1}}$ (background). Green gates encode the coupling between impurity and environment; in this work they are chosen to be equivalent to gray gates. Blue gates represent impurity interactions $e^{-i\delta t H_{\mathcal{S}}}$ (foreground) and $e^{+i\delta t H_{\mathcal{S}}}$ (background). The labels on the external legs of the right IM define the impurity's trajectory (analogously for the left IM).
    b) Floquet phase diagram of the trotterized XY model $(\varphi = 0)$. Phase boundaries (diagonals) are associated with the quasienergy gap closing, and the labels refer to the types of edge modes in the corresponding phase.
    }
    \label{fig:circuit}
\end{figure*}
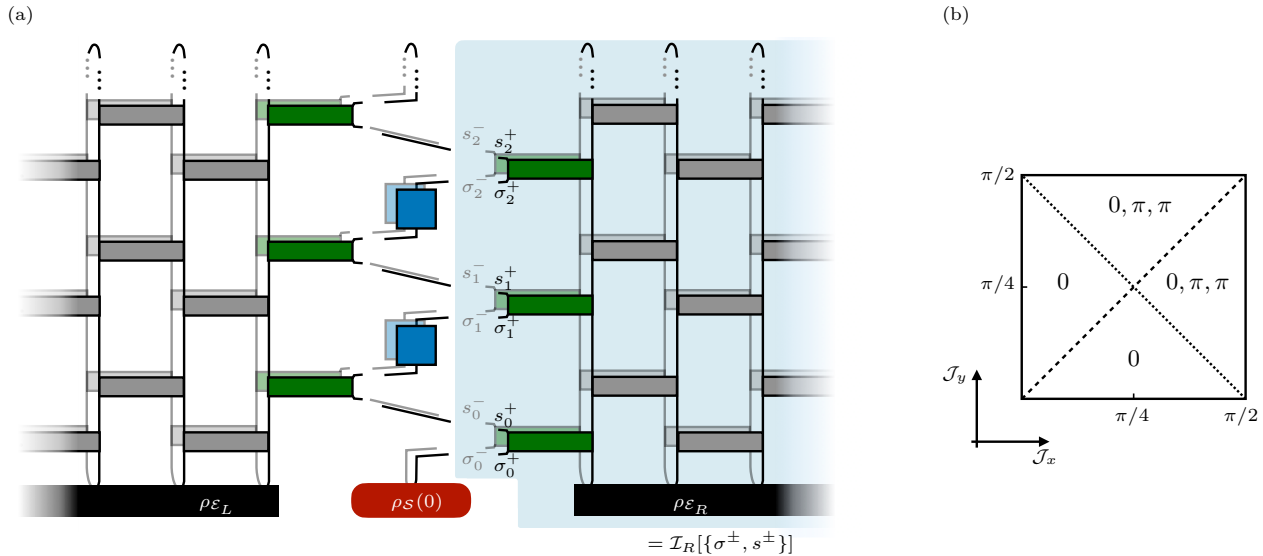

We will analyze IF properties for a lattice model (Fig.~\ref{fig:circuit}), with an impurity situated at site $j=0$ and left/right reservoirs $\mathcal{E}_{L/R}$ situated at $j<0$ and $j>0$, respectively. (As will be clear shortly, the analysis of multiterminal geometries with more than two reservoirs requires no extra effort.) 
For simplicity, we consider Hamiltonians with nearest-neighbor interactions,
\begin{equation}
    H=H_{\mathcal{S}}+\sum_{j} H_{j,j+1}.
\end{equation}
 Here $H_{\mathcal{S}}$ denotes part of the Hamiltonian acting on impurity only (e.g. Hubbard interaction in the Anderson model), while terms $H_{0,1}, H_{-1,0}$ describe the coupling between the impurity and the reservoirs $\mathcal{E}_{L/R}$.  

As a first step, we represent
time evolution of the full impurity + reservoirs system as a unitary circuit, by discretizing Hamiltonian dynamics with a small Trotter step $\delta t$
\begin{equation}
\label{eq_trotterization}
    e^{-itH}
    \simeq
    \bigg[
    e^{-i\delta t H_{\mathcal{S}}}
    \Big(\prod_{j}   
    e^{-i\delta t H_{2j-1,2j}}
    \Big) 
    \Big(\prod_{j}   e^{-i\delta t H_{2j,2j+1}}
    \Big)
    \bigg]^{t/{\delta t}} \, .
\end{equation}
The Keldysh path integral representation of a time-evolved impurity observable $\hat O$, 
\begin{equation}
    \Big\langle O_{j=0} (t) \Big\rangle = \mathrm{Tr}\Big(
O_{j=0} \, e^{-itH} \rho_{\mathcal{E}_L} \otimes \rho_{\mathcal{S}}\otimes \rho_{\mathcal{E}_R}
\, e^{+itH}
\Big) \, ,
\end{equation}
is obtained by inserting resolutions of identity between each operator multiplication in the two unitary circuits associated with the two branches of time evolution, given by Eq.~\eqref{eq_trotterization} and its Hermitian conjugate.
The resulting sum over paths can be interpreted as the contraction of a $(1+1)$-dimensional two-sheet tensor network, as illustrated in  Fig.~\ref{fig:circuit}a.

In this formulation, the time-discretized influence functional -- or {\it influence matrix} (IM) $\mathcal{I}_{\mathcal{E}_{L/R}}$ of the left/right environment -- is the tensor with indices associated with the Keldysh impurity trajectory $\sigma_\tau^\pm, s_\tau^\pm$, $\tau=0,\dots,T-1$ ($+/-$ referring to the forward/backward branch, and $T\equiv t/\delta t$), obtained by summing over all environment indices. For example, the right IM reads:
  \begin{widetext}
\begin{equation}
\label{eq:IM_operator}
\mathcal{I}_{R}[\sigma_\tau^\pm,s_\tau^\pm] \; = \; \mathrm{Tr}_{\mathcal{E}_R}\Big(\mathcal{U}_{\mathcal{E}_R}[\mathcal{U}_{\mathcal{S}\mathcal{E}_R}]_{s_{T-1}^+,\sigma_{T-1}^+}\hdots \mathcal{U}_{\mathcal{E}_R}[\mathcal{U}_{\mathcal{S}\mathcal{E}_R}]_{s_0^+,\sigma_0^+} \, \rho_{\mathcal{E}_R}  \,
 [\mathcal{U}^\dagger_{\mathcal{S}\mathcal{E}_R}]_{\sigma_0^-,s_0^-}\mathcal{U}_{\mathcal{E}_R}^\dagger\hdots [\mathcal{U}^\dagger_{\mathcal{S}\mathcal{E}_R}]_{\sigma_{T-1}^-,s_{T-1}^-}\mathcal{U}_{\mathcal{E}_R}^\dagger\Big).
\end{equation} 
\end{widetext}
where one step of the environment's evolution is
\begin{equation}
    \mathcal{U}_{\mathcal{E}_R} =
    \mathcal{U}_{\text{even}}
    \mathcal{U}_{\text{odd}} = 
    \Big(\prod_{j>0}   
    e^{-i\delta t H_{2j-1,2j}}
    \Big)
    \Big(\prod_{j>0}   e^{-i\delta t H_{2j,2j+1}}
    \Big)
\end{equation}
and 
\begin{equation}
    \mathcal{U}_{\mathcal{S}\mathcal{E}_R} = 
       e^{-i\delta t H_{0,1}},  
\end{equation}
represents the system-environment interaction. We have also defined the partial matrix elements $[\mathcal{U}_{\mathcal{S}\mathcal{E}_R}]_{s,\sigma} \equiv \braket{s|\mathcal{U}_{\mathcal{S}\mathcal{E}_R}|\sigma}$, where $\ket{\sigma}$ and $\ket{s}$ are impurity basis states, such that $[\mathcal{U}_{\mathcal{S}\mathcal{E}_R}]_{\sigma,s}$ is an operator acting only on site $j=1$ of the right environment.
The IM of the left environment $\mathcal{I}_{L}$ is defined similarly. %\al{do we need the following sentence here? what are the "IM tensors"?} Arbitrary temporal impurity correlators can be expressed as contractions of the IM tensors with $(q\times q)$-dimensional tensors associated with the impurity evolution. 

The impurity dynamics can be computed by contracting the left and right IMs with the impurity evolution operators. A crucial observation is that each IM is a property of the associated reservoir (left or right), and is completely independent of the impurity Hamiltonian and of the other reservoir. Hence, in our approach, a highly non-equilibrium problem with strongly imbalanced reservoirs prepared in different equilibrium states involves exactly the same computational cost as a local impurity quench with balanced, equilibrated reservoirs.
The problem of computing QIM dynamics is thus reduced to the problem of computing the individual reservoir's IMs; this fact directly generalizes to multiterminal geometries with more than two reservoirs. For this reason, in the forthcoming analysis, we will exclusively focus on the properties of a single reservoir.

Previous works that considered dynamics in extended interacting systems~\cite{Banuls09,lerose2020,Chan21} approximated $\mathcal{I}_{L,R}[\sigma_\tau^\pm,s_\tau^\pm]$ as a MPS by numerically contracting corresponding tensor networks. This approach can be made efficient~\cite{lerose2022overcoming,frias2022light} provided temporal entanglement of the IM remains low. In certain $1d$ Floquet spin models, the low TE of the IM was established, via either  exact solutions or  numerical simulations~\cite{Piroli2020,lerose2020,lerose2021,sonner2021influence,klobas2021exact,giudice2021temporal,sonner22characterizing}. 
Here, building on Ref.~\cite{lerose2021}, we take a different approach to representing the IM, utilizing the fact that in QIMs the reservoirs are non-interacting. This allows us to obtain an explicit form of the IM, followed by an efficient conversion to a MPS form.

\subsection{Models}
\label{subsec_models}

Below, we will focus on a family of one-dimensional reservoirs (or ``leads'') of spinless fermions governed by a tight-binding Hamiltonian with  $p$-wave superconducting pairing~\cite{Kitaev_2001}:
\begin{equation}
\label{eq_kitaev}
    H_{j,j+1} = -t (c^\dagger_j c_{j+1} - c_{j} c^\dagger_{j+1})
    - \Delta
    (c^\dagger_j c^\dagger_{j+1} - c_{j} c_{j+1}) 
     %+ g (1-2
     -\mu c_j^\dagger c_j
     %)
\end{equation}
Site $j=0$ describes a possible multi-orbital impurity with Hilbert space dimension $q$ and an additional interacting Hamiltonian $H_{\mathcal{S}}$, which can be arbitrary. We note that generalization to the case of spinful fermions is straightforward and will be discussed below.

In view of practical implementations of this model in classical or quantum simulations, it can be convenient to consider an equivalent chain of spins-$1/2$ or qubits,  
\begin{equation}
\label{eq_xychain}
     H_{j,j+1} = -J \bigg(\frac{1+\gamma}{2} \sigma^x_{j} \sigma^x_{j+1}
    + \frac{1-\gamma}{2}
    \sigma^y_{j} \sigma^y_{j+1} 
    \bigg) + g \sigma^z_j
\end{equation}
The identification between the two models as $t=J$, $\Delta=J\gamma$, $\mu=2g$ is obtained via a Jordan-Wigner transformation.

We will consider a discretized version of the dynamics, Eq.~\eqref{eq_trotterization}, with a finite Trotter step~$\delta t$. Such dynamics exactly correspond to a unitary circuit, facilitating classical simulations and quantum-computer implementations. Moreover, by making $\delta t$ sufficiently small, the Floquet dynamics will approximate the Hamiltonian dynamics with an arbitrarily high precision.

To simplify the notation, we absorb the finite time step in the dimensionless parameters
\begin{equation}
\label{eq_gate_param}
    \mathcal{J}_x \equiv \delta t \, J \frac {1+\gamma} 2  , \;\;
    \mathcal{J}_y \equiv \delta t \, J \frac {1-\gamma} 2 , \;\;
    \varphi \equiv \delta t  \, g \, .
\end{equation}
We focus on the dynamics of the right reservoir, governed by repeated applications of unitary gates,
\begin{equation}
\label{eq_UER}
    \mathcal{U}_{\mathcal{E}_R} =
    \Big(\prod_{j>0}   
    U_{2j-1,2j}
    \Big)
    \Big(\prod_{j>0}   U_{2j,2j+1}
    \Big)
\end{equation}
with~\footnote{Note that we have chosen here a more convenient, slightly different trotterization compared to Eqs.~\eqref{eq_trotterization} and~\eqref{eq_xychain} concerning the placement of the transverse field; this choice is inconsequential for our results.}
\begin{align}\label{eq:two_site_gate}
U_{2j,2j+1}&=\exp ( i\mathcal{J}_x \sigma_{2j}^x\sigma_{2j+1}^x +  i\mathcal{J}_y \sigma_{2j}^y\sigma_{2j+1}^y), \\
\label{eq:onsite_gate}
U_{2j-1,2j}&=
\exp (i\varphi\sigma^z_{2j-1} )
\exp (i\varphi\sigma^z_{2j} )
\\ & \quad\quad \times
\exp ( i\mathcal{J}_x \sigma_{2j-1}^x\sigma_{2j}^x +  i\mathcal{J}_y \sigma_{2j-1}^y\sigma_{2j}^y) \, ; \nonumber
\end{align} 
accordingly,
\begin{equation}
\label{eq_xyintgate}
    \mathcal{U}_{\mathcal{S}\mathcal{E}_R} = U_{0,1} =
    \exp ( i\mathcal{J}_x \sigma_{0}^x\sigma_{1}^x +  i\mathcal{J}_y \sigma_{0}^y\sigma_{1}^y) \, .
\end{equation}
Analogous expressions hold for the dynamics of the left bath $\mathcal{U}_{\mathcal{E}_L}$. 
Due to simple symmetry relations, we can restrict the parameter ranges to $\mathcal{J}_x, \mathcal{J}_y,\varphi \in [0,\pi/2]$.

Away from the Trotter limit $\delta t \to0$, the unitary circuit evolution corresponds to sequential application of interactions between even and odd pairs of sites as illustrated in Fig.~\ref{fig:circuit}a.
Integrability of the model gives rise to a well-defined quasilocal effective  Hamiltonian for the Floquet dynamics (see, e.g., Ref.~\cite{arze20floquetresonances}), thus there is no indefinite heating characteristic of chaotic Floquet systems~\cite{Alessio14,Ponte14,Lazarides14}.
Therefore, even in the Floquet setting, we can conveniently think of our reservoirs as conventional Hamiltonian reservoirs, and we will be able to explore the role of their temperature and spectral properties of quasiparticles on the impurity dynamics. 

The different phases realized by the effective Floquet Hamiltonian are characterized by the presence/absence of strong zero-/$\pi$-modes and quasienergy gap closures at critical lines~\cite{dutta}.
 For $\mathcal{J}_x,\mathcal{J}_y,\varphi \ll \pi/4$, 
 our Floquet model has a phase diagram similar to that of the Hamiltonian model (\ref{eq_xychain}). 

Below we consider two cases discussed in Sec.~\ref{Sec:Impurity_entanglement_Ising} and Sec.~\ref{subsec_kxy}, respectively:
\begin{align}
&(i) \quad\; \text{ kicked Ising model: }
\mathcal{J}_y  = 0
\label{Eq:KIC_model}
\\ 
&(ii) \quad \text{ trotterized XY model: } \varphi = 0 
\label{Eq:XY_model}
\end{align}
In the former case there are two critical lines $\varphi=\mathcal{J}_x$ and $\varphi=\pi/2-\mathcal{J}_x$ belonging to the Ising universality class (free Majorana fermions), which separate Floquet phases with and without strong zero and $\pi$-modes, respectively~\cite{dutta}. In the latter case, there are two critical lines $\mathcal{J}_x=\mathcal{J}_y$ and $\mathcal{J}_x=\pi/2-\mathcal{J}_y$, belonging to the XY universality class (free Dirac fermions). The pattern of Floquet phases in this case is shown in Fig.~\ref{fig:circuit}b. 
The two cases are related, as the trotterized XY chain can be mapped to a pair of decoupled kicked Ising chains, with parameters $(\mathcal{J}_x \mapsto \mathcal{J}_x,\varphi \mapsto \mathcal{J}_y)$ and $(\mathcal{J}_x \mapsto \mathcal{J}_y,\varphi \mapsto \mathcal{J}_x)$, respectively. 
This link becomes manifest when the models are rewritten in terms of Majorana fermions, and was already noted for the corresponding Hamiltonians~\cite{peschel1984time,Turban_1985}. 
We will be particularly interested in analyzing IM behavior at the critical lines, as this corresponds to a standard setup for QIMs.

\subsection{Exact IM} 

\label{subsec_method}

We will analyze the IM of the right environment, referring to it by $\mathcal{E}$ rather than $\mathcal{E}_R$ to simplify notations. For non-interacting baths, the IM in Eq.~(\ref{eq:IM_operator}) can be derived in closed form in terms of fermionic degrees of freedom --- we will hence use the fermionic representation. 

\begin{figure} 
\center
\begin{overpic}[width=0.4\textwidth]{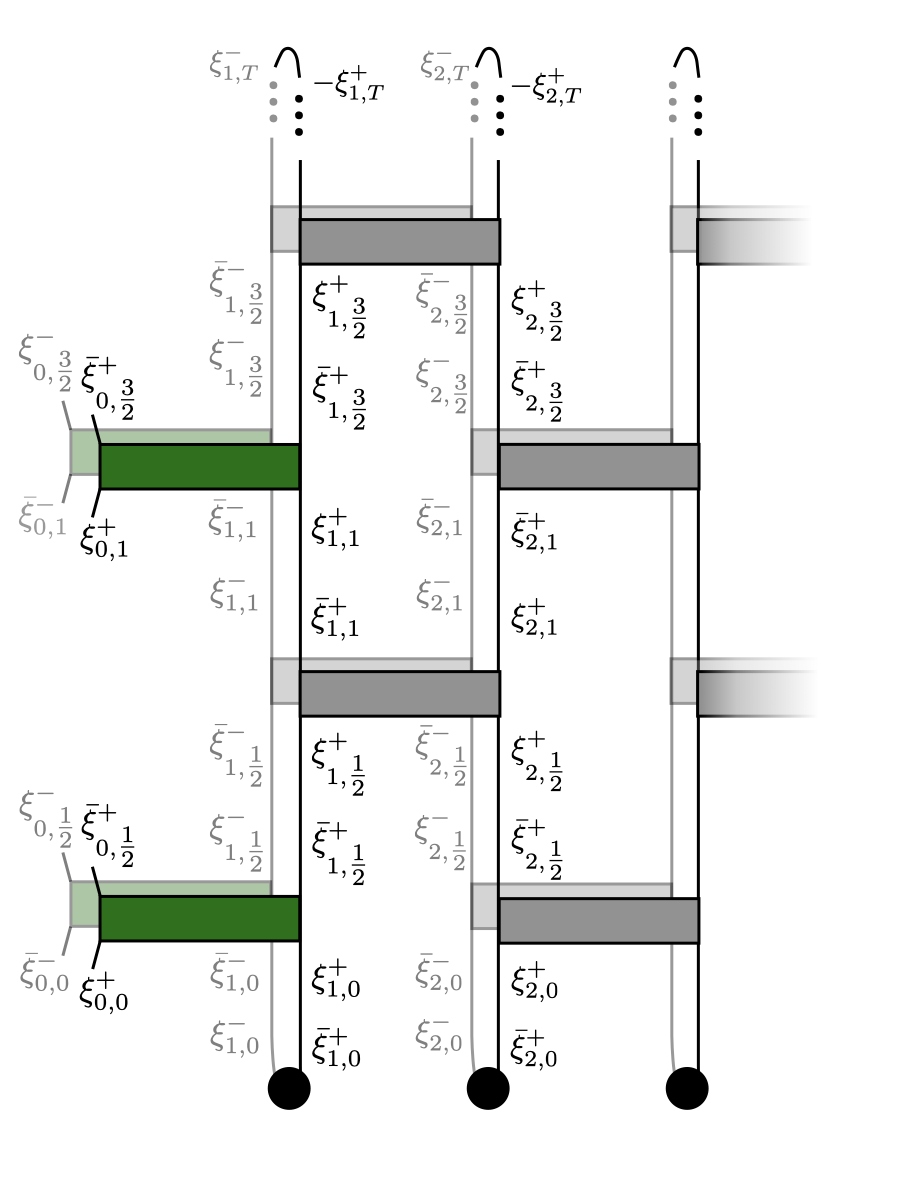}
\put(21,5){\scriptsize $j=1$}
\put(37.5,5){\scriptsize $j=2$}
\put(54,5){\scriptsize $j=3$}
\end{overpic}
\caption{By inserting resolutions of identity in terms of fermionic coherent states into Eq. (\ref{eq:IM_operator}), the IM can be written as Grassmann path integral. The leg at site $j$ and time $\tau$ in the tensor network is associated with Grassmann variables $\bar{\xi}_{j,\tau},\xi_{j,\tau}$.}
\label{Fig:Grassmann_labels}
\end{figure}

To express the IM as a path integral, we insert Grassmann resolutions of identity between each multiplication of operators in Eq.~(\ref{eq:IM_operator}). In each identity resolution, we label the Grassmann variables with indices specifying their position in Eq.~(\ref{eq:IM_operator}):
$$\mathds{1}=\int d(\bar{\xi}^\pm_{j,\tau},\xi^\pm_{j,\tau})\ket{\xi^\pm_{j,\tau}}\bra{\bar{\xi}^\pm_{j,\tau}}e^{-\bar{\xi}^\pm_{j,\tau}\xi^\pm_{j,\tau}} \,.$$ This associates a pair of Grassmann variables $\bar{\xi}^\pm_{j,\tau},\xi^\pm_{j,\tau}$ with each leg of the tensor network in Fig.~\ref{fig:circuit}a, as illustrated in Fig.~\ref{Fig:Grassmann_labels}. The operators corresponding to initial states and gates become Grassmann kernels of these variables. For the non-interacting models considered in this work, all kernels of the gates are Gaussian. Their derivation is given in Appendix~\ref{app:gate_kernels}.

The initial state of the environment $\rho_{\mathcal{E}}$ is also assumed to be Gaussian, as is natural in relevant QIM setups: one often considers stationary ensembles of their Hamiltonian -- in our setting, their (quasilocal) Floquet Hamiltonian.
{\it Nonequilibrium} Gaussian states also naturally arise following a global quantum quench; examples of such states will be considered below. 

The Grassmann form of the IM arises from the integration over environment trajectories $\bm{\xi}_j=\{(\bar\xi^\pm_{j,\tau+1/2},\xi^\pm_{j,\tau})\}$, $j\geq 1$, resulting in a functional of the impurity's Grassmann trajectory $\bm{\xi}_0=\{(\bar\xi^\pm_{0,\tau+1/2},\xi^\pm_{0,\tau})\}$.
For future convenience, we make a transformation of impurity variables, 
\begin{gather}
\label{eq_zeta}
\bm{\zeta} = \{{\zeta}_\tau\}, \quad
{\zeta}_\tau = (\zeta_{\tau}^{\uparrow+}, \zeta_{\tau}^{\uparrow-},\zeta_{\tau}^{\downarrow+}, \zeta_{\tau}^{\downarrow-})^T, \\
\zeta^{\uparrow\pm}_{\tau} = \tfrac{1}{\sqrt{2}} (\xi_{0,\tau}^\pm + \bar{\xi}_{0,\tau+1/2}^\pm),  \quad 
\zeta^{\downarrow\pm}_{\tau} = \tfrac{1}{\sqrt{2}} (\xi_{0,\tau}^\pm - \bar{\xi}_{0,\tau+1/2}^\pm). \nonumber
\end{gather}
We emphasize that the labels $\uparrow$, $\downarrow$ do not refer a spin variable -- fermions are spinless in the models considered. 

The IM path integral expression takes the following general form:
\begin{equation}
    \label{eq:grassmannpi}
    \mathcal{I}[ \bm{\zeta}]
    =
    c \;
    e^{\frac 1 2 \bm{\zeta}^T
    \bm{\mathcal{A}}_{\mathcal{S}}\bm{\zeta}
    }
    \,
    \int 
    D\bm{\xi}
    \, 
    e^{\bm{\zeta}^T
    \bm{\mathcal{A}}_{\mathcal{S}\mathcal{E}}\bm{\xi}_1
    }
    \, 
    e^{\frac 1 2 \bm{\xi}^T
    \bm{\mathcal{A}}_{\mathcal{E}}\bm{\xi}
    } \, ,
\end{equation}
where the array $\bm{\xi}=(\bm{\xi}_1,\bm{\xi}_2,\dots)$ contains all the environment trajectories variables, $c$ is a c-number, and $\bm{\mathcal{A}}_{\mathcal{S}}$, $\bm{\mathcal{A}}_{\mathcal{E}}$, $\bm{\mathcal{A}}_{\mathcal{S}\mathcal{E}}$ are the blocks of the Gaussian Keldysh action arising from Eq.~\eqref{eq:IM_operator} involving the impurity trajectory only, the environment trajectories only, and both of them, respectively.
(Note that $\bm{\mathcal{A}}_{\mathcal{S}\mathcal{E}}$ involves only the first site of the environment in our one-dimensional geometry.)
As these matrices arise from the Gaussian kernels of unitary gates, they only couple neighboring points in space and time.
Performing integration, one obtains a Gaussian {\it influence action}, 
\begin{align}
\label{eq:IM_free_fermions}
\mathcal{I}[ \bm{\zeta}] &= C \, %\exp\Bigg[
e^{\frac{1}{2}
%\sum_{\tau,\tau^\prime=0}^t 
\bm{\zeta}
%_{\tau}
^T
%(
\bm{\mathcal{B}}
%^{-1}
%)_{\tau,\tau^\prime}  
\bm{\zeta}
%_{\tau^\prime}
}
%\Bigg] %= \dots
\end{align} 
with $C=c \, \det \bm{\mathcal{A}}_{\mathcal{E}}$
and
\begin{equation}
\label{eq_influenceaction}
    \bm{\mathcal{B}}
    =
    \bm{\mathcal{A}}_{\mathcal{S}}
    +
    \bm{\mathcal{A}}_{\mathcal{S}\mathcal{E}}^T
    \bm{\mathcal{A}}_{\mathcal{E}}^{-1}
    \bm{\mathcal{A}}_{\mathcal{S}\mathcal{E}}
    \equiv 
    \bm{\mathcal{A}}_{\mathcal{S}}
    +
    \bm{\mathcal{A}}_{\mathcal{S}}^{\mathrm{eff}}
    \, .
\end{equation}
The structure of this equation is particularly clear:
while $\bm{\mathcal{A}}_{\mathcal{S}}$ is by construction local in time,  $\bm{\mathcal{A}}_{\mathcal{S}}^{\mathrm{eff}}$ encodes non-local-in-time ``self-interactions'' of the impurity trajectory mediated by the reservoir. 

We compute the self-interactions described by $\bm{\mathcal{A}}_{\mathcal{S}}^{\mathrm{eff}}$ by 
relating them to certain Keldysh correlations of the reservoir: viewing the system variables $\bm{\zeta}$ as source fields in the path integral in Eq.~\eqref{eq:grassmannpi}, we can write
\begin{equation}
\label{Eq:relation_exponent_derivatives}
 \big(\bm{\mathcal{A}}_{\mathcal{S}}^{\mathrm{eff}}\big)_{\tau,\tau^\prime}^{ab,a^\prime b^\prime} 
 \propto
 \frac{\delta^2}{\delta \zeta_{\tau^\prime}^{ab} \, \delta \zeta_{\tau}^{a^\prime b^\prime}}
 \int 
    D\bm{\xi}
    %\big[D\bar\xi%_{j,\tau}
    %D\xi%_{j,\tau}
    %\big]_{j>0,\tau}
    %_{\substack{{j=1,2,\dots,\infty}\\ {\tau=1,\dots,N}}}
    \, 
    e^{\bm{\zeta}^T
    \bm{\mathcal{A}}_{\mathcal{S}\mathcal{E}}\bm{\xi}_1
    }
    \, 
    e^{\frac 1 2 \bm{\xi}^T
    \bm{\mathcal{A}}_{\mathcal{E}}\bm{\xi}
    } 
 %= \frac 1 {\mathcal{I}( \bm{0})} \frac{\delta^2}{\delta \zeta_{\tau^\prime}^{ab} \, \delta \zeta_{\tau}^{a^\prime b^\prime}}\mathcal{I}( \bm{\zeta})
 \Big|_{\bm{\zeta}=\bm{0}} \, ,
\end{equation}
where $a,a^\prime=\, \uparrow$ or $\downarrow$ and $b,b^\prime=+$ or $-$.
Applying the functional derivatives and setting the sources to zero,  one 
obtains linear combinations of Keldysh correlation functions 
of the boundary site ($j=1$) of the reservoir governed by the Keldysh action $\bm{\mathcal{A}}_{\mathcal{E}}$.
We conclude that
$ \bm{\mathcal{A}}_{\mathcal{S}}^{\mathrm{eff}}$ can be identified with a matrix of Keldysh correlation functions, 
\begin{equation}
    \bm{\mathcal{A}}_{\mathcal{S}}^{\mathrm{eff}} \leftrightarrow \bm{\mathcal{G}}
\end{equation}
which we can compute and analyze independently by solving the dynamics of the reservoir.

It is worth mentioning here an important technical detail, discussed at length in Apps.~\ref{app:gate_kernels}-\ref{app:generalized_keldysh_greens_functions}.
The two-fermion Grassmann kernel associated with a general impurity-environment interaction gate, $\langle \bar\xi_0,\bar\xi_1 | U_{0,1}| \xi_0,\xi_1 \rangle$, involves not only bilinear terms in $(\bar\xi_0,\xi_0)$ and $(\bar\xi_1,\xi_1)$, but also quadratic terms $\bar\xi_0\xi_0 $ and $\bar\xi_1\xi_1 $.
For example, the Grassmann kernel of the XY gate in Eq.~\eqref{eq_xyintgate} reads (see App.~\ref{app:gate_kernels})
\begin{multline}
\label{eq_xyintgatekernel}
\bra{\bar{\xi}_0,{\bar{\xi}_1}} U_{0,1}\ket{\xi_0,{\xi}_1} 
\; = \;   \cos(\mathcal{J}_x-\mathcal{J}_y) \;
e^{\bar{\xi}_0\xi_0+{\bar{\xi}}_1{\xi}_1} \\
\exp\Big[ i\frac{t_x+t_y}{1+t_xt_y}  (\bar{\xi}_0 \xi_{1}-\xi_0\bar{\xi}_{1}) + i\frac{t_y-t_x}{1+t_xt_y} ( \xi_0 \xi_{1} -\bar{\xi}_0\bar{\xi}_{1}) \\
- 2\frac{t_xt_y}{1+t_xt_y} (\bar{\xi}_0\xi_0+\bar{\xi}_{1}\xi_{1})\Big] \, ,
\end{multline}
where we introduced a shorthand notation $ t_{x,y}\equiv \tan(\mathcal{J}_{x,y})$.
Terms in the third line in this equation provide additional quadratic contributions entering $\bm{\mathcal{A}}_{\mathcal{S}}$ and $\bm{\mathcal{A}}_{\mathcal{E}}$
in Eq.~\eqref{eq:grassmannpi}.
Thus, in particular, $\bm{\mathcal{A}}_{\mathcal{E}}$ differs from the physical Keldysh action of the environment, i.e. from the one that arises from the interactions between environment degrees of freedom only. 
The additional single-site gate acting on the boundary site is generally non-unitary~\footnote{This can be understood considering that 
$\bm{\mathcal{A}}_{\mathcal{E}}$
is obtained by setting $(\bm{\zeta}=\bm{0})$ in the full impurity+bath system, which corresponds to projecting the impurity fermion to its vacuum state $(\bm{\zeta}=\bm{0})$ at all times.
The resulting evolution of the environment thus becomes non-unitary.
}.
{While this occurrence does not affect the possibility of an efficient numerical evaluation of the {\it exact} IM, it however considerably complicates its analytical analysis via Keldysh correlation functions from Eq.~(\ref{Eq:relation_exponent_derivatives}).}

We note that this complication disappears in two important cases. First, when the interaction Hamiltonian is the tensor product of an operator acting only on the impurity and one acting only on the environment: in our model, this is realized in the Ising limit $\mathcal{J}_y=0$.
In this case, terms in the third line of Eq.~\eqref{eq_xyintgatekernel} vanish~\footnote{In other words, the non-unitary gate acting on site $j=1$ of the bath reduces to a trivial c-number $\cos(\mathcal{J}_x) \mathds{1}_1$.}.
The exact IM of this model has been analyzed in Ref.~\cite{lerose2021}, for the special case $\rho_{\mathcal{E}}\propto\mathds{1}$; below, we extend this study with particular focus on the quantum-critical state of the environment.
Second, for arbitrary interactions, the above complication disappears in the Trotter limit $\delta t\to0$.
In our model, Eq.~\eqref{eq_xyintgatekernel}, terms in the third line are of higher order in $\delta t$, and in the Trotter limit one recovers the standard textbook path-integral expression 
\begin{equation}
    \bra{\bar{\xi}_0,{\bar{\xi}_1}} U_{0,1}\ket{\xi_0,{\xi}_1} 
  =    
e^{\bar{\xi}_0\xi_0+{\bar{\xi}}_1{\xi}_1 - i \delta t H_{0,1}(\bar{\xi}_0,\bar{\xi}_1,\xi_0,{\xi}_1) }
+ \mathcal{O}(\delta t^2).
\end{equation}
 This is associated with the continuous-time dynamics of the fermionic local Hamiltonian~\eqref{eq_kitaev}, which by construction only couples $(\bar\xi_0,\xi_0)$ and $(\bar\xi_1,\xi_1)$ [second line in Eq.~\eqref{eq_xyintgatekernel}].
 In these two cases, the exact influence action $\bm{\mathcal{G}}$ is composed of Keldysh correlation functions of the boundary site of the bare unitarily-evolving environment.
 Away from these limits, one finds $\bm{\mathcal{G}}$ by solving for Keldysh correlation functions of a modified non-unitary environment Floquet evolution $\widetilde{\mathcal{U}}_\mathcal{E}$, which differs from Eq.~\eqref{eq_UER} by the presence of an extra non-unitary gate acting on site $j=1$ [see the illustration in Fig.~\ref{fig:correlation_diagram} in App.~\ref{app:fd_to_cf}]. 
 
 The resulting exact expression of the influence action of our model reads
\begin{widetext}
\begin{equation}
\label{eq:influence_form_maintext}
\mathcal{I}[\bm{\zeta}]  \; = \;  \mathcal{N}\; e^{
\gamma\sum_{\tau}
(\zeta^{\uparrow +}_\tau \zeta^{\downarrow +}_\tau + \zeta^{\uparrow -}_\tau \zeta^{\downarrow -}_\tau)}\times \exp\Bigg( \frac 1 2  \sum_{0\leq\tau^\prime ,\tau < T}
{\zeta}_\tau^T
{\mathcal{G}}_{\tau,\tau^\prime}
{\zeta}_{\tau^\prime}\Bigg),
\end{equation}
\end{widetext}
where
$\mathcal{G}_{\tau,\tau^\prime}\equiv \big(
\bm{\mathcal{G}}
\big)_{\tau,\tau^\prime}^{\uparrow/\downarrow \pm,\uparrow/\downarrow \pm}$ 
are $4\times 4$ matrix blocks of Keldysh correlation functions, $T$ denotes the number of discrete time steps,
and
\begin{align}
\gamma &= %\frac{1+\cos{2\mathcal{J}_x} \cos{2\mathcal{J}_y}}{\cos{2\mathcal{J}_x} + \cos{2\mathcal{J}_y}}, 
\frac{1-t_x t_y}{1+t_x t_y}
= \frac{\cos(\mathcal{J}_y+\mathcal{J}_x)}{\cos(\mathcal{J}_y-\mathcal{J}_x)} \, ,
\\
\mathcal{N} &=\Big(\cos(\mathcal{J}_y+\mathcal{J}_x)\cos(\mathcal{J}_y-\mathcal{J}_x)\Big)^{2T}\, .
\end{align}

We finally remark that the Grassmann form of the IM can be straightforwardly connected to its tensor form in a computational basis, as defined e.g. in Eq.~\eqref{eq:IM_operator}.
Namely, the $2^{4T}$ coefficients of the functional $\mathcal{I}[\bm{\zeta}]$ in the basis of Grassmann monomials can be directly identified with the $2^{4T}$ coefficients of the tensor in a  computational basis, corresponding to a given ordering on the Keldysh contour.

\section{Temporal Entanglement}
\label{Sec:temporal_entanglement}

In this Section, we study the scaling of temporal entanglement entropy (TE), which, following Refs.~\cite{lerose2020,lerose2021}, we define as the bipartite von Neumann entanglement entropy of the IM viewed as a fictitious wavefunction on the Keldysh contour. This quantity gives a measure of the efficiency associated with the MPS representation of the IM, similar to how conventional spatial entanglement determines the feasibility of approximating  many-body wave functions with MPS.  We will be concerned with large enough or infinite-size reservoirs to avoid the high temporal entanglement of finite environments associated with recurrences~\cite{lerose2022overcoming}.

As a central result in this Section, we show that TE scales favorably throughout the parameter space for a broad set of Gaussian initial states of the reservoir. Specifically, for finite-temperature as well as generic non-equilibrium initial quasiparticle distributions, TE saturates to a finite value at long evolution times ({\it area law}). For quantum-critical states at zero temperature we find a logarithmic violation of the area law.

We compute scaling of TE by mapping the exact form of the IM in Eq.~(\ref{eq:influence_form_maintext}) to a many-body wave function in a temporal Fock space and using standard Gaussian techniques~\cite{Latorre_2009}. 
The exact form of the IM can be either obtained through a direct evaluation of the matrix in Eq.~\eqref{eq_influenceaction}, or through the computation of Keldysh correlation functions of the bath, as described in Sec.~\ref{subsec_method}.
In order to obtain insights into the relation between the temporal and spatial correlations (see below), we follow here the latter approach. This is furthermore advantageous since it allows for computations with critical initial states (such as the Fermi sea) and is, in principle, applicable in the thermodynamic limit~\footnote{Certain Gaussian states relevant to this work, such as the Fermi sea,  have a degenerate representation as Gaussian Grassmann kernels, which hinders a direct application of the approach relying on a direct evaluation of the path-integral, Eq.~({\ref{eq:influence_form_maintext}}).}.

Temporal entanglement entropy of the IM in Eq.~\eqref{eq:influence_form_maintext} associated with a bipartition of degrees of freedom in time (usually into two intervals $A=[0,\tau]$, $\bar A=[\tau+1,T-1]$), is defined as $$S(\tau,T)=-\mathrm{Tr}_{ A} \rho_A \log \rho_A.$$ Here, $\rho_A$ is the reduced density matrix $\rho_A=\mathrm{Tr}_{\bar A} 
\frac{|\mathcal{I}\rangle\langle \mathcal{I}|}{\langle\mathcal{I}| \mathcal{I}\rangle} $ in region $A$, computed from the (properly normalized) fermionic IM wavefunction. This wavefunction is obtained by expressing the Grassmann kernel from Eq.~\eqref{eq:influence_form_maintext} in terms of fermionic operators acting on a Fock space that is defined in the temporal domain [cf. Eq.~(\ref{eq:influence_form_maintext})]:
\begin{widetext}
\begin{equation}
\label{eq:BCS_wavefunction}
\ket{\mathcal{I}} \; \propto \;
e^{\gamma\sum_{\tau=0}^{T-1}\big((c_\tau^{\uparrow+})^\dagger(c_\tau^{\downarrow+})^\dagger+(c_\tau^{\uparrow -})^\dagger(c_\tau^{\downarrow -})^\dagger\big)}
\exp \Bigg( \frac 1 2
    \sum_{0\le\tau^\prime,\tau < T}
   \big(c_\tau^\dagger\big)^T \mathcal{G}_{\tau,\tau^\prime}  c_{\tau^\prime}^\dagger
   \Bigg)
    \Big| \emptyset \Big\rangle\,.
\end{equation}
\end{widetext}
Here, $c_\tau=(c_{\tau}^{\uparrow+}, c_{\tau}^{\uparrow-},c_{\tau}^{\downarrow+}, c_{\tau}^{\downarrow-})^T$ [cf. Eq.~\eqref{eq_zeta}] are canonical operators of four fermionic ``species'' per temporal lattice site. The species are associated with forward/backward and input/output legs in the tensor network in Fig.~\ref{fig:circuit}, respectively.

In the following we will write $S(T)$ referring to the maximum of $S(\tau,T)$ over $\tau$, which we have observed to be around $\tau \simeq T/2$. 
In the remainder of this section, we will study $S(T)$ for the two models introduced in Eqs.~(\ref{Eq:KIC_model},\ref{Eq:XY_model}), comparing in particular the qualitative features at and away from quantum criticality.

\subsection{Kicked Ising Model}
\label{Sec:Impurity_entanglement_Ising}

Substituting Eq.~\eqref{eq_zeta} into Eq.~\eqref{eq_xyintgatekernel} and setting $t_y=0$, we see that the dependence of the influence action on the variables $\zeta^{\downarrow}$ is trivial.
Furthermore, the non-trivial part of the influence action assumes the following simple form: For $\tau \geq \tau^\prime,$ one has
\begin{equation}\label{eq_influenceaction_Ising}
\mathcal{G}_{\tau,\tau^\prime} = 2\tan^2\mathcal{J}_x\begin{pmatrix}
0 & 0 & 0 & 0 \\
0 & 0 & 0 & 0\\
0 & 0 &  g_{\tau,\tau^\prime} & -g^*_{\tau,\tau^\prime}  \\
0 & 0 & g_{\tau,\tau^\prime} & -g^*_{\tau,\tau^\prime}   \\

\end{pmatrix},\end{equation} and $$\mathcal{G}_{\tau^\prime,\tau} = -\mathcal{G}_{\tau,\tau^\prime}^T.$$ We note that the IM is determined by a single function, similarly to a bath of bosonic oscillators coupling to a single impurity operator~\cite{FeynmanVernon}. Here, we find:
\begin{equation}\label{eq:trace_correlation_function_Ising}
   g_{\tau,\tau^\prime} =  \begin{cases} \mathrm{Tr}_{\mathcal{E}}\Big(\mathcal{U}_\mathcal{E}^{T-\tau}\hat{O}\mathcal{U}_\mathcal{E}^{\tau-\tau^\prime}\hat{O}\mathcal{U}_\mathcal{E}^{\tau^\prime}\rho_\mathcal{E} (\mathcal{U}_\mathcal{E}^\dagger)^T\Big) & \text{for }\tau>\tau^\prime,\\
    0 & \text{for }\tau = \tau^\prime,
    \end{cases}
    \end{equation}
    with $\hat{O} = c_1+c_1^\dagger$ and $\mathcal{U}_\mathcal{E}$ given by Eq.~\eqref{eq_UER}. 
    We stress that $g_{\tau,\tau^\prime}$ represents a temporal correlation function of the decoupled reservoir, and does not represent any actual temporal correlation of the coupled impurity+reservoir system after the impurity quench.

Next, we express the response function $g_{\tau,\tau^\prime}$ for a reservoir of $L$ sites via the eigenmodes of the Floquet operator of the kicked Ising model, $\{\big| \phi_1 \big),\dots,\big| \phi_L \big),\big|-\phi_1 \big),\dots,\big| -\phi_L \big)\}$, with quasienergies $\{\pm\phi_m\}$, as well as via the initial Gaussian density matrix $\rho_{\mathcal{E}}$ of the environment. The eigenmode operator $d_m$, which obeys
$\mathcal{U}_\mathcal{E}^\dagger d_m \mathcal{U}_\mathcal{E}=e^{-i\phi_m} d_m$, is a linear combination of operators $c_j,c_j^\dagger$, and $|\phi_m)$ is a $2L$-dimensional vector in a space spanned by  $c_j,c_j^\dagger$. We can rewrite Eq.~(\ref{eq:trace_correlation_function_Ising}) as 
\begin{equation}
\label{eq:general_KIC_correlations}
g_{\tau,\tau^\prime} 
=
\sum_{m,n}
{\bm{\mathcal{C}}}_m^* 
\Lambda_{mn}
{\bm{\mathcal{C}}}_n
e^{-i\phi_m \tau+i\phi_n\tau^\prime },
\end{equation} 
 where $\Lambda_{mn} =  \big( \phi_m| \Lambda |\phi_n\big)$ is the correlation matrix of $\rho_{\mathcal{E}}$ in the eigenmode basis, %fermion basis where the effective Hamiltonian of the kicked Ising model is diagonal: 
 \begin{equation}
     \Lambda =  \mathrm{Tr}\big(\bm{d} \cdot \bm{d}^\dagger\rho_\mathcal{E}\big),\, \bm{d}=(d_1,\hdots,d_L,d_1^\dagger,\hdots,d_L^\dagger)^T \, ,
 \end{equation}
  and $\bm{\mathcal{C}} = (\mathcal{C}_1^*,\hdots,\mathcal{C}^*_L,\mathcal{C}_1,\hdots \mathcal{C}_L)^T,$ where $\{\mathcal{C}_{m}\}$
 are coefficients of the eigenmode operators on site $j=1$ that couples to the impurity (see App.~\ref{App:Ising_corr_funcs_evaluation} for details). We note that the eigenmodes of our model form a continuous band $\phi_k$ in the thermodynamic limit $L\to\infty$, associated with fermionic quasiparticles with momentum $k\in[-\pi,0]$ scattering at the edge at $j=1$, and, in part of the parameter space, include edge modes with  $\phi_e=0,\pi$~\cite{dutta}. For simplicity, we will report data for parameter regions without edge modes, so we will write $\phi_m \mapsto \phi_k$.
 
\begin{figure}[h]
\centering
    \hspace{0mm}\subfloat{
    \includegraphics{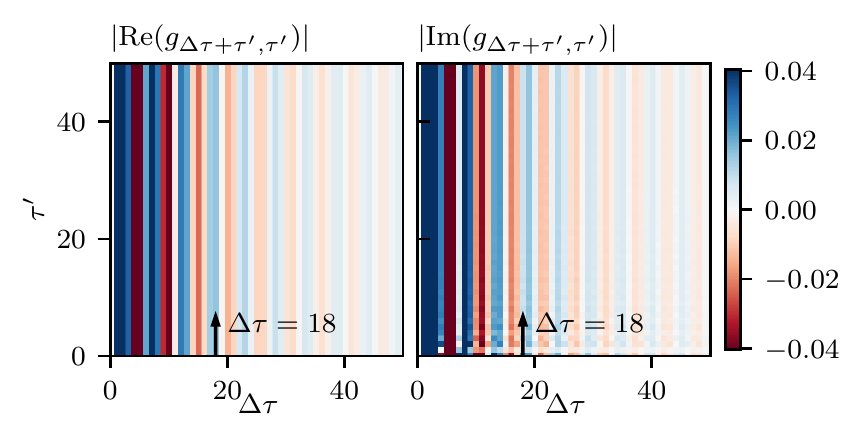}
   \label{fig:0.31_pi4}}%\hfill
    \vskip\baselineskip\vspace{-0.9cm}
     \subfloat{
    \includegraphics{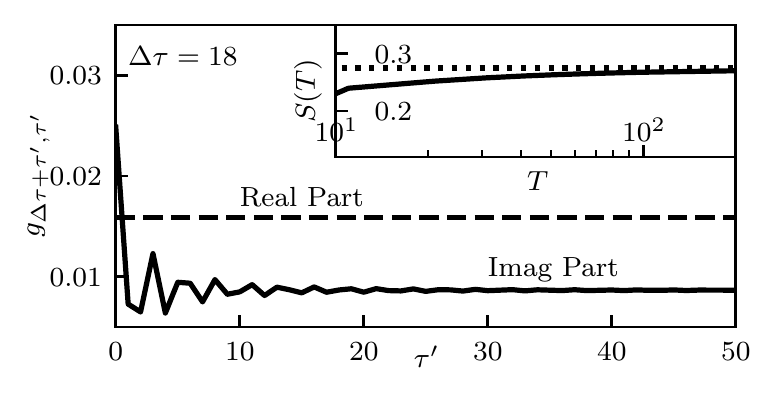}
    \label{fig:0.3_0.3}}
    \caption{Upper panels: Real and imaginary part of the response function $g_{\tau,\tau^\prime}$ for the kicked Ising model with parameters $\mathcal{J} =\varphi = 0.3$ and non-equilibrium initial state $\rho_\mathcal{E}=\prod_j \rho_j,\; \rho_j \propto e^{-\beta  \sigma^z_j},\, \beta = 10$.  Lower panel: Response function $g_{\Delta\tau+\tau^\prime,\tau^\prime}$ for fixed $\Delta\tau = 18,$ corresponding to the direction indicated by the arrows in the upper panels.
    Inset: scaling of temporal entanglement entropy.
    }
    \label{fig:non_equil}
\end{figure}

 Equation~\eqref{eq:general_KIC_correlations} allows us to analyze the asymptotic behavior of $g_{\tau,\tau^\prime}$ for $\tau,\tau^\prime\to\infty$, which determines the scaling of TE.
Let us first consider a nonequilibrium initial state $\rho_{\mathcal{E}}$ of the reservoir, resulting e.g. from a global quench.
In integrable systems like the non-interacting fermionic baths considered in this work, %Eqs.~(\ref{Eq:KIC_model},\ref{Eq:XY_model}), 
the steady state locally reached after a quench from a non-equilibrium initial state $[\rho_{\mathcal{E}},\mathcal{U}_{\mathcal{E}}]\neq0$ is  captured by a  statistical ensemble  $\rho^{GGE}_{\mathcal{E}}$ called a generalized Gibbs ensemble,
which is a nontrivial stationary state  $[\rho^{GGE}_{\mathcal{E}},\mathcal{U}_{\mathcal{E}}]=0$. It is specified by a set of generalized temperatures $\{\beta_m\}$ associated with the eigenmodes of the bath. 
As $\tau,\tau^\prime\to\infty$, the function $g_{\tau,\tau^\prime}$ approaches a time-translational invariant form $g_{\tau-\tau^\prime}\equiv g_{\Delta\tau}$ associated with $\rho^{GGE}_{\mathcal{E}}$, resulting from the inhomogeneous dephasing of off-diagonal contributions $m\neq n$ in Eq.~\eqref{eq:general_KIC_correlations}. We demonstrate this in Fig.~\ref{fig:non_equil}, where we consider time evolution following a global quench in the reservoir, from the initial state \begin{equation}
\label{eq_neqinitialstate}
    \rho_\mathcal{E} = \prod_{j>0} \rho_j,\quad \rho_j =
\frac {\exp(-\beta \, \sigma_j^z)} {2 \cosh \beta}.
\end{equation}
For the kicked Ising model, we demonstrate the saturation of $g_{\Delta\tau+\tau^\prime,\tau^\prime}$ to a finite value after a few Floquet periods. Note that information on the initial state --- and hence the full dependence on $\tau^\prime$ --- is only carried by the imaginary part of $g_{\Delta\tau+\tau^\prime,\tau^\prime}$, as can be derived from Eq.~(\ref{eq:general_KIC_correlations}) by considering the symmetries of $\Lambda_{mn},$ see App.~\ref{App:Ising_corr_funcs_evaluation} for details.
The corresponding temporal entanglement obeys an area law (see inset in Fig.~\ref{fig:non_equil}, bottom panel). Below, we will study the conditions for the presence of an area law and its violations in detail.

In light of the above discussion, we will consider initial states $\rho_{\mathcal{E}}$ that are stationary in the following.
Then $
\Lambda_{mn}= \delta_{m,n} \frac 1 {e^{\beta_m\phi_m}+1}$ in Eq.~\eqref{eq:general_KIC_correlations}  becomes a diagonal matrix. Thus, in the thermodynamic limit,

\begin{multline}\label{eq:G_lesser_finiteT}
g_{\tau,\tau^\prime} = g_{\Delta\tau}
= \int_{-\pi}^0 \frac{dk}{2\pi} |\mathcal{C}_k|^2 \Bigg[\cos\big(\phi_k\Delta\tau\big)\\ -i\sin\big(\phi_k\Delta\tau\big) \tanh\bigg(\frac{\beta_k\phi_k}{2}\bigg)\Bigg].
\end{multline}

{The influence action, defined by Eqs.~(\ref{eq_influenceaction_Ising},\ref{eq:G_lesser_finiteT}), is thus determined by the spectral density $J(\omega)=\int \frac{dk}{2\pi} |\mathcal{C}_k|^2 \delta(\omega-\phi_k)$ and by generalized temperatures $\beta_k$. This action has a form reminiscent of the Feynman-Vernon  continuous-time IF for a bosonic bath in thermal equilibrium~\cite{FeynmanVernon}.}

\begin{figure}[h]
    \centering
    \includegraphics{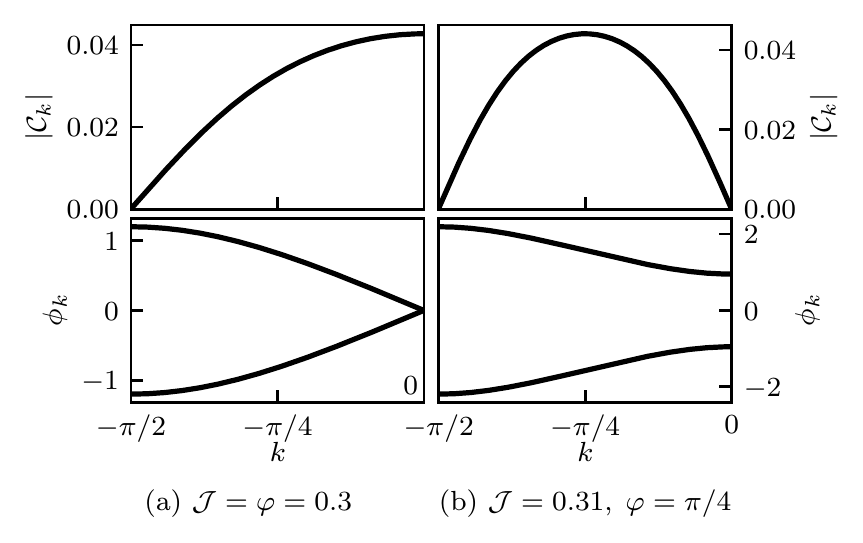}
    \caption{Spectrum $\phi_k$ of the Floquet Hamiltonian and function $|\mathcal{C}_k|$ for the kicked Ising model in the (a) gapless and (b) gapped phase. }
    \label{fig:environment_scaling}
\end{figure}

\begin{figure}[h]\center
 \subfloat{
    \includegraphics{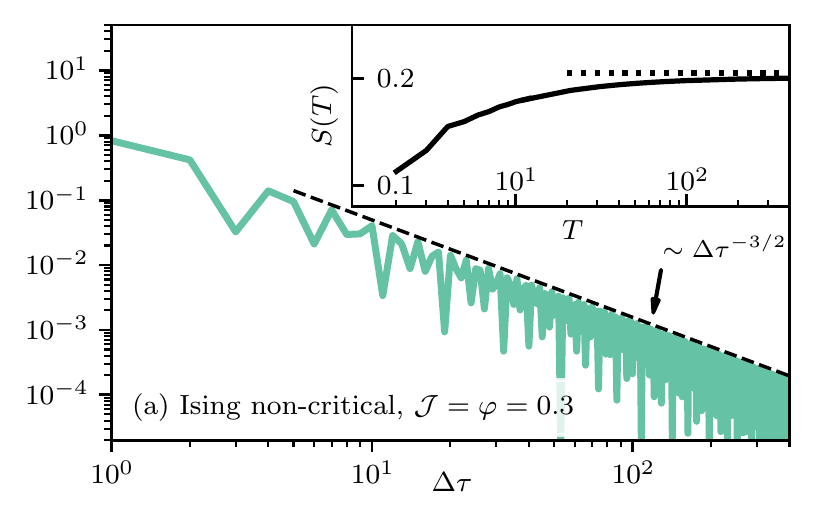}
       \label{fig:scaling_Ising_IT}}
     \vskip\baselineskip\vspace{-0.9cm}
 \subfloat{
        \includegraphics{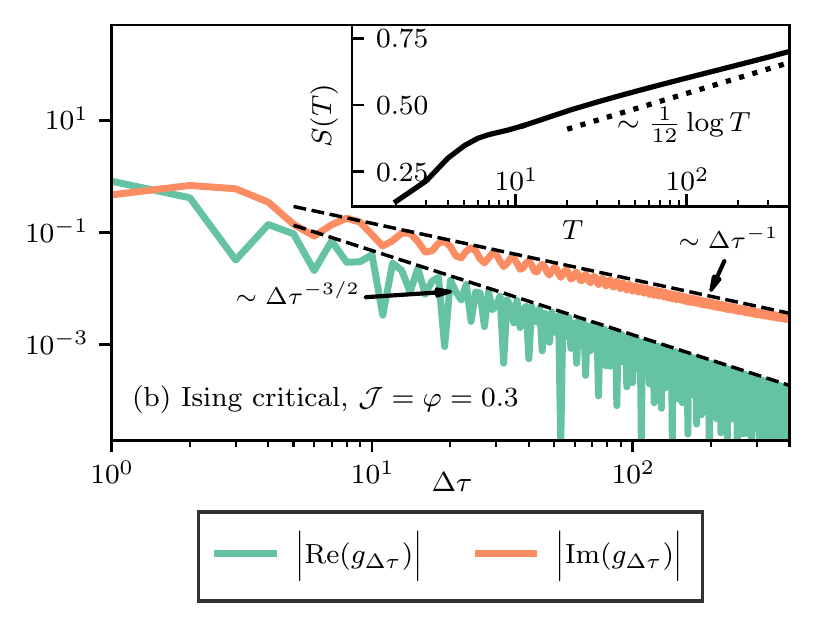}
         \label{fig:scaling_Ising_bog}}
   \caption{Response functions and temporal entanglement entropy scaling for the {kicked Ising model} with parameters $\mathcal{J}=\mathcal{\varphi} = 0.3$ for (a) the infinite temperature initial state and (b) the critical initial state. Dashed and dotted lines serve as guide to the eye to identify power laws and logarithmic/saturation behavior, respectively.
   }
   \label{fig:Ising_scaling}
\end{figure}

{\sl Long-time scaling}---
Next, we investigate the long-time behavior of the function (\ref{eq:G_lesser_finiteT}) which controls the TE properties, starting with the case when parameters $\mathcal{J}_x$, $\varphi$ correspond to a gapped phase. The simplest state, previously considered in Ref.~\cite{lerose2021} is that of
an infinite-temperature initial state, $\rho_\mathcal{E} = \otimes_{j\ge 1}\rho_j$, $\rho_j = \mathds{1}_j/2,$
that corresponds to $\beta_k=0$. In this case Eq.~\eqref{eq:G_lesser_finiteT} is purely real, reducing to 
%corresponds to $\kappa_i = 0\quad  \forall i,$ such that: 
\begin{equation}
    \label{eq:G_lesser_IT}
%-iG^<_{\Theta\Theta} (\tau_2,\tau_1)
g_{\Delta\tau}=\int_{-\pi}^0 \frac{dk}{2\pi}|\mathcal{C}_k|^2\cos\Big(\phi_k \Delta\tau\Big).
\end{equation}
In the absence of edge modes this function oscillates and decays as $|\Delta\tau|^{-3/2}$ as $|\Delta\tau|\to\infty$~\cite{lerose2021}. 
These  temporal correlations arise from the quasiparticles residing near the band edges $k\simeq0$ or $-\pi$.
In the vicinity of e.g. $k=0$ point, we expand $\phi_k \sim \phi_0 + a k^2$ and use the fact that $\mathcal{C}_k\sim |k|$, as illustrated in Fig.~\ref{fig:environment_scaling}a. Then, the contribution of the $k\simeq0$ region can be computed using the saddle point approximation: 
\begin{equation}
\label{eq_gasymptoticnoncritical}
g_{\Delta\tau} \, \sim \,
    \int_{-\infty}^0 \frac {dk} {2\pi}
    k^2 \cos[(\phi_0+ak^2)\Delta\tau] \, \sim \,
    \frac {\cos(\phi_0 \, \Delta\tau+\tfrac{\pi}{4})}{|a\Delta\tau|^{3/2}}  .
\end{equation}
An analogous contribution is found for $k\simeq -\pi$.
This  decay is fast enough to produce a long-time saturation of TE, as illustrated in Fig.~\ref{fig:Ising_scaling}a. The area law of TE has been established in Ref.~\cite{lerose2021} by explicitly constructing gapped quasilocal parent Hamiltonians for the IM wavefunction~\cite{Its08,Its09}.
The effect of edge modes is to add a non-decaying contribution to $g_{\Delta\tau}$, associated with long-lived local memory at the edge of the environment. Such contributions change the saturation value of TE, which is found to be discontinuous at phase boundaries.
The TE area-law, however, is obeyed throughout the phase diagram.

In gapped phases, the asymptotic behavior (\ref{eq_gasymptoticnoncritical}) holds also at finite temperature $\beta_k\le\infty$, as the imaginary part of $g_{\Delta\tau} $ in Eq.~\eqref{eq:G_lesser_finiteT} behaves similarly to the real part. 
Thus TE area-law persists for arbitrary initial states where $\beta_k$ is a smooth function of quasienergy; in particular, this includes the ground state of the Floquet Hamiltonian, $\beta_k\to\infty$, where {Eq.~\eqref{eq:G_lesser_finiteT}}
reduces to
\begin{equation}\label{eq:G_lesser_eigenstate:Ising}
%-iG^<_{\Theta\Theta} (\tau_2,\tau_1)=
g_{\Delta\tau}=
\int_{-\pi}^0
\frac{dk}{2\pi} |\mathcal{C}_k|^2 e^{-i\phi_k\Delta\tau}. 
\end{equation}
Saddle-point approximation gives the behavior analogous to Eq.~\eqref{eq_gasymptoticnoncritical}, with the cosine is replaced by the corresponding complex phase.

Turning to the critical lines, $\mathcal{J}_x=g$ or $\pi/2-g$, we note that quasienergy gap closes at either $k=0$ or $k=-\pi$ (or both), implying that the corresponding saddle point in the integral (\ref{eq:G_lesser_finiteT}) disappears. The other band edge still leads to the asymptotic behavior of $g_{\Delta\tau}$ described above. To analyze the contribution of the gapless point at $k=0$ to the imaginary part of $g_{\Delta\tau}$, we first note that  $\mathcal{C}_k\sim \mathrm{const}$ and $\phi_k \sim  v |k|$, as illustrated in Fig.~\ref{fig:environment_scaling}b. At non-zero temperature, this, combined with the vanishing of the argument of the hyperbolic tangent {as $\phi\to 0$}, results in an asymptotic 
scaling $g_{\Delta \tau} \sim \Delta \tau^{-2}$, which is subleading compared to the contribution $ \sim \Delta \tau^{-3/2}$ of the gapped band edge.

However, a qualitatively different behavior is found at zero temperature, $\beta_k\to\infty$. In this case, the hyperbolic tangent equals $1$, and hence the contribution of the gapless point of the spectrum $k\simeq 0$ takes the form
\begin{equation}
\label{eq_gasymptoticcritical}
g_{\Delta\tau} \, \sim \,
    \int_{-\infty}^0 \frac {dk} {2\pi} \; 
    \mathrm{const} \; e^{iv|k|\Delta\tau} \, \sim \,
    \frac {i}{v\Delta\tau}  
\end{equation}
(an analogous behavior is found when the gapless point is $k\simeq -\pi$).
This slower decay of the correlation function is analogous to that of {\it spatial} correlations in the critical initial state.
Thus, one expects a logarithmic violation of the area law. 

The occurrence of this logarithmic scaling is supported by the following universality argument.
The leading behavior of the spatiotemporal correlation functions of a critical lattice model at low temperature, such as the model investigated here, is captured by conformal field theory (CFT)~\cite{mussardo2010statistical}.
In a $(1+1)$-dimensional CFT, local operators $O(x,t)$ can be split as $O_L(x,t)+O_R(x,t)$, where the two  components belong to the two chiral sectors of the CFT (left and right movers), and hence satisfy $O_L(x,t)=e^{itH}O_L(x,0)e^{-itH}=O_L(x+vt,0)$ and $O_R(x,t)=e^{itH}O_R(x,0)e^{-itH}=O_R(x-vt,0)$, where $v$ is the ``speed of light'' of the model. Due to the open boundary condition at $x=0$, left movers scatter into right mover partners as they hit the boundary.
Thus, the Keldysh temporal correlation functions of interest for the IM, can be recast to corresponding spatial correlation functions in the initial state:
\begin{equation}
    \langle O(0,t) O(0,t^\prime) \rangle = 
    \langle \tilde{O}(vt,0) \tilde{O}(vt^\prime,0) \rangle  \, ,
\end{equation}
where $\tilde O$ is the local operator obtained from $O$ by flipping the right-mover component $R \mapsto L$.
For the operators and critical initial state relevant to Eq.~\eqref{eq:trace_correlation_function_Ising}, we obtain
\begin{equation}
    \langle O(0,t) O(0,t^\prime) \rangle  
 \propto \frac 1 {v(t-t^\prime)} \, ;
\end{equation}
more generally, arbitrary multi-time correlation functions entering the influence action can be recast to corresponding multi-point correlation functions in the initial state~\footnote{Note that operators located on opposite branches of the Keldysh contour are mapped to operators acting on the same point in space at $t=0$.}.
It is thus natural to conclude that the temporal entanglement content of the IM is equivalent to the spatial entanglement content of the critical initial state, which is, to leading order, universal and solely fixed by the central charge of the CFT~\cite{Vidal03,Calabrese04} (here $c=1/2$): 
\begin{equation}
\label{eq_TElogIsing}
S(T)\sim \frac 1 {12} \log T  \, .
\end{equation}
Our numerical results confirm this prediction, see Fig.~\ref{fig:Ising_scaling}b.
We envision that this temporal entanglement scaling law could be shown more rigorously by applying the CFT techniques developed by Calabrese and Cardy~\cite{Calabrese04} to the boundary CFT emerging in the present IM context.

\begin{figure}[h] \center
  \subfloat{
        \includegraphics{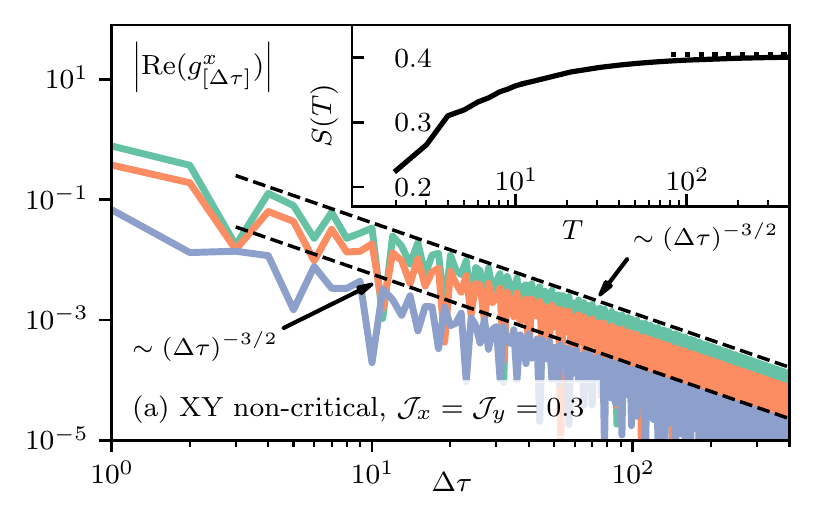}
        \label{fig:scaling_xy_IT}}
     \vskip\baselineskip\vspace{-0.9cm}
         \subfloat{
        \includegraphics{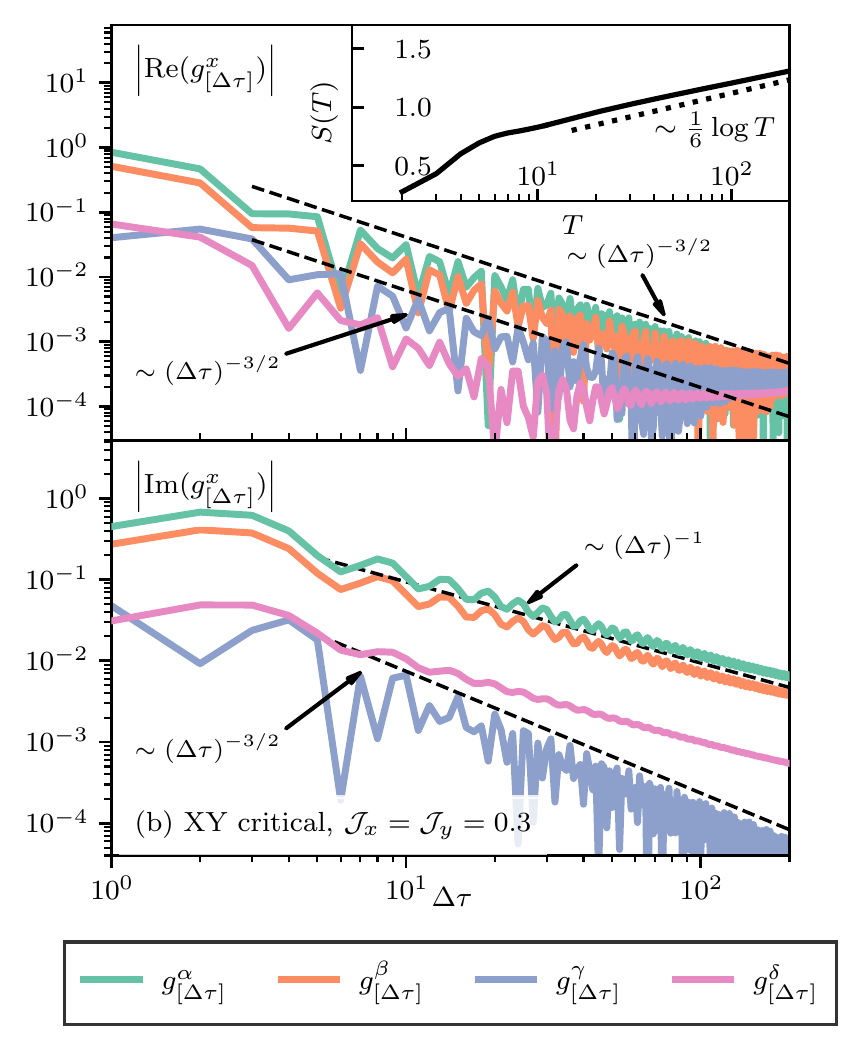}
        \label{fig:scaling_xy_bog_imag}}
   \caption{``Temporal bulk'' response functions and temporal entanglement entropy scaling for the {trotterized XY model} with parameters $\mathcal{J}_x=\mathcal{J}_y=0.3,\, T=800,$ for (a) the infinite temperature initial state and (b) the critical initial state. For the infinite temperature initial state in (a), all four functions $g^x_{[\Delta\tau]}$ are purely real. For better distinguishability of the curves, we have multiplied the functions $g^{\alpha,\beta,\gamma,\delta}$ with factors of $1.0,0.5,0.25,0.125,$ respectively. Dashed and dotted lines serve as guide to the eye to identify power laws and logarithmic/saturation behaviour, respectively.
   }
   \label{fig:xy_scaling}
\end{figure}

\subsection{Trotterized XY model} \label{subsec_kxy}

In the above analysis of the kicked Ising model, where no complication related to effective non-unitary environment evolution arises, the non-trivial part of the influence action, $\mathcal{G}_{\tau,\tau^\prime}$, 
has a simple form, see Eq.~\eqref{eq_influenceaction_Ising} and Eq.~(\ref{eq:general_KIC_correlations}).
The generalized form of the influence action for the trotterized XY model ($\varphi = 0$) reads for $\tau > \tau^\prime$:
\begin{equation}
\label{eq_influenceaction_XY}
\mathcal{G}_{\tau,\tau^\prime}
= \frac{2}{T_{xy}^2}
\begin{pmatrix}
t_yt_y \begin{pmatrix}
g_{1} & -g_{2}^* \\
g_2 & -g_1^*
\end{pmatrix} &
t_x t_y \begin{pmatrix}
g_{3} & \ g_{4}^* \\
g_4 & \ g_3^*
\end{pmatrix}\\
t_xt_y \begin{pmatrix}
g_{5} & \ g_{6}^* \\
g_6 & \ g_5^*
\end{pmatrix} &
t_xt_x \begin{pmatrix}
g_{7} & -g_{8}^* \\
g_8 & -g_7^*
\end{pmatrix}
\end{pmatrix}
\Bigg|_{\tau,\tau^\prime},
\end{equation}
and $\mathcal{G}_{\tau^\prime,\tau} = - \mathcal{G}^T_{\tau,\tau^\prime}$ as before.
Tuning parameters to the critical line $\mathcal{J}_x=\mathcal{J}_y$, where $U(1)$ symmetry is realized, Eq.~(\ref{eq_influenceaction_XY}) reduces to only four independent functions:
\begin{align}
g^\alpha &\equiv -g_1 =  g_7\\
g^\beta &\equiv g_2 =  g_8\\
g^\gamma &\equiv g_3 =  g_5\\
g^\delta &\equiv -g_4 =  g_6.
\end{align}

As a consequence of the effective non-unitarity of the environment dynamics, the four functions $g^{\alpha,\beta,\gamma,\delta}$ cannot be expressed as compactly as Eq.~(\ref{eq:G_lesser_IT}) and do not allow an analytical analysis as simple.  We have therefore performed a numerical study of their behavior. 
We further note that the effective non-unitarity spoils time-translational invariance of the influence action $\mathcal{G}_{\Delta\tau+\tau^\prime,\tau^\prime}$, which now explicitly depends on both $\Delta\tau$ and $\tau^\prime$, as it is evident from the interpretation of $\mathcal{G}$ as Keldysh correlation functions: the initial state $\rho_{\mathcal{E}}$ is not stationary under the effective non-unitary evolution. 
For large $T$, however, approximate time-translation invariance is restored in the ``temporal bulk'', $0\ll \tau,\tau^\prime\ll T.$ Since we are ultimately interested in the long-time limit $T\to \infty$, and since the correlations across a cut in the bulk are expected to be dominated by these bulk parts of the influence action, we focus our attention on the temporal bulk response functions
$$\mathcal{G}_{[\Delta\tau]}\equiv \mathcal{G}_{\frac{T+\Delta\tau}{2},\frac{T-\Delta\tau}{2}},
$$ which minimize finite-size effects due to the temporal boundaries.
In App.~\ref{Sec:temp_finite_size}, we present a more detailed discussion of temporal bulk vs boundary effects, supported by an analysis of our data.

As in the study of the kicked Ising model above (Fig.~\ref{fig:Ising_scaling}), we report in Fig.~\ref{fig:xy_scaling}
the results for infinite- and zero-temperature initial states.
For the infinite-temperature initial state, all response functions exhibit a power-law decay of $\sim | \Delta\tau|^{-3/2},$ resulting in an area-law scaling of temporal entanglement $S(T)$ throughout the phase diagram, see e.g. Fig.~\ref{fig:xy_scaling}a. 
Tuning the reservoir to quantum criticality, i.e. taking $\rho_{\mathcal{E}}$ as the ground state of the (gapless) Floquet Hamiltonian for $\mathcal{J}_x=\mathcal{J}_y$ (or $\mathcal{J}_x=\pi/2-\mathcal{J}_y$), the long-time scaling of correlation functions and TE changes qualitatively.
Similarly to the kicked Ising model, the imaginary parts of the response functions have components that decay as $\sim|\Delta\tau|^{-1},$ while all real parts show the same type of decay $\sim|\Delta\tau|^{-3/2}$ as for the infinite temperature initial state. (The temporal finite size effect in panel Fig.~\ref{fig:scaling_xy_bog_imag} are expected from the analysis in App.~\ref{Sec:temp_finite_size} and could be remedied by time-evolving the state to much larger values of $T.$)
As we remarked above for the kicked Ising model, these ``critical-like'' long-ranged correlations in time in the imaginary part are responsible for a logarithmic violation of the TE area-law.
In this case the emergent CFT (free Dirac fermions) has central charge $c=1$, leading to
\begin{equation}
S(T)\sim \frac 1 {6} \log T  \, .
\end{equation}
Our numerical results confirm this prediction, see Fig.~\ref{fig:xy_scaling}b.

We observed that the absolute value of $S(T)$ for the trotterized XY chain is exactly twice as large as that of the kicked Ising model with corresponding parameters $\mathcal{J}_y \leftrightarrow \varphi$ throughout the parameter space. 
This relation between temporal entanglement entropies of the two models appears to parallel the analogous known relation between their spatial entanglement entropies~\cite{Igl_i_2008}, and can be traced back to the known exact relation between the models described in Sec.~\ref{subsec_models}.

\section{MPS Representation}
\label{sec:mps_conversion}

The IM wave function described by Eq.~(\ref{eq:BCS_wavefunction}) has a BCS-like form, regardless of the reservoirs properties and size. In this Section, we outline how it can be converted to  MPS form. 
The results of the previous Section on TE suggest that an efficient conversion procedure is possible for general infinite-size reservoirs. Once the IM is represented as a MPS with a moderate bond dimension, we can analyze dynamics of an interacting impurity by contracting the time-local impurity evolution operators with the IM MPS, cf. Fig.~\ref{fig:intro}b. 

The presented algorithm is an extension of the method proposed by Fishman and White in Ref. \cite{Fishman2015MPS} for particle-conserving fermionic systems. Our aim is to construct $|\mathcal{I}\rangle$ by acting with local two-site unitary operations on the vacuum $|\emptyset\rangle$.
A set of suitable operations can be found at the single-particle level by rotating the system to a basis of almost localized natural orbitals. Each such rotation can be decomposed into a finite sequence of local rotations involving pairs of neighboring (temporal) sites.
Interpreting these local rotations as two-body local unitary gates in the Fock space, one obtains an approximate expression of $|\mathcal{I}\rangle$ as a finite-depth unitary circuit applied to $|\emptyset\rangle$.
The contraction of such a circuit can then be performed with standard MPS techniques, and the resulting maximum bond dimension can be upper bounded in terms of the maximum circuit depth, which, in turn, is determined by the maximum localization length of the approximate natural orbitals.

\subsection{Approximate diagonalization of the correlation matrix}
\label{Sec:Appr_diag}

In the following, it will be convenient to relabel the fermions in the temporal Fock space as \begin{equation}
\label{eq:chain_def}
    c_1=c_0^{\uparrow+},c_2=c_0^{\uparrow-},c_3=c_0^{\downarrow+},\hdots, c_{4T}=c_{T-1}^{\downarrow-}. \end{equation}
We will refer to the new indices as site indices of a ficticious chain of spinless fermions of length $4T$.
The Gaussian IM is uniquely identified by the two-point correlations between these sites,
\begin{equation}\label{eq:original_corr_matrix}
  \Lambda =
  \frac{\langle \mathcal{I}| \bm{\eta}\bm{\eta}^\dagger|\mathcal{I}\rangle}{\braket{\mathcal{I}|\mathcal{I}}}
  ,\quad \bm{\eta}
  = \big(c_1, c_1^\dagger,\hdots,c_{4T}, c_{4T}^\dagger\big)^T,\end{equation}

The correlation matrix $\Lambda$ has size $(8T)\times (8T).$
Its entries can be computed from Eq.~\eqref{eq:BCS_wavefunction} using standard techniques. The subblocks corresponding to a pair of sites $(i,j)$ have the structure
\begin{equation}
    \Lambda_{ij}=\begin{pmatrix}
\frac{\bra{\mathcal{I}} c_ic_j^\dagger\ket{\mathcal{I}}}{\braket{\mathcal{I}|\mathcal{I}}} & \frac{\bra{\mathcal{I}} c_ic_j\ket{\mathcal{I}}}{\braket{\mathcal{I}|\mathcal{I}}}
\\
\frac{\bra{\mathcal{I}} c^\dagger_ic^\dagger_j\ket{\mathcal{I}}
}{\braket{\mathcal{I}|\mathcal{I}}}
& \frac{\bra{\mathcal{I}} c^\dagger_ic_j\ket{\mathcal{I}}
}{\braket{\mathcal{I}|\mathcal{I}}}
    \end{pmatrix} = \begin{pmatrix}
    a & b \\
    -b^* & \delta_{i,j} - a
    \end{pmatrix}.
    \label{eq:struc_corr_matrix}
\end{equation}
Since $\ket{\mathcal{I}}$ is a pure state, half of the eigenvalues of $\Lambda$ are equal to $0$, and the other half are equal to $1$.\footnote{ 
This fact can be used to enforce normalization of $\ket{\mathcal{I}}$ and restore its original norm at the end.}

\begin{figure*}
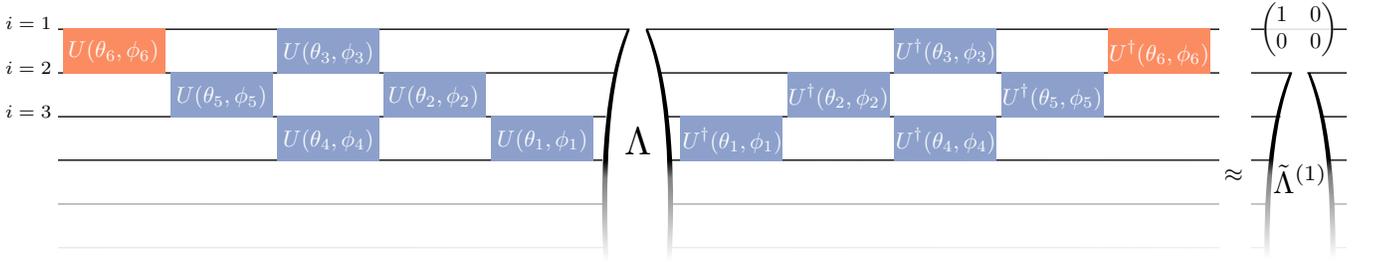

\centering
     \begin{overpic}[width=\textwidth]{figures/algo}
    \put(0,17.3){\scriptsize $i=1$}
     \put(0,14){\scriptsize $i=2$}
     \put(0,10.7){\scriptsize $i=3$}
      \put(90.4,6){$\approx$}
    \end{overpic}
    \caption{By applying the unitary two-site single-body rotations from one cycle of the algorithm (Steps 1-2), the original correlation matrix $\Lambda$ can locally be brought into diagonal form, where it describes the vacuum state. Each horizontal line corresponds to one site of the chain defined in Eq.~(\ref{eq:chain_def}). Blue gates symbolize Givens rotations, orange gates represent Bogoliubov rotations.}
 \label{Fig:corr_diag} 
\end{figure*}

Let us first summarize the approximate diagonalization procedure.
The central idea is to iteratively identify a basis of almost-localized ``natural orbitals'' for the state $\ket{\mathcal{I}}$, and hence to reconstruct $\ket{\mathcal{I}}$ through a sequence of local rotations (i.e., a low-depth quantum circuit) applied to a product state --- similarly to previous work on simulating fermionic many-body systems with quantum computers~\cite{ortiz01quantumalgorithms,verstraete09quantumcircuits,wecker15solving,jiang18quantumalgorithms}.
We thus start by finding a subsystem containing $n_1 \leq 4T$ sites, whose reduced correlation matrix  \begin{equation}
    C^{(1)}\equiv\Lambda\big|_{1,\hdots,2n_1}^{1,\hdots,2n_1},
    \label{eq:subcorr_matrix}
    \end{equation}
    consisting of the first $2n_1$ rows and columns of $\Lambda$, has its smallest eigenvalue $\mu_1$ fulfilling $\mu_1 < \epsilon$ for a fixed small threshold value $\epsilon \ll 1$.  This is done by starting with a small tentative subsystem size $n_1$ and progressively increasing it until the criterion is met. 
Thus, the considered subsystem contains an approximate eigenmode of the full state $\Lambda$ that is created by the fermionic operator
\begin{equation}\label{eq:eigenmode}
\gamma_1 = u_1^1 c_1 + v_1^1 c_1^\dagger + \hdots + u_1^{n_1} c_{n_1}  + v_1^{n_1} c_{n_1}^\dagger.\end{equation} 
The coefficients are the entries of the normalized eigenvector that corresponds to the identified eigenvalue $\mu_1\approx 0$ of the reduced correlation matrix $C^{(1)}:$ 
\begin{equation*}
\bm{b}_1 =(u_1^1,v_1^1,\hdots,u_1^{n_1},v_1^{n_1})^T.\end{equation*}

In order to isolate the localized mode, we rotate into a new basis, in which only one of the first two coefficients in Eq.~(\ref{eq:eigenmode}) is nonzero, and all other coefficients are zero.
This unitary change of basis ${U}_1$ can be determined at the single-particle level in terms of a sequence of $2n_1-2$ nearest-neighbor rotations.
In particular, the BCS-form of Eq.~({\ref{eq:BCS_wavefunction}}) implies that, by performing an orthogonal rotation, the IM can be written as a tensor product of Bell pairs: $ \ket{\mathcal{I}} \propto \otimes_{k>0} (1+\omega_k \gamma_k^\dagger\gamma_{-k}^\dagger)\ket{\emptyset}.$ In order to single out and localize one mode, one thus needs to combine a real (number conserving) rotation with a Bogoliubov transformation to disentangle its Bell partner.
The real transformation can be expressed as a set of so-called Givens rotations~\cite{Troyer2015},  $$e^{-\theta(c^\dagger_jc_{j+1}-h.c.)}$$ acting on pairs of neighboring sites. 
The Bogoliubov transformation takes the form $$e^{\theta (c_j^\dagger c_{j+1}^\dagger-h.c.)}.$$ We assemble ${U}_1$ using these elementary rotation matrices in such a way that 
\begin{equation}
{U}_1\bm{b}_1 = (0,v,0,\hdots,0)^T,
\label{eq:U_effect}
\end{equation} with $|v|=1.$
Having obtained ${U}_1$, one finds that ${U}_1 \Lambda {U}_1^\dagger$ --- where it is understood that ${U}_1$ is extended such that it acts trivially on the complement of the subsystem --- is, up to an error $\epsilon$, diagonal in the first two upper left rows/columns, with diagonal entries $1-\mu_1 \approx 1$ and $\mu_1 \approx 0$. 

We repeat all steps to the remaining system of $4T-1$ fermions, obtaining two more diagonal entries $1-\mu_2 \approx 1$ and $\mu_2 \approx 0$ through a sequence of $2n_2-2$ nearest-neighbor rotations, collected in the unitary ${U}_2$, and so on. At the end, we have constructed a transformation in terms of ``few'' two-site rotations ($\le 2n_{\rm{max}}-2$ per bond)  that fully diagonalizes $\Lambda$. 

In detail, every cycle that diagonalizes one subsystem is carried out in two steps:\\

{\it Step 1 -- }
We first iteratively determine a set of nearest-neighbor Givens rotations that, acting on $\bm{b}_1,$ map the vector entries $v_1^2,v_1^3,\hdots, v_1^{n_1} $ to zero, corresponding to a new mode: $$\gamma_1 = u_1^1 c_1 + v_1^1 c_1^\dagger + \sum_{l=2}^{n_1} u_1^l c_l. $$
We start by constructing a Givens rotation on the two sites $(n_1-1,n_1)$,  which has the form:
\begin{equation}
U^{G}_{n_1-1,n_1} = G(\theta_{n_1-1,n_1})D(\phi_{n_1-1,n_1}),
\end{equation} with:
\begin{align}
\label{eq:Givens_single_body}
G(\theta) &=
\begin{pmatrix}
\cos\theta & 0 & \sin\theta& 0\\
0 & \cos\theta & 0 & \sin\theta \\
-\sin\theta & 0 & \cos\theta& 0 \\
0 & -\sin\theta & 0 & \cos\theta \\
\end{pmatrix}, \\
\label{eq:phase_single_body}
D(\phi)  &=\text{diag}(1,1,e^{-i\phi},e^{i\phi}).
\end{align}
The appropriate angles are determined by 
\begin{align}
\tan\theta_{n_1-1,n_1} &= \left|\frac{v_1^{n_1}}{v_1^{n_1-1}}\right |,\\
e^{i\phi_{n_1-1,n_1}} &= \frac{v_1^{n_1-1}}{v_1^{n_1}}\cdot\left|\frac{v_1^{n_1}}{v_1^{n_1-1}}\right|.
\end{align} With this, the last element of $$\bm{b}_1^{(1)} = U^G_{n_1-1,n_1}\bm{b}_1 $$ is zero.
Next, we repeat this procedure and determine a matrix $U^G_{n_1-2,n_1-1}$ acting on the two sites $(n_1-2,n_1-1)$, parametrized by the new angles $\theta_{n_1-2,n_1-1}$, $\phi_{n_1-2,n_1-1}$ determined from the entries of $\bm{b}_1^{(1)}$, to obtain a transformed  $\bm{b}_1^{(2)}$ with the last and third-to-last elements equal to zero. This step is iterated until we have constructed $n_1-1$ different rotation matrices, such that 
\begin{multline}
\bm{b}_1^{\prime}\equiv \bm{b}_1^{(n_1-1)}=U^G_{1,2}U^G_{2,3}\hdots U^G_{n_1-1,n_1} \bm{b}_1\\
=(u_1^{\prime 1},v_1^{\prime 1},u_1^{\prime 2},0,u_1^{\prime 3},0,\hdots,u_1^{\prime \, n_1},0),
\end{multline}with updated coefficients $\{u^{\prime j}_1\}$ and $v^{\prime 1}_1$.\\

{\it Step 2 -- } For the coefficients of annihilation operators, we now proceed analogously: From the angles
\begin{align}
\tan\theta^\prime_{n_1-1,n_1} &= \left|\frac{u_1^{\prime \, n_1}}{u_1^{\prime \, n_1-1}}\right |,\\
e^{-i\phi^\prime_{n_1-1,n_1}} &= \frac{u_1^{\prime \, n_1-1}}{u_1^{\prime \, n_1}}\cdot\left|\frac{u_1^{\prime \, n_1}}{u_1^{\prime \, n_1-1}}\right|,
\end{align} 
and so on, we determine another set of $n_1-2$ Givens rotations $U^{\prime \,  G}_{i-1,i},$ such that 
\begin{multline}
\bm{b}_1^{\prime \prime} =U^{\prime \, G}_{2,3}\hdots U^{\prime \, G}_{n_1-1,n_1} \bm{b}^{\prime}_1
=(u_1^{\prime\prime 1},v_1^{\prime\prime 1},u_1^{\prime\prime 2},0,\hdots,0),
\end{multline} corresponding to a rotated fermionic mode $$\gamma_1^{\prime\prime}=u_1^{\prime\prime 1}c_1+v_1^{\prime\prime 1}c_1^\dagger + u_1^{\prime\prime 2}c_2.$$ The fermionic relation $(\gamma_1^{\prime\prime})^2=0$ dictates that either $u_1^{\prime\prime 1}=0$ or $v_1^{\prime\prime 1}=0$. Since we have only applied number-conserving Givens rotations so far, the remaining correlation with the rest of the system must be of Bell-type, i.e. $u^{\prime\prime \, 1}_1=0.$ To disentangle the eigenmode fully, we conclude by applying a Bogoliubov transformation acting on sites $(1,2)$:
\begin{equation}
U^{\prime \prime \, B}_{1,2}=B(\theta_{1,2}^{\prime\prime})D(\phi_{1,2}^{\prime\prime}),
\end{equation} where
\begin{equation}
\label{eq:Bog_single_body}
B(\theta)=\begin{pmatrix}
\cos\theta & 0 & 0& -\sin\theta \\
0 & \cos\theta & - \sin \theta  & 0\\
0& \sin\theta & \cos\theta & 0 \\
\sin\theta & 0 & 0 & \cos\theta  \\
\end{pmatrix},
\end{equation}
and 
\begin{align}
    \tan\theta_{1,2}^{\prime\prime} &= -\left|\frac{u_1^{\prime\prime 2}}{v_1^{\prime\prime 1}}\right|,\\
    e^{-i\phi_{1,2}^{\prime\prime}} &= \frac{v_1^{\prime\prime 1}}{u_1^{\prime\prime 2}}\cdot\left|\frac{u_1^{\prime\prime 2}}{v_1^{\prime\prime 1}}\right|.
    \end{align}

We collect all rotation matrices from Steps 1-2 in a unitary $${U}_1 =\underbrace{U^{\prime \prime \, B}_{1,2}U^{\prime \, G}_{2,3}\hdots U^{\prime \, G}_{n_1-1,n_1}}_{\text{Step }2} \underbrace{U^G_{1,2}U^G_{2,3}\hdots U^G_{n_1-1,n_1}}_{\text{Step }1},$$ that acts on $\bm{b}_1$ as:
$$
\bm{b}_1^{\prime\prime\prime} \equiv {U}_1\bm{b}_1 = (0,v_1^{\prime\prime\prime \, 1},0,\hdots,0)^T,
$$ 
with $|v_1^{\prime\prime\prime \, 1}|=1.$

The structure of fermionic correlation matrices, Eq.~(\ref{eq:struc_corr_matrix}), ensures that the subcorrelation matrix $C^{(0)}$ from Eq.~(\ref{eq:subcorr_matrix}) contains another localized mode $\bm{b}_1^h$ with eigenvalue $1 - \mu_1$ that has entries:
\begin{equation}
\bm{b}_1^h =\big((v_{1}^1)^*, (u_{1}^1)^*,\hdots ,(v_{1}^{n_1})^*,(u_{1}^{n_1})^*\big)^T.
\end{equation} 
It can easily be checked that 
$${U}_1 \bm{b}_{1}^h = \big((v^{\prime\prime\prime \, 1}_1)^*,0,0,\hdots,0\big)^T.$$ 

In the new basis, we have thus isolated a pair of eigenvectors, as well as a nontrivial correlation matrix describing the orthogonal space of $4T-1$ fermionic modes: Extending $\mathcal{U}_1$ to act as the identity on the complement of the considered block of $n$ sites, we get 
\begin{equation}
\Lambda^{(1)} ={U}_1 \Lambda {U}_1^\dagger =\begin{pmatrix}
D_1^{2\times 2} & 0\\
0& \tilde{\Lambda}^{(1)}
\end{pmatrix},
\end{equation}
with $$D_1^{2\times 2} \approx \begin{pmatrix}
1 & 0 \\
0 & 0
\end{pmatrix}.$$ This is the end of the two-step cycle that diagonalizes one subsystem. This cycle is illustrated in Fig.~\ref{Fig:corr_diag}. \\

We repeat the above procedure for the remainder of the system, i.e. we extract a new subcorrelation matrix $C^{(2)}=\tilde{\Lambda}^{(1)}\big|^{1,\hdots,2n_2}_{1,\hdots,2n_2}$ for the smallest $n_2$  such that the smallest eigenvalue $\mu_2$ satisfies $\mu_2 <\epsilon$, and determine the sequence of two-site rotation matrices acting on this block of $n_2$ sites, that push the corresponding eigenvector to the first site, analogously to the above. From these, we can construct a new unitary matrix ${U}_2$ which transforms $\Lambda^{(1)}$ to:
\begin{equation}\nonumber
\Lambda^{(2)} = {U}_2\Lambda^{(1)} {U}_2^\dagger
=\begin{pmatrix}
D_1^{2\times 2} & 0 & 0\\
0 & D_2^{2\times 2} & 0\\
0 & 0& \tilde{\Lambda}^{(2)}\end{pmatrix},
\end{equation}
with $$D_2^{2\times 2} \approx \begin{pmatrix}
1 & 0 \\
0 & 0
\end{pmatrix}.
$$ 
After $4T-1$ repetitions of the cycle, each applied to a different subsystem, we assemble the unitary ${U}={U}_{4T-1} \cdots {U}_2{U}_1$,
such that we arrive at:
\begin{equation}
\label{eq:vacuum_correlation}
\Lambda^{(4T-1)} = 
{U}
\Lambda
{U}^\dagger
\approx \diag\Big(1,0,1,0,\hdots,1,0\Big).
\end{equation} 
The resulting transformed correlation matrix uniquely identifies the vacuum state $\ket{\emptyset}$ in Fock space, which is a product state.

\subsection{Constructing the MPS}
Since the approximate vacuum correlation matrix in Eq.~(\ref{eq:vacuum_correlation}) has been obtained by a sequence of unitary rotations of the initial correlation matrix, Eq.~(\ref{eq:original_corr_matrix}), we can invert this procedure: We start from the vacuum correlation matrix $\Lambda^{vac} =\diag\Big(1,0,1,0,\hdots,1,0\Big),$ to which we apply the reversed sequence of inverse rotations. 
This leads us to a correlation matrix close to the one of the IM wavefunction, Eq.~(\ref{eq:original_corr_matrix}).
Since Gaussian states are uniquely determined by their correlation matrix, the transformation constructed above between the two correlation matrices determines a corresponding transformation between the two states $\ket{\emptyset}$ and $\ket{\mathcal{I}}$.

 To obtain the wave-function transformation, we convert the single-body rotation matrices to Gaussian unitary gates acting on the fermionic Fock space. For this, we note that the quadratic generators of the single-body rotation matrices in Eqs.~(\ref{eq:Givens_single_body}),~(\ref{eq:phase_single_body}), and~(\ref{eq:Bog_single_body}), are:
\begin{align}
\hat{g}^{G}_{j,j+1}&=-i(c_j^\dagger c_{j+1} - h.c.) \rightarrow \frac{1}{2}(\sigma_j^x\sigma_{j+1}^y - \sigma_j^y\sigma_{j+1}^x),\\
\hat{g}^{B}_{j,j+1}&=i(c_j^\dagger c_{j+1}^\dagger - h.c.) \rightarrow \frac{1}{2}(\sigma_j^x\sigma_{j+1}^y + \sigma_j^y\sigma_{j+1}^x),\\
\hat{g}^{D}_{i}&=(c_j c_{j}^\dagger - c_{j}^\dagger c_{j}) \rightarrow \sigma^z_i.
\end{align} 
The unitary gates are found by exponentiating these generators:
\begin{align}
\label{eq:quantum_gate_Givens}
\hat{V}_{j,j+1}^{G}(\theta,\phi)&= e^{-i\theta\hat{g}_{j,j+1}^{G}} e^{-i\phi\hat{g}_{j+1}^{D}},\\
\label{eq:quantum_gate_Bog}
\hat{V}_{j,j+1}^{B}(\theta,\phi)&= e^{-i\theta\hat{g}_{j,j+1}^{B}} e^{-i\phi\hat{g}_{j+1}^{D}}.
\end{align}
By expressing ${U}$ explicitly as product of matrices $U^G$ and $U^B$ and replacing each of them by their corresponding many-body unitary gate, thus promoting $${U}\rightarrow\hat{{V}},$$ we obtain the sequence of Gaussian unitary gates in Fock space which transforms $\ket{\mathcal{I}}$ to $\ket{\emptyset}$.
In particular, $\hat V$ is a {\it unitary circuit} of depth determined by $n_{\mathrm{max}}=\max(n_1,\dots,n_{4T-1})$.

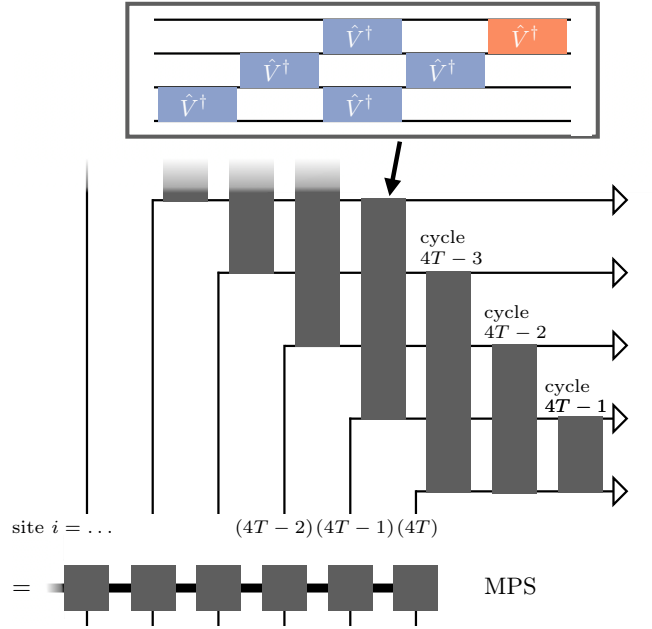
\begin{figure} 
\center
\begin{overpic}[width=0.5\textwidth]{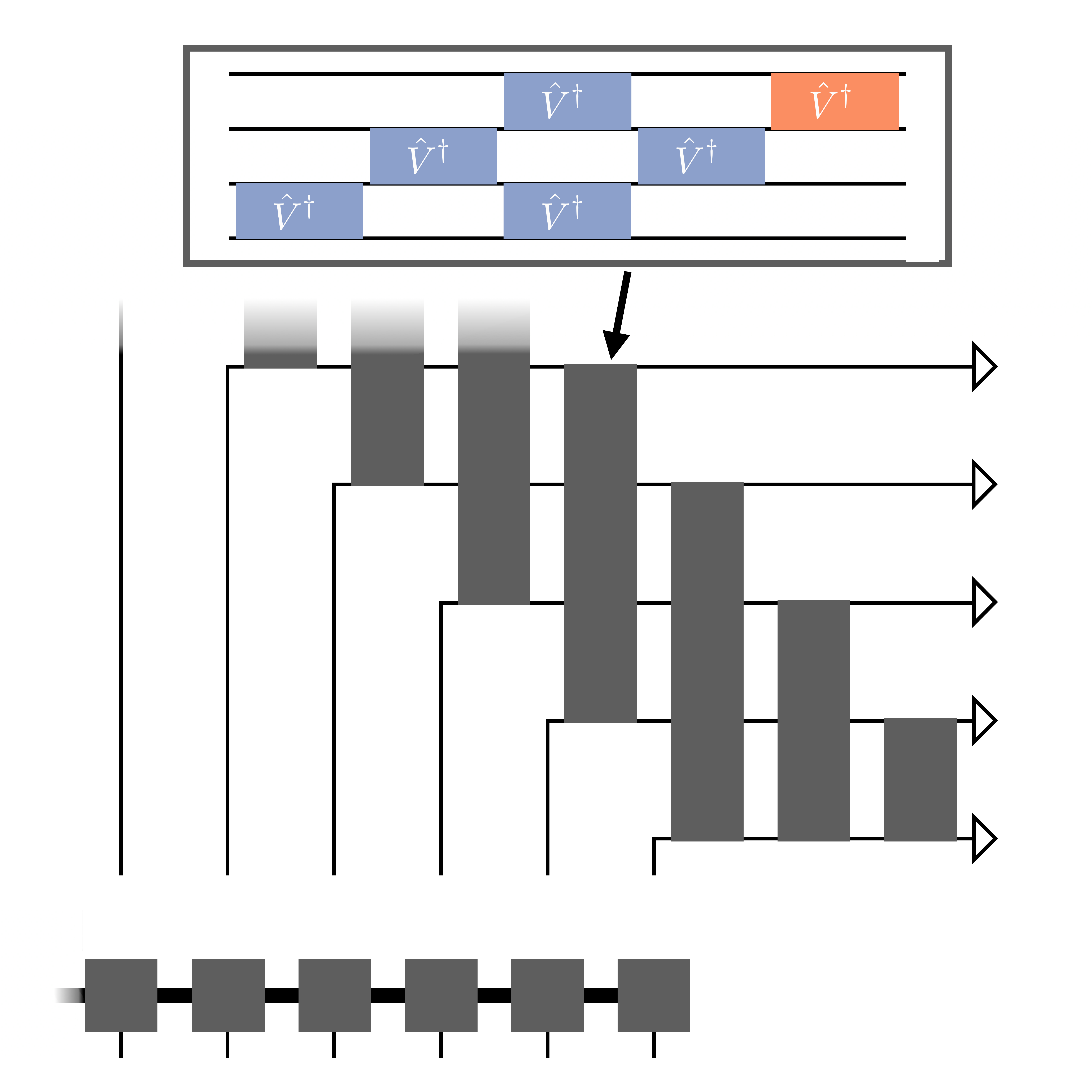}

\put(79,35){\scriptsize $4T-1$}
\put(79,38){\scriptsize cycle}
\put(79,35){\scriptsize $4T-1$}
\put(70,49){\scriptsize cycle}
\put(70,46){\scriptsize $4T-2$}
\put(60.5,60){\scriptsize cycle}
\put(60.5,57){\scriptsize $4T-3$}
\put(57,17){\scriptsize $(4T)$}
\put(45,17){\scriptsize $(4T-1)$}
\put(33,17){\scriptsize $(4T-2)$}
\put(0,17){\scriptsize site $i=\hdots$}
\put(0,8){$=$}
\put(70,8){MPS}
\end{overpic}
\caption{Illustration of the MPS construction: After the single body-rotations are converted to quantum gates, they are applied to the vacuum many body state in reverse order and conjugated, which provides a MPS representation of the IM. From this illustration one can deduce the bound on the local bond dimension $\chi_{(j,j+1)}\le 2^{n_{j}-1}$.
}
\label{Fig:MPS_circuit}
\end{figure}

Now we have set the stage to construct a MPS representation of the state $\ket{\mathcal{I}}$. We represent the vacuum as a product state and use standard MPS methods to contract the reverse conjugated unitary circuit $\hat{{V}}^\dagger$ 
illustrated in Fig.~\ref{Fig:MPS_circuit}. 
For a maximum subsystem size $n_{\rm{max}}$, this circuit consists of $ \le 8 n_{\rm{max}} T$ gates. 
It can be deduced graphically from Fig.~\ref{Fig:MPS_circuit} (see also  Ref.~\cite{Fishman2015MPS}) that the bond dimension of the resulting exact MPS is upper-bounded by $\chi = 2^{n_{\rm{max}}-1}$: Starting from the bond $(j,j+1)$ of the outcome chain (bottom), the cut passing through that bond that disconnects the circuit in two separate networks cuts $n_{j}-1$ (two-dimensional) legs.
The efficiency of the MPS is thus related to the optimal values of $n$, which, in turn, depend on both the total size (evolution time) $T$ and the error threshold $\epsilon$. In the following section we will investigate the typical scaling of the circuit depth  for the IMs considered in this work.

\subsection{Efficiency of the MPS}
\label{sec:subsystem_scaling_ising}
Next, we report the results of applying the algorithm from in Sec.~\ref{Sec:Appr_diag} to the IM presented in Fig.~\ref{fig:Ising_scaling}. We focus here on the kicked Ising model since, in this case, we have an analytical formula for the response functions, Eq.~(\ref{eq:G_lesser_finiteT}), that allows us to work in the thermodynamic limit. 

\begin{figure} 
\center
\includegraphics[width=.5\textwidth]{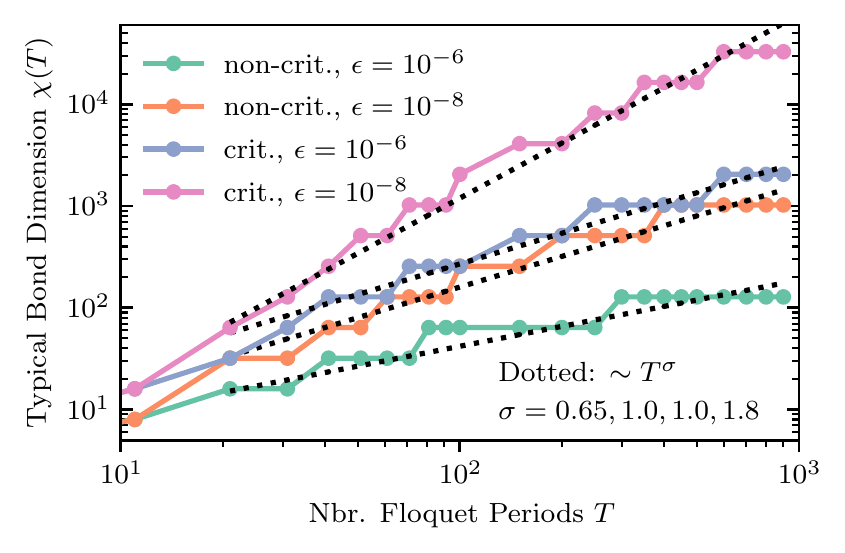}
\caption{Scaling of the typical bond dimension upper bound $\chi(T)=2^{n_{\mathrm{av}}-1}$, associated with the average subsystem block size $n_{\mathrm{av}}=\sum_{j=1}^{4T-1} n_j/(4T-1)$ 
%and maximum $n_{\mathrm{max}}=\max(n_1,\dots,n_{4T-1})$
needed to diagonalize the correlation matrix $\Lambda$ of the IM of the kicked Ising model over $T$ Floquet periods. The IM is calculated analytically in the thermodynamic limit from Eq.~(\ref{eq:G_lesser_finiteT}) for the same non-critical (infinite temperature) and critical initial states and for the same  parameters as in Fig.~\ref{fig:Ising_scaling}. Dotted lines are power-law fits (notice the horizontal logarithmic scale).
}
\label{fig:subsystem_scaling_Ising}
\end{figure}

As shown in the previous subsection, a strict upper bound on the local bond dimension of the resulting MPS is $\chi_{(j,j+1)}\le 2^{n_{j}-1}$.
In Fig.~\ref{fig:subsystem_scaling_Ising}, we show the {\it typical} bond dimension upper bound associated with the average block size $n_{\mathrm{av}}=\sum_{j=1}^{4T-1} n_j/(4T-1)$ needed to fully diagonalize the correlation matrix $\Lambda$ of the IM over $T$ Floquet periods, with a fixed tolerance $\epsilon$ on the localization of the IM's eigenmodes.
(We have checked that the  bond dimension associated with the {\it maximum} block size  $n_{\mathrm{max}}=\max(n_1,\dots,n_{4T-1})$ behaves similarly.) % to the maximal one.)
We plot them as a function of $T$ for two different fixed values of $\epsilon$. 
We observe that in both the non-critical and critical case, the scaling of subsystem size $n$ is {\it at most} logarithmic in $T$, corresponding to a polynomial scaling of the bond dimension upper bound $\chi \sim T^\sigma$. 
The exponents $\sigma$ obtained from fitting the data in Fig.~\ref{fig:subsystem_scaling_Ising} are: $\sigma = 0.65$, $1$, $1$, $1.8$ (lowermost to topmost line).
Subsystem sizes needed for the critical initial state are generally larger than for the non-critical one, indicating that eigenmodes are more localized in the latter. In both cases, both the average and the maximal block size increase as the tolerance $\epsilon$ is decreased.

In App.~\ref{Sec:block_size_XY}, we report analogous results for the trotterized XY model. %While 
The general conclusions coincide with the ones presented here.
Overall, these results, combined with the results of Sec.~\ref{Sec:temporal_entanglement} on temporal entanglement scaling, establish the efficiency of our method for studying real-time dynamics of relevant quantum impurity problems after a local or global quantum quench.

\section{Conclusions and discussion}

% summary of motivation
To summarize, we have introduced an approach to analyzing dynamics of (arbitrary) interacting impurities in contact with (arbitrary) non-interacting fermionic environments. 
The central ingredient of our approach is representing the environment's influence functionals as MPS in the temporal domain. 
This is expected to be efficient whenever the non-local-in-time ``interactions'' encoded by the influence functional, equivalent to Keldysh correlation functions of the environment, generate a moderate amount of {\it temporal entanglement}.
The main advantage of our approach is that an efficient MPS encoding of {\it individual} reservoirs' IMs directly guarantees an efficient computation of the dynamics of an arbitrary interacting impurity simultaneously coupled to all of them. 

% summary of results
By analyzing temporal entanglement entropy scaling of the IM for a class of one-dimensional reservoir models, we demonstrated that in many cases of interest, including Fermi-sea-like, superconducting, and non-equilibrium initial states following a quantum quench in the reservoir, MPS encoding is provably efficient. 
In particular, temporal entanglement is finite (area law) for reservoirs with short-range correlations --- due to a gap and/or to a finite temperature.
For quantum-critical (gapless {\it and} zero-temperature) reservoirs we have found a logarithmic violation of the area law, and linked this TE scaling law to the celebrated scaling of spatial entanglement in conformal field theories.
Further, by extending the work of Fishman and White, we have proposed an algorithm for converting the BCS-like IM wave function to a MPS.
The efficiency of this algorithm is related to the localization length of IM's ``natural orbitals'' in the temporal domain --- a quantity related to the range of temporal correlations and temporal entanglement pattern in the IM.

% computational complexity
We have investigated the scaling of the computational resources required by this algorithm, demonstrating that our approach is indeed efficient in all cases considered.
Similarly to previous iterative path-integral~\cite{EggerIterative2008,MillisImp2010} or Monte Carlo~\cite{Muhlbacher08Realtime,Schiro09realtime,Werner09Diagrammatic,gull10boldline,gull11NumericallyExact,cohen11memory,cohen13neqkondo,cohen15taming} approaches to real-time dynamics of QIMs, our method is numerically exact. 
However, while their  computational complexity generally grows exponentially as the control parameters of the method are scaled up, our method only requires {\it polynomial} resources, cf.~Fig.~\ref{fig:subsystem_scaling_Ising}.

%generalizations
Our results can be generalized to a number of settings, arising in the context of QIMs in materials, mesoscopic devices and cold atoms.
First, we note that while above we studied models of reservoirs that host spinless fermions, the generalization to the spinful case does not present any conceptual difficulty, only implying  a doubling of the degrees of freedom in the IMs. Likewise, generalization to  multi-band reservoirs and/or multi-orbital interacting impurities is also straightforward and entails a similar overhead.
%
% higher dim
Changing the environment's geometry is also feasible: for higher-dimensional reservoirs, which are being studied experimentally,
the IM can still be viewed as a many-body wavefunction in one temporal dimension, and the influence action can be computed by exploiting the solvability of the environment's dynamics, as we described in the $1d$ case.
Since the efficiency of the proposed IM method is controlled by the decay of local temporal correlations in the reservoirs, we expect that the method is equally efficient for higher-dimensional 
%gapless (gapped) 
fermionic environments.

Further, our method can be straightforwardly generalized to describe global quenches starting from entangled states of impurity and reservoirs, realized, e.g., by including a segment of imaginary time evolution in the definition of the IM~\cite{Banuls09,Tang20cmps}. 
A further interesting  extension of our approach involve the effects of dissipative channels localized on the impurity -- a setup motivated by cold-atom experiments that recently received considerable attention~\cite{Zezyulin12Macroscopic,viciani15observation,Lebrat19quantized,maier19environmentassisted,syassen2008strong,tonielli19oc,froml20ultracold,dolgirev20nongaussian}.
Crucially, at the level of computational complexity, all these extensions entail at most polynomial overheads.

%comparison to recent OQS literature
Our work complements recently proposed tensor-network approaches to dynamics of open quantum systems such as the spin-boson model~\cite{TEMPO,Jorgensen19Exploiting,Luchnikov2019,bose2021tensor,Chan21}, which proved efficient for thermal-equilibrium initial states of the bosonic baths, and of homogeneous strongly interacting spin chains~\cite{Banuls09,lerose2020,Chan21}.
Here, we studied general fermionic reservoirs relevant in condensed matter, mesoscopic physics or cold atom setups, in particular for (highly) non-equilibrium transport problems. 
Accordingly, we focused on the role of the reservoirs' states, including (lack of) equilibrium, temperature, and gap. In addition, we provided a detailed analysis of the IM wave function in a range of regimes, establishing its temporal entanglement entropy scaling. 
Furthermore, our algorithm is complementary to those proposed for spin-boson models: it avoids both explicit memory-range cutoffs and long-range interacting gates.
We note that further improvements are likely possible: one possibility is to exploit the approach of Schuch and Bauer~\cite{schuch19fermionicgaussianMPS}, which uses Gaussian MPS representations of Gaussian states;  in a different direction, chain mappings through orthogonal polynomials~\cite{woods14ChainMapping} developed in the context of open quantum systems could be combined with efficient versions of the IM approach for one-dimensional systems~\cite{lerose2022overcoming,frias2022light}.

% errors
A challenging task left to future work is a more rigorous investigation of how errors in constructing the MPS representation of the IM --- i.e., truncation of singular values, tolerance $\epsilon$ in natural orbital localization --- relate to errors in the computation of impurity observables.
In particular, it would be interesting to compare the impact of the localization error $\epsilon$ with that of a more standard hard memory-range cutoff~\cite{TEMPO}.
A rigorous theory of errors in IM MPSs, similar to the well-established theory of ground state approximations by MPSs, would give an insight into the computational complexity of non-equilibrium QIMs~\cite{bravyi2017complexity}.
In a related direction, elimination of time-discretization errors using continuous-MPS approaches in the real-time domain, generalizing what recently achieved for the imaginary-time domain~\cite{Tang20cmps}, also represents a promising direction for a conceptual improvement of our method.

In the future work, the approach introduced here will be applied to non-equilibrium QIMs that may be challenging for other methods.  
An immediate application of our approach~\cite{UsToBe} is to analyze highly non-equilibrium impurity quenches.
Transport statistics, in particular, can be studied using counting fields, which can be easily incorporated within the IM approach. 
Further, our method can be naturally embedded into dynamical mean-field theory loops as an impurity solver, and  its performance compared with state-of-the-art techniques. The TE scaling properties suggest that the approach described above may be advantageous, especially when highly non-equilibrium settings are considered. 

\section{Acknowledgements}

We thank M. Sonner and G. Mazza for insightful discussions and collaboration on related projects, and J. von Delft, O. Scarlatella, M. Schir\'o, and M. Stoudenmire for feedback on this work.
This work was supported by the Swiss National Science Foundation (AL, DA) and by the European Research Council (ERC) under the European Union's Horizon 2020 research and innovation program (grant agreement No. 864597) (JT, DA). 

%\nocite{*}
\bibliography{impurity}

\appendix
\clearpage
\newpage

\section{Path integral representation of IM}
\label{app:path_integral}

In order to determine the path integral representation of the IM,  the gates $U_{j,j+1}$ from Eq.~(\ref{eq:two_site_gate}) need to be mapped onto fermionic gates by means of a Jordan-Wigner transformation: \begin{align}
\sigma_j^+ &= e^{i\pi \sum_{l<j}c^\dag_l c_l}c_j\\
\sigma_j^- &= c_j^\dag e^{-i\pi \sum_{l<j}c^\dag_l c_l}\\
\sigma_j^z &= (1-2c_j^\dag c_j).
\end{align}
Next, we introduce Grassmann variables $\bar{\xi}_j,\xi_j$ at every site $i=0,1,\hdots,L,$ i.e. $c_j \ket{\xi_j} = \xi_j \ket{\xi_j},$ and define the vectors of Grassmanns in the environment: $\bm{\bar{\xi}}=(\bar{\xi}_1,\bar{\xi}_2,\hdots,\bar{\xi}_L), \bm{\xi}=(\xi_1,\xi_2,\hdots,\xi_L).$ Inserting a resolution of identity of such a set of Grassmann variables at every time step $\tau=\{0,\tfrac{1}{2},1,\hdots,T\}$ 
on both Keldysh branches, the trace expression, Eq.~(\ref{eq:IM_operator}), is converted to a path integral which reads (see Fig.~\ref{Fig:Grassmann_labels}):
\begin{widetext}
\begin{align}
\nonumber
\mathcal{I}\left[\{\bar{\xi}^\pm_{0,\tau},\xi^\pm_{0,\tau}\}\right]=&\bigintsss \Big[\prod\limits_{\tau=0,1/2,}^t d\bar{\bm{\xi}}_\tau^{\pm}d\bm{\xi}_\tau^{\pm}\Big]\\
\nonumber
&\Bigg(e^{-\bar{\bm{\xi}}^+_t\bar{\bm{\xi}}^-_t}\bra{-\bar{\bm{\xi}}^+_{t}}\mathcal{U}_\text{even} \ket{\bm{\xi}^+_{t-1/2}}e^{-\bar{\xi}^+_{0,t}\xi^+_{0,t-1/2}}\cdot e^{-\bar{\bm{\xi}}^+_{t-1/2}\bm{\xi}^+_{t-1/2}}\cdot\bra{\bar{\xi}^+_{0,t-1/2},\bar{\bm{\xi}}^+_{t-1/2}}\mathcal{U}_\text{odd}\mathcal{U}_{\mathcal{SE}}\ket{\xi^+_{0,t-1},\bm{\xi}^+_{t-1}} \\
\nonumber
&\cdots \bra{\bar{\bm{\xi}}^+_{1}}\mathcal{U}_\text{even} \ket{\bm{\xi}^+_{1/2}}e^{+\bar{\xi}^+_{0,1}\xi^+_{0,1/2}}\cdot e^{-\bar{\bm{\xi}}^+_{1/2}\bm{\xi}^+_{1/2}}\cdot\bra{\bar{\xi}^+_{0,1/2},\bar{\bm{\xi}}^+_{1/2}}\mathcal{U}_\text{odd}\mathcal{U}_{\mathcal{SE}}\ket{\xi^+_{0,0},\bm{\xi}^+_{0}}\\
\nonumber
& \cdot e^{-\bar{\bm{\xi}}^+_{0}\bm{\xi}^+_{0}}\cdot \bra{\bar{\bm{\xi}}^+_{0}}\rho_{\mathcal{E}}\ket{\bm{\xi}^-_{0}}\cdot e^{-\bar{\bm{\xi}}^-_{0}\bm{\xi}^-_{0}}\\
\nonumber
&\cdot \bra{\bar{\xi}^-_{0,0},\bar{\bm{\xi}}^-_{0}}\mathcal{U}_{\mathcal{SE}}^\dagger \mathcal{U}^\dagger_\text{odd} \ket{\xi^-_{0,1/2},\bm{\xi}^-_{1/2}}\cdot e^{-\bar{\bm{\xi}}^-_{1/2}\bm{\xi}^-_{1/2}}\cdot e^{+\bar{\xi}^+_{0,1/2}\xi^+_{0,1}}\bra{\bar{\bm{\xi}}^-_{1/2}}\mathcal{U}^\dagger_{\text{even}} \ket{\bm{\xi}^-_{1}}\\
&\cdots \bra{\bar{\xi}^-_{0,t-1},\bar{\bm{\xi}}^-_{t-1}}\mathcal{U}_{\mathcal{SE}}^\dagger \mathcal{U}^\dagger_\text{odd} \ket{\xi^-_{0,t-1/2},\bm{\xi}^-_{t-1/2}}\cdot e^{-\bar{\bm{\xi}}^-_{t-1/2}\bm{\xi}^-_{t-1/2}}\cdot e^{+\bar{\xi}^+_{0,t-1/2}\xi^+_{0,t}}\bra{\bar{\bm{\xi}}^-_{t-1/2}}\mathcal{U}^\dagger_{\text{even}} \ket{\bm{\xi}^-_{t}}\Bigg).\label{eq:GM_integral}
\end{align}
\end{widetext}

\section{Grassmann kernels of gates}
\label{app:gate_kernels}

Evaluation of the path integral, Eq.~(\ref{eq:GM_integral}), requires the evaluation of Grassmann kernels of the type $\bra{\bar{\xi}_0,\bm{\bar{\xi}}} U_{j,j+1}\ket{\xi_0,\bm{\xi}}.$
The fermionic version of the two-site gates $U_{j,j+1}$ from Eqs.~(\ref{eq:two_site_gate}, \ref{eq:onsite_gate}) read:
\begin{align}
U_{j,j+1} &= \exp\big[i \mathcal{J}_x (c_j^\dag - c_j)(c_{j+1}+c_{j+1}^\dag)\\
&+ i \mathcal{J}_y (c_j^\dag + c_j)(c_{j+1}-c_{j+1}^\dag)\big] .
\end{align} Note that the two terms in the exponent commute.\\
Using $(c_j^\dag \mp c_j)(c_{j+1}\pm c_{j+1}^\dag) \in \{-1,1\}$ to rewrite the exponential in terms of trigonometric functions and defining the abbreviations $c_\alpha\equiv \cos(\mathcal{J}_\alpha), t_\alpha\equiv \tan(\mathcal{J}_\alpha),T_{xy}\equiv 1+t_xt_y,$ this can be rewritten as: 
\begin{multline}
 U_{j,j+1}=c_xc_y  T_{xy}\Bigg[ 1+ i\frac{t_x+t_y}{T_{xy}} (c_j^\dag c_{j+1} - c_jc_{j+1}^\dag)\\+i\frac{t_y-t_x}{T_{xy}} (-c_j^\dag c_{ij+1}^\dag + c_jc_{j+1})\\-\frac{t_x t_y}{T_{xy}} \Big( 2(c_j^\dag c_j + c_{j+1}^\dag c_{j+1}) - 4 c_j^\dag c_jc_{j+1}^\dag c_{j+1} \Big)\Bigg],
\end{multline} where we have brought all operators at every site to normal order. With this, it is straightforward to evaluate the Grassmann kernel:
\begin{multline}
\bra{\bar{\xi}_j,
\bar{\xi}_{j+1}} U_{j,j+1}\ket{\xi_j,\xi_{j+1}}=\mathcal{F}_{j,j+1}[\bar{\xi}_j,
\bar{\xi}_{j+1},\xi_j,\xi_{j+1}] \\
=  c_xc_yT_{xy}\exp\Big[ i\frac{t_x+t_y}{T_{xy}}  (\bar{\xi}_j \xi_{j+1}-\xi_j\bar{\xi}_{j+1}) \\+ i\frac{t_y-t_x}{T_{xy}} (-\bar{\xi}_j\bar{\xi}_{j+1} + \xi_j \xi_{j+1}) \\
- 2\frac{t_xt_y}{T_{xy}} (\bar{\xi}_j\xi_j+\bar{\xi}_{j+1}\xi_{j+1})\Big] e^{\bar{\xi}_j\xi_j + \bar{\xi}_{j+1}\xi_{j+1}}.
\label{eq:Kernel_schr}
\end{multline}
We notice that for $\mathcal{J}_x=\mathcal{J}_y$ there are no couplings between barred and between non-barred variables, reflecting $U(1)$ symmetry; for $\mathcal{J}_x=0$ or $\mathcal{J}_y=0$  couplings between variables on the same site vanish (a fact that greatly simplifies the analysis, see the main text).  Furthermore, at the self-dual point $\mathcal{J}_x = \mathcal{J}_y=\pi /4$, only the terms in the first line survive, manifestly exhibiting spacetime duality.

Due to the structure of interaction, it is convenient to rotate the Grassmann variables of the system,
$\zeta^{\uparrow} = \frac{1}{\sqrt{2}}(\xi_0 + \bar{\xi}_0),$ $\zeta^\downarrow = \frac{1}{\sqrt{2}}(\xi_0 - \bar{\xi}_0), $
such that the interaction gate between the system and the first site of the environment reads:
\begin{multline}
\label{eq:interaction_GM}
\mathcal{F}_{0,1}[\zeta^\uparrow,\zeta^\downarrow,\bar{\xi}_1,\xi_1] = 
 c_xc_y  T_{xy}\exp\Big[i \frac{\sqrt{2}t_y}{T_{xy}} \zeta^\uparrow(\xi_1-\bar{\xi}_{1}) \\
 + i\frac{-\sqrt{2}t_x}{T_{xy}} \zeta^\downarrow(\xi_1+\bar{\xi}_{1}) -
2\frac{t_xt_y}{T_{xy}} (\zeta^\uparrow\zeta^\downarrow+\bar{\xi}_{1}\xi_{1})\Big] e^{\zeta^\uparrow\zeta^\downarrow+\bar{\xi}_1\xi_1}.
\end{multline}

The fermionic version of the local kick operator is: $$\exp(i\varphi\sigma_j^z)\rightarrow \exp\big(i\varphi(1-2c_j^\dagger c_j)\big).$$ Evaluating the Grassmann Kernel analogously to the above yields:
\begin{equation}
    \bra{\bar{\xi}_j} \exp\big(i\varphi(1-2c_j^\dagger c_j)\big) \ket{\xi_j}= e^{i\varphi}\exp\big(e^{-2i\varphi}\bar{\xi}_j\xi_j \big).
\end{equation}

\section{Setting source fields $\zeta^\uparrow_\tau, \zeta^\downarrow_\tau$ to zero}
\label{App:source_fields_zero}

Using Eq.~\eqref{Eq:relation_exponent_derivatives}, the influence action is reconstructed by computing correlation functions in the environment.
We obtain this identification by applying functional derivatives to the expression in Eq.~(\ref{eq:GM_integral}) and setting all ``source fields'' $\zeta_\uparrow,\zeta_\downarrow$ to zero.
The resulting expression is then interpreted as a Keldysh correlation function involving the environment only. 

In our discrete time formalism, the interaction gate kernels generally involve quadratic terms in the bath degrees of freedom, which modify the quadratic Keldysh action of the bath.
Setting ``source fields'' to zero corresponds to projecting the fermionic trajectory at site $j=0$ onto the vacuum at all times, resulting in a {\it non-unitary} operator acting on site $j=1$.
In fact,
\begin{multline}\nonumber
\mathcal{F}_{0,1}[\zeta^\uparrow,\zeta^\downarrow,\bar{\xi}_1,\xi_1]\Bigg|_{\zeta^\uparrow,\zeta^\downarrow = 0} \\
=  c_xc_y  T_{xy}\exp\Big[ \frac{\cos(\mathcal{J}_x + \mathcal{J}_y)}{\cos(\mathcal{J}_x - \mathcal{J}_y)}\bar{\xi}_{1}\xi_{1}\Big] e^{\bar{\xi}_1\xi_1},
\end{multline} is the kernel of a new operator:
\begin{equation}
\mathcal{F}_{0,1}[\zeta^\uparrow,\zeta^\downarrow,\bar{\xi}_1,\xi_1]\Bigg|_{\zeta^\uparrow,\zeta^\downarrow = 0} \rightarrow e^{-\tilde{\beta}}c_xc_y  T_{xy} F_1,
\end{equation} 
with
\begin{align}
F_1 &\equiv e^{\tilde{\beta}(c_1 c_1^\dagger-c_1^\dagger c_1)}\\
\tilde{\beta} &\equiv \frac{1}{2}\log\Big[ \frac{\cos(\mathcal{J}_x - \mathcal{J}_y)}{\cos(\mathcal{J}_x + \mathcal{J}_y)}\Big].
\end{align}
Thus
\begin{multline}
\mathcal{I}[\{\zeta^\uparrow,\zeta^\downarrow\}]\Bigg|_{\{\zeta^\uparrow_\tau\}=\{ \zeta^\downarrow_\tau\} = 0}\\
\rightarrow \Bigg(e^{-\tilde{\beta}}c_xc_y  T_{xy}\Bigg)^{2T} \text{Tr}_\mathcal{E}\Bigg( (\tilde{\mathcal{U}}_\mathcal{E})^{T}  \rho_\mathcal{E} (\tilde{\mathcal{U}}_\mathcal{E}^{\dagger})^{T}  \Bigg),
\label{eq:Trace_without_derivative}
\end{multline}
where we defined the {\it non-unitary} environment Floquet operator $\tilde{\mathcal{U}}_\mathcal{E} \equiv \mathcal{U}_\mathcal{E} F_1.$ 

\section{From functional derivatives to correlation functions}
\label{app:fd_to_cf}
Next, we determine the operator expression that corresponds to a kernel after taking a functional derivative.
Exemplarily, we consider the derivative $\frac{\delta}{\delta\zeta^\uparrow}$ and evaluate:
\begin{align}
\frac{\delta}{\delta\zeta^\uparrow}\mathcal{F}_{0,1}[\zeta^\uparrow,\zeta^\downarrow,\bar{\xi}_1,\xi_1]\Bigg|_{\zeta^\uparrow,\zeta^\downarrow = 0}&\\
=\frac{i\sqrt{2}t_y}{T_{xy}}c_xc_y  T_{xy}&\exp\Big[ \frac{\cos(\mathcal{J}_x + \mathcal{J}_y)}{\cos(\mathcal{J}_x - \mathcal{J}_y)}\bar{\xi}_{1}\xi_{1}\Big]\\
&\cdot\Big(\xi_{1} - \bar{\xi}_{1}\Big)
%e^{\sum_{i=2}^L\bar{\xi}_i\xi_i} 
\\
\rightarrow   \frac{i\sqrt{2}t_y}{T_{xy}}e^{-\tilde{\beta}}c_xc_y  T_{xy} \ F_1 &\Big(c_1-e^{2\tilde{\beta}}c_1^\dagger\Big),
\end{align} where we have used:
\begin{equation}
c^\dagger F_1 = e^{2\tilde{\beta}}F_1 c^\dagger.
\end{equation}

With this, we can derive relations of the type:
\begin{multline}
    \frac{\delta^2\mathcal{I}[\{\zeta^\uparrow,\zeta^\downarrow\}]}{\delta\zeta_{\tau_1}^{\uparrow+}\delta\zeta_{\tau_2}^{\uparrow-}}\Bigg|_{\{\zeta^\uparrow_\tau\}=\{ \zeta^\downarrow_\tau\} = 0}=\Bigg(e^{-\tilde{\beta}}c_xc_y  T_{xy} \Bigg)^{2T}\frac{2t_y^2}{T_{xy}^2}\\
 \times \text{Tr}_\mathcal{E}\Bigg( \underbrace{\tilde{\mathcal{U}}_\mathcal{E} \hdots \tilde{\mathcal{U}}_\mathcal{E}}_{(T -\tau_1) \text{ factors}} \Big(c_1 - e^{+\tilde{2\beta}}c_1^\dagger\Big)\underbrace{\tilde{\mathcal{U}}_\mathcal{E}\hdots \tilde{\mathcal{U}}_\mathcal{E}}_{\tau_1\text{ factors}}\rho_\mathcal{E}\\
\times\underbrace{\tilde{\mathcal{U}}_\mathcal{E}^{\dagger}\hdots \tilde{\mathcal{U}}_\mathcal{E}^{\dagger}}_{\tau_2 \text{ factors}}\Big(e^{+2\tilde{\beta}} c_1 - c_1^\dagger\Big)\underbrace{\tilde{\mathcal{U}}_\mathcal{E}^{\dagger}\hdots \tilde{\mathcal{U}}_\mathcal{E}^{\dagger}}_{T  - \tau_2 \text{ factors}}\Bigg).
\label{eq:trace_ex}
\end{multline}

The trace expression, Eq.~(\ref{eq:trace_ex}), is depicted in Fig.~\ref{fig:circuit}b.

\begin{figure}
    \centering
       \begin{overpic}[width=0.28\textwidth]{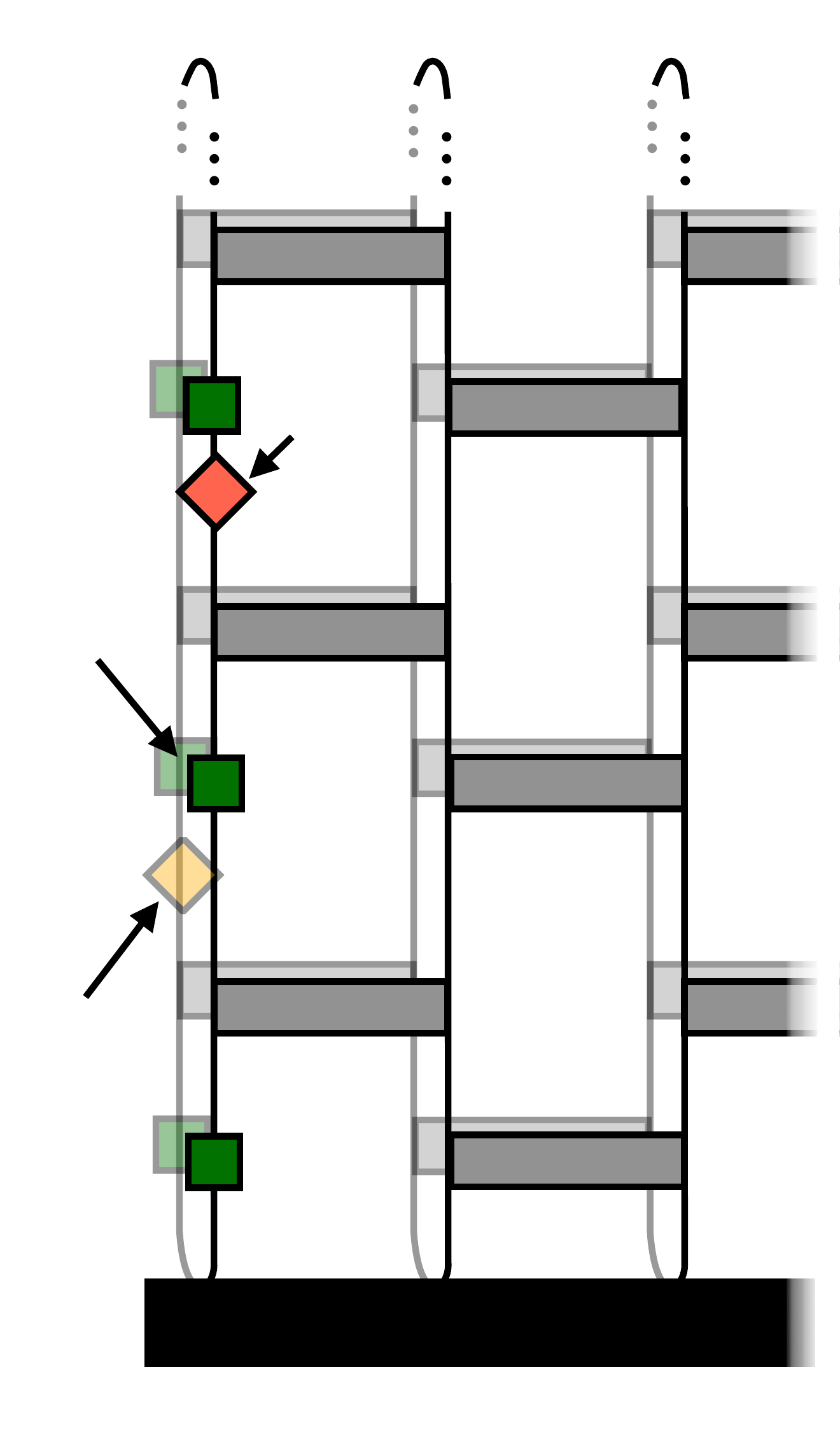}
       \put(3,55){\scriptsize $F_1$}
          \put(3,28){\scriptsize $\hat{O}_2$}
           \put(20,70){\scriptsize $\hat{O}_1$}
       \end{overpic}
    \caption{
 Diagrammatic representation of Eq.~(\ref{eq:general_correlations}). Here, $\hat{O}_1 = \hat{O}_2=c_1-c_1^\dagger.$ The non-unitary gate $F_1$ results from $\mathcal{F}_{0,1}$ with the source fields set to zero.
    }
    \label{fig:correlation_diagram}
\end{figure}

Introducing identities of the form: $$\mathds{1}=\big(\tilde{\mathcal{U}}_\mathcal{E}\big)^{-\tau}\big(\tilde{\mathcal{U}}_\mathcal{E}\big)^\tau=\big(\tilde{\mathcal{U}}_\mathcal{E}^{\dagger}\big)^{\tau}\big(\tilde{\mathcal{U}}_\mathcal{E}^{\dagger}\big)^{-\tau},$$ and defining time evolution of the fermionic operators as:
\begin{align}
c^{(\dagger)}_j[\tau]_+&=\big(\tilde{\mathcal{U}}_\mathcal{E}\big)^\tau c^{(\dagger)}_j \big(\tilde{\mathcal{U}}_\mathcal{E}\big)^{-\tau},\label{eq:time_evolution_fw}\\
c^{(\dagger)}_j[\tau]_-&=\big(\tilde{\mathcal{U}}_\mathcal{E}^{\dagger}\big)^{-\tau} c^{(\dagger)}_j \big(\tilde{\mathcal{U}}_\mathcal{E}^{\dagger}\big)^{\tau},
\label{eq:time_evolution_bw}
\end{align} on the forward and backward branch, respectively, the correlation function in Eq.~(\ref{eq:trace_ex}) can be rewritten as:
\begin{multline}
\frac{\delta^2\mathcal{I}[\{\zeta^\uparrow,\zeta^\downarrow\}]}{\delta\zeta_{\tau_1}^{\uparrow +}\delta\zeta_{\tau_2}^{\uparrow -}}\Bigg|_{\{\zeta^\uparrow_\tau\}=\{ \zeta^\downarrow_\tau\} = 0}=\Bigg(e^{-\tilde{\beta}}c_xc_y  T_{xy} \Bigg)^{2T}\frac{2t_y^2}{T_{xy}^2} \\
\times\text{Tr}_\mathcal{E}\Bigg(\Big(e^{+2\tilde{\beta}} c_1[T-\tau_2]_- - c_1^\dagger[T-\tau_2]_-\Big) \\
\times\Big( c_1[T-\tau_1]_+ - e^{+2\tilde{\beta}}c_1^\dagger[T-\tau_1]_+\Big)\rho_\mathcal{E}^{(T)}\Bigg),
\label{eq:general_correlations}
\end{multline} 
where we defined the dressed density matrix 
\begin{equation}
    \rho_\mathcal{E}^{(T)} = (\tilde{\mathcal{U}}_\mathcal{E})^{T}\rho_\mathcal{E} (\tilde{\mathcal{U}}_\mathcal{E}^{\dagger})^{T}.\end{equation}
In Fig.~(\ref{fig:correlation_diagram}), this correlation function is presented diagramatically.

\section{``Generalized Keldysh'' Greens functions}
\label{app:generalized_keldysh_greens_functions}

In this section, we outline how to compute correlation functions as in Eq.~(\ref{eq:trace_ex}) that contain operators that following non-unitary time evolution.

Let us first evaluate the time evolved fermion operators from Eqs.~(\ref{eq:time_evolution_fw},\ref{eq:time_evolution_bw}). For convenience, we collect all creation and annihilation operator in a vector $\bm{c} = (c_1,\hdots,c_L,c_1^\dagger,\hdots,c_L^\dagger)^T.$ We thus wish to compute $\bm{c}_l[\tau]_{\pm}.$ \\
Analogously to the unitary case,  we express the Floquet operator in terms of generators, 

\begin{align}\label{eq:def_eff_ham_fw}
\tilde{\mathcal{U}}_\mathcal{E} = \mathcal{U}_\mathcal{E} F_1 & =  \exp(\tfrac{i}{2}\bm{c}^\dagger\mathcal{G}_\mathcal{E}\bm{c}) \exp(\tfrac{1}{2}\bm{c}^\dagger\mathcal{G}_1\bm{c})\\
&=  \exp(\tfrac{i}{2}\bm{c}^\dagger\mathcal{G}_{\text{eff},+}\bm{c}),\\
\text{ with }&\mathcal{G}_{\text{eff},+} = -i\log\Big( \exp( i\mathcal{G}_\mathcal{E}) \exp(\mathcal{G}_1)\Big),
\end{align} and 
\begin{align}
\label{eq:def_eff_ham_bw}
\tilde{\mathcal{U}}_\mathcal{E}^{\dagger} = F_1^\dagger \mathcal{U}_\mathcal{E}^\dagger &= \exp(\tfrac{1}{2}\bm{c}^\dagger\mathcal{G}_1\bm{c}) \exp(-\tfrac{i}{2}\bm{c}^\dagger\mathcal{G}_\mathcal{E}\bm{c})\\
&=  \exp(-\tfrac{i}{2}\bm{c}^\dagger\mathcal{G}_{\text{eff},-}\bm{c}),\\
\text{ with }&\mathcal{G}_{\text{eff},-} = +i\log\Big( \exp(\mathcal{G}_1)\exp( -i\mathcal{G}_\mathcal{E}) \Big).
\end{align} 

Here, $\mathcal{G}_1 =2 \diag(\tilde{\beta},\underbrace{\hdots}_{(L-2)},-\tilde{\beta},\underbrace{\hdots}_{(L-2)})$ and $\mathcal{G}_\mathcal{E}$ is the generator of the environment evolution operator $\mathcal{U_{\mathcal{E}}}$, see Eq.~(\ref{eq_UER}). Note the absence of the imaginary unit $i$ in front of $\mathcal{G}_1$ in the exponential. For the kicked Ising model, i.e. $\mathcal{J}_x=0$ or $\mathcal{J}_y=0$, one has $\mathcal{G}_{\text{eff},\pm} = \mathcal{G}_\text{eff}.$

The time evolved fermionic operators read:
\begin{equation}\label{eq:time_evolution_vector}
\bm{c}[\tau]_\pm = \exp\Big(-i \tau\mathcal{G}_{\text{eff},\pm}\Big)\bm{c}.
\end{equation}
For later convenience, we define
\begin{equation}
T^\tau_\pm \equiv  \exp\Big(-i \tau\mathcal{G}_{\text{eff},\pm}\Big) \rightarrow T^\tau_\pm \bm{c} = \bm{c}[\tau]_\pm.
\end{equation}
The matrix $\mathcal{G}_{\text{eff},\pm}$ is of the form $$\mathcal{G}_{\text{eff},\pm}=\begin{pmatrix}
\mathcal{G}_{\text{eff},\pm}^A & \mathcal{G}_{\text{eff},\pm}^B\\
\mathcal{G}_{\text{eff},\pm}^C & (-\mathcal{G}_{\text{eff},\pm}^A)^T
\end{pmatrix},$$ where $\mathcal{G}_{\text{eff},\pm}^A,\mathcal{G}_{\text{eff},\pm}^B,$ and $\mathcal{G}_{\text{eff},\pm}^C$ are complex $(L-1)\times (L-1)$ matrices and $\mathcal{G}_{\text{eff},\pm}^B$ and $\mathcal{G}_{\text{eff},\pm}^C$ are antisymmetric. In the special case of a unitary evolution, one has the additional structure $\mathcal{G}_{\text{eff},\pm}^B = (\mathcal{G}_{\text{eff},\pm}^C)^\dagger$ and $(\mathcal{G}_{\text{eff},\pm}^A)^T = \mathcal{G}_{\text{eff},\pm}^A.$ The matrix $\mathcal{G}_{\text{eff},\pm}$ can be diagonalized by a complex matrix $\mathcal{M}_{\pm}:$ $\mathcal{M}^{-1}_\pm\mathcal{G}_{\text{eff},\pm}\mathcal{M}_\pm = \diag(\phi_{1}^{\pm},\hdots \phi_{L}^{\pm}, -\phi_{1}^{\pm},\hdots,-\phi_{L}^{\pm})\equiv \mathcal{D}_{\text{eff},\pm}$ where $\{\phi_{k}^{\pm}\}$  are the (complex) eigenvalues of $\mathcal{G}_{\text{eff},\pm}$ (the mentioned general properties of $\mathcal{G}_{\text{eff},\pm}$ are the minimal requirement for this structure of $\mathcal{D}_{\text{eff},\pm}$). 

Suppose we know the ``bare'' density matrix in the $\{\bm{c}_k\}$-basis (where it is not necessarily diagonal): $$\rho_\mathcal{E}=\tfrac{1}{\mathcal{Z}_0}\exp\Big(-\bm{c}^\dagger\tilde{\rho}_\mathcal{E}\bm{c}\Big).$$
From this, we can compute the ``dressed'' density matrix:
\begin{align}
\rho_\mathcal{E}^{(T)} =& \underbrace{\exp(\tfrac{iT}{2}\bm{c}^\dagger\mathcal{G}_{\text{eff},+}\bm{c})}_{(\tilde{\mathcal{U}}_\mathcal{E})^T}\tfrac{1}{\mathcal{Z}_0}\exp\Big(-\bm{c}^\dagger\tilde{\rho}_\mathcal{E}\bm{c}\Big)\\
&\times \underbrace{ \exp(-\tfrac{iT}{2}\bm{c}^\dagger\mathcal{G}_{\text{eff},-}\bm{c})}_{(\tilde{\mathcal{U}}_\mathcal{E}^{\dagger})^T}\\
=&\tfrac{1}{\mathcal{Z}_0}\exp\Big(-\bm{c}^\dagger\tilde{\rho}_\mathcal{E}^T\bm{c}\Big), 
\end{align}with:
\begin{equation}
\tilde{\rho}_\mathcal{E}^T = -\tfrac{1}{2}\log \Big( \exp(iT\mathcal{G}_{\text{eff},+})\exp(-2\tilde{\rho}_\mathcal{E})\exp(-iT\mathcal{G}_{\text{eff},-})\Big).
\label{eq:dress_rho_def}
\end{equation}
Now, we can find the diagonalizing basis for the dressed density matrix which we call $\{\hat{\delta}_k\},$ i.e. we can determine the transformation $\mathcal{T}_T^{-1}\tilde{\rho}_\mathcal{E}^T\mathcal{T}_T = \diag(\tilde{\rho}_1^T,\hdots,\tilde{\rho}_{2L}^T)\equiv \mathcal{D}_{\tilde{\rho}}^T,$ such that $\hat{\bm{\delta}}=\mathcal{T}_T^{-1}\bm{c}.$ In order to evaluate the trace expression, we express $\bm{c}_l[\tau]_\pm$ also in the $\{\hat{\delta}_k\}$-basis:
$$\bm{c}_l[\tau]_\pm = \sum_{k,k^\prime=1}^{2L} T^\tau_{\pm,(l,k)} \mathcal{T}_{T,(k,k^{\prime})}\hat{\delta}_{k^\prime}.$$
Let us evaluate trace expression as examplarily shown in Eq.~(\ref{eq:general_correlations}). 
We want to compute auto-correlations of the following operator:
\begin{align}
&e^{(1\mp1)\tilde{\beta}}c_l[\tau]_\pm +s^\tau e^{(1\pm1)\tilde{\beta}}c^\dagger_l[\tau]_\pm\\
&=\sum_{k,k^\prime=1}^{2L}\underbrace{\Big(\tilde{T}^{\tau}_{\pm,(l,k)}+s^\tau\tilde{T}^{\tau}_{\pm,(l+L,k)}\Big)}_{\equiv Z^{\tau,\pm,s^\tau}_{(l,k)}}\mathcal{T}_{t,(k,k^\prime)}\hat{\delta}_{k^\prime}, 
\end{align}
where $s^\tau$ takes the value $+1\, (-1)$ if the correlation function arises from a functional derivative with respect to the Grassmann field $\zeta^\downarrow (\zeta^\uparrow)$ at time $\tau.$ Moreover, we defined: $$\tilde{T}^{\tau}_\pm \equiv \diag\Big(e^{(1\mp1)\tilde{\beta}},\hdots,e^{(1\pm1)\tilde{\beta} },\hdots\Big)T^\tau_\pm.$$
In the case of interest, $l=1$. Hence, we can write every trace expression as:
\begin{multline}
\text{Tr}_\mathcal{E}\Big(\hdots\Big) =\sum_{k^\prime,\tilde{k}^\prime=1}^{2L} \Big(\sum_{k=1}^{2L}Z^{\tau_2,\pm,s^{\tau_2}}_{(1,k)}\mathcal{T}_{T,(k,k^\prime)}\Big)\\
\times \Big(\sum_{\tilde{k}=1}^{2L}Z^{\tau_1,\pm,s^{\tau_1}}_{(1,\tilde{k})}\mathcal{T}_{T,(\tilde{k},\tilde{k}^\prime)}\Big)\times n_\text{F}\Big(2\cdot\text{sgn}(k^\prime - L)\cdot\tilde{\rho}^T_{k^\prime}\Big).
\end{multline}
The above derivations can be generalized for functional derivatives on both Keldysh branches. For two derivatives on the same branch, one needs to take into account the correct time ordering in the operator expression.  
If $\tau_1 \neq \tau_2$, one finds for two functional derivatives on the same Keldysh branch:
\begin{multline}
    \frac{\delta^2\mathcal{I}[\{\zeta^\uparrow,\zeta^\downarrow\}]}{\delta\zeta^{\uparrow+}_{\tau_1}\delta\zeta^{\downarrow+}_{\tau_2}}\Bigg|_{\{\zeta^\uparrow_\tau\}=\{ \zeta^\downarrow_\tau\} = 0}= \Bigg(e^{-\tilde{\beta}}c_xc_yT_{xy}\Bigg)^{2T}\\
\times \frac{-2 t_y^2}{T^2_{xy}}\text{Tr}_\mathcal{E}\Bigg[\tilde{\mathcal{T}}\Bigg( \Big(c_1[T-\tau_1]_+ - e^{+2\tilde{\beta}}c_1^\dagger[T-\tau_1]_+\Big)\\
\times \Big( c_1[T-\tau_2]_+ - e^{+2\tilde{\beta}}c_1^\dagger[T-\tau_2]_+\Big)\Bigg)\rho_\mathcal{E}^{(T)}\Bigg]. \label{eq:Greenfunc_def_equalbranch}
\end{multline}
Here, we have {\sl anti-}time ordering since the time evolution of the operators is defined in backward direction, see Eqs.~(\ref{eq:time_evolution_fw},\ref{eq:time_evolution_bw}). 

Special care needs to be taken only for functional derivatives with $\tau_1 = \tau_2$ on the same branch and with respect to two different types of Grassmann variables -- this case is explained in App. \ref{appendix:equal_time_derivatives}. \\

\subsection{Equal-time functional derivatives}
\label{appendix:equal_time_derivatives}
Special care needs to be taken for functional derivatives with respect to two different variables when $\tau_1 = \tau_2$ on the same branch, i.e. for instance $\frac{\delta^2\mathcal{I}[\{\zeta^\downarrow,\zeta^\uparrow\}]}{\delta\zeta^{\downarrow+}_{\tau}\delta\zeta^{\uparrow+}_{\tau}}\Bigg|_{\{\zeta^\uparrow_\tau\}=\{ \zeta^\downarrow_\tau\} = 0}.$ Evaluating these derivatives on the corresponding Grassmann Kernel of the interaction gate (see Eq.~\ref{eq:interaction_GM}), one finds
\begin{align}
&\frac{\delta^2\mathcal{F}_{0,1}[\zeta^\uparrow,\zeta^\downarrow,\bm{\bar{\xi}},\bm{\xi}]}{\delta\zeta^\uparrow\delta\zeta^\downarrow}\Bigg|_{\{\zeta^\uparrow\}=\{ \zeta^\downarrow\} = 0} \\
 &= c_xc_yT_{xy}  \frac{-2t_xt_y}{T_{xy}^2}(2\bar{\xi}_1\xi_1)\exp\Big[ \frac{\cos(\mathcal{J}_x + \mathcal{J}_y)}{\cos(\mathcal{J}_x - \mathcal{J}_y)}\bar{\xi}_{1}\xi_{1}\Big] .
\end{align}
The corresponding operator is:
$$c_xc_yT_{xy}  \frac{2t_xt_y}{T_{xy}^2}e^{-\tilde{\beta}}F_1(-2e^{2\tilde{\beta}}c^\dagger_1c_1).$$
Hence,
\begin{multline}\nonumber
\frac{\delta^2\mathcal{I}[\{\zeta^\downarrow,\zeta^\uparrow\}]}{\delta\zeta^{\uparrow+}_{\tau}\delta\zeta^{\downarrow+}_{\tau}}\Bigg|_{\{\zeta^\uparrow_\tau\}=\{ \zeta^\downarrow_\tau\} = 0}=\Big(e^{-\tilde{\beta}}c_xc_yT_{xy}\Big)^{2T}\frac{2t_xt_y}{T_{xy}^2}\\
\times \big(-2e^{2\tilde{\beta}}\big)\text{Tr}_{\mathcal{E}}\Bigg[c^\dagger_1[T-\tau]_+c_1[T-\tau]_+\rho_{\mathcal{E}}^{T}\Bigg].\label{eq:equal_time_deriv}
\end{multline} 

\section{Reconstructing the IM}
\label{App:Reconstructing_IM}
With the above derivations, we are now in a position to reconstruct the IM from correlation functions.  Looking at Eq.~(\ref{eq:general_correlations}), one sees that the prefactor $(e^{-\tilde{\beta}}c_xc_y  T_{xy})^{2T}$ appears also in the trace expression that we obtain without taking functional derivatives, Eq.~(\ref{eq:Trace_without_derivative}). Therefore, it is a global prefactor of the IM and does not enter the exponent. The prefactor $\frac{2t_y^2}{T_{xy}^2}$, however, comes from the functional derivatives and must therefore be included in the exponent.

This type of reasoning must be repeated for every possible combination of functional derivatives, i.e. on both branches and with respect to both source fields $\zeta^{\uparrow,\downarrow}$. Here, we list the operator-level correlation functions that correspond to the entries of the influence action in Eq.~(\ref{eq_influenceaction_XY}). For $\tau \geq \tau^\prime,$ one has:
\begin{widetext}
\begin{align}
g_1\big|_{\tau,\tau^\prime} &= -\frac{T_{xy}^2}{2t_y^2}\frac{\partial^2}{\partial\zeta^{\uparrow +}_{\tau}\partial\zeta^{\uparrow +}_{\tau^\prime}}\mathcal{I}[\{\zeta^\downarrow,\zeta^\uparrow\}]\Bigg|_{\{\zeta^\uparrow_\tau\}=\{\zeta^\downarrow_\tau\}=0}\\
&= \text{Tr}_{\mathcal{E}}\Bigg[ \Big( c_1[T-\tau]_+ - e^{2\tilde{\beta}}c_1^\dagger[T-\tau]_+\Big)\Big( c_1[T-\tau^\prime]_+ - e^{2\tilde{\beta}}c_1^\dagger[T-\tau^\prime]_+\Big)\rho_{\mathcal{E}}^{T}\Bigg]
\\
g_2\big|_{\tau,\tau^\prime} &= -\frac{T_{xy}^2}{2t_y^2}\frac{\partial^2}{\partial\zeta^{\uparrow -}_{\tau}\partial\zeta^{\uparrow +}_{\tau^\prime}}\mathcal{I}[\{\zeta^\downarrow,\zeta^\uparrow\}]\Bigg|_{\{\zeta^\uparrow_\tau\}=\{\zeta^\downarrow_\tau\}=0}\\
&=\text{Tr}_{\mathcal{E}}\Bigg[ \Big(e^{2\tilde{\beta}} c_1[T-\tau]_- - c_1^\dagger[T-\tau]_-\Big)\Big( c_1[T-\tau^\prime]_+ - e^{2\tilde{\beta}}c_1^\dagger[T-\tau^\prime]_+\Big)\rho_{\mathcal{E}}^{T}\Bigg]
\\
g_3\big|_{\tau,\tau^\prime} &= -\frac{T_{xy}^2}{2t_xt_y}\frac{\partial^2}{\partial\zeta^{\uparrow +}_{\tau}\partial\zeta^{\downarrow +}_{\tau^\prime}}\mathcal{I}[\{\zeta^\downarrow,\zeta^\uparrow\}]\Bigg|_{\{\zeta^\uparrow_\tau\}=\{\zeta^\downarrow_\tau\}=0}\\
&= - \Bigg(\text{Tr}_{\mathcal{E}}\Bigg[ \Big( c_1[T-\tau]_+ - e^{2\tilde{\beta}}c_1^\dagger[T-\tau]_+\Big)\Big( c_1[T-\tau^\prime]_+ + e^{2\tilde{\beta}}c_1^\dagger[T-\tau^\prime]_+\Big)\rho_{\mathcal{E}}^{T}\Bigg] - \delta_{\tau,\tau^\prime} e^{2\tilde{\beta}}\text{Tr}_\mathcal{E}(\rho_\mathcal{E}^{(T)})\Bigg)
\\
g_4\big|_{\tau,\tau^\prime} &= -\frac{T_{xy}^2}{2t_xt_y}\frac{\partial^2}{\partial\zeta^{\uparrow -}_{\tau}\partial\zeta^{\downarrow +}_{\tau^\prime}}\mathcal{I}[\{\zeta^\downarrow,\zeta^\uparrow\}]\Bigg|_{\{\zeta^\uparrow_\tau\}=\{\zeta^\downarrow_\tau\}=0}\\
&= - \text{Tr}_{\mathcal{E}}\Bigg[ \Big( e^{2\tilde{\beta}}c_1[T-\tau]_- - c_1^\dagger[T-\tau]_-\Big)\Big( c_1[T-\tau^\prime]_+ + e^{2\tilde{\beta}}c_1^\dagger[T-\tau^\prime]_+\Big)\rho_{\mathcal{E}}^{T}\Bigg]
\\
g_5\big|_{\tau,\tau^\prime} &= -\frac{T_{xy}^2}{2t_xt_y}\frac{\partial^2}{\partial\zeta^{\downarrow +}_{\tau}\partial\zeta^{\uparrow +}_{\tau^\prime}}\mathcal{I}[\{\zeta^\downarrow,\zeta^\uparrow\}]\Bigg|_{\{\zeta^\uparrow_\tau\}=\{\zeta^\downarrow_\tau\}=0}\\
&= - \Bigg(\text{Tr}_{\mathcal{E}}\Bigg[ \Big( c_1[T-\tau]_+ + e^{2\tilde{\beta}}c_1^\dagger[T-\tau]_+\Big)\Big( c_1[T-\tau^\prime]_+ - e^{2\tilde{\beta}}c_1^\dagger[T-\tau^\prime]_+\Big)\rho_{\mathcal{E}}^{T}\Bigg] + \delta_{\tau,\tau^\prime} e^{2\tilde{\beta}}\text{Tr}_\mathcal{E}(\rho_\mathcal{E}^{(T)})\Bigg)
\\
g_6\big|_{\tau,\tau^\prime} &= -\frac{T_{xy}^2}{2t_xt_y}\frac{\partial^2}{\partial\zeta^{\downarrow -}_{\tau}\partial\zeta^{\uparrow +}_{\tau^\prime}}\mathcal{I}[\{\zeta^\downarrow,\zeta^\uparrow\}]\Bigg|_{\{\zeta^\uparrow_\tau\}=\{\zeta^\downarrow_\tau\}=0}\\
&= - \text{Tr}_{\mathcal{E}}\Bigg[ \Big( e^{2\tilde{\beta}}c_1[T-\tau]_- + c_1^\dagger[T-\tau]_-\Big)\Big( c_1[T-\tau^\prime]_+ - e^{2\tilde{\beta}}c_1^\dagger[T-\tau^\prime]_+\Big)\rho_{\mathcal{E}}^{T}\Bigg]
\\
g_7\big|_{\tau,\tau^\prime} &= -\frac{T_{xy}^2}{2t_x^2}\frac{\partial^2}{\partial\zeta^{\downarrow +}_{\tau}\partial\zeta^{\downarrow +}_{\tau^\prime}}\mathcal{I}[\{\zeta^\downarrow,\zeta^\uparrow\}]\Bigg|_{\{\zeta^\uparrow_\tau\}=\{\zeta^\downarrow_\tau\}=0}\\
&= \text{Tr}_{\mathcal{E}}\Bigg[ \Big( c_1[T-\tau]_+ + e^{2\tilde{\beta}}c_1^\dagger[T-\tau]_+\Big)\Big( c_1[T-\tau^\prime]_+ + e^{2\tilde{\beta}}c_1^\dagger[T-\tau^\prime]_+\Big)\rho_{\mathcal{E}}^{T}\Bigg]
\\
g_8\big|_{\tau,\tau^\prime} &= -\frac{T_{xy}^2}{2t_x^2}\frac{\partial^2}{\partial\zeta^{\downarrow -}_{\tau}\partial\zeta^{\downarrow +}_{\tau^\prime}}\mathcal{I}[\{\zeta^\downarrow,\zeta^\uparrow\}]\Bigg|_{\{\zeta^\uparrow_\tau\}=\{\zeta^\downarrow_\tau\}=0}\\
&= \text{Tr}_{\mathcal{E}}\Bigg[ \Big( e^{2\tilde{\beta}}c_1[T-\tau]_- + c_1^\dagger[T-\tau]_-\Big)\Big( c_1[T-\tau^\prime]_+ + e^{2\tilde{\beta}}c_1^\dagger[T-\tau^\prime]_+\Big)\rho_{\mathcal{E}}^{T}\Bigg]
\end{align}
\end{widetext}

\section{Temporal correlation functions in the kicked Ising model}
\label{App:Ising_corr_funcs_evaluation}
In this section, we sketch how to arrive at Eq.~(\ref{eq:G_lesser_finiteT}) for the correlation function $g_{\tau,\tau^\prime}$ from Eq.~(\ref{eq:trace_correlation_function_Ising}). For the kicked Ising model, Eq.~(\ref{eq:time_evolution_vector}) yields equivalent prescriptions on both Keldysh branches and we can write:
$$\bm{c}_j[\tau] = \mathcal{M}_{j,k}e^{-i\phi_k\tau}(\mathcal{M}^\dagger)_{k,l} \bm{c}_l,$$ where $\mathcal{M}$ is the matrix that diagonalizes the effective Hamiltonian $\mathcal{G}_\text{eff}$ as defined in Eqs.~(\ref{eq:def_eff_ham_fw},\ref{eq:def_eff_ham_bw}). The function $g_{\tau,\tau^\prime}$ can thus be written as:
\begin{align}
g_{\tau,\tau^\prime}=&\text{Tr}_{\mathcal{E}}\Bigg[ \Big( c_1[\tau] + c_1^\dagger[\tau]\Big)\Big( c_1[\tau^\prime] + c_1^\dagger[\tau^\prime]\Big)\rho_{\mathcal{E}}\Bigg]\\
=& \sum_{k,k^\prime,l,l^\prime}^{2L}\big(\mathcal{M}_{1,k^\prime}+\mathcal{M}_{1+L,k^\prime}\big) \mathcal{M}^\dagger_{k^\prime,l^\prime}  e^{-i\phi_{k^\prime}\tau}\langle \bm{c}_{l^\prime}\bm{c}_l\rangle\\
&\times e^{-i\phi_{k}\tau^\prime} \mathcal{M}_{l,k}^*\big( \mathcal{M}_{1,k} + \mathcal{M}_{1+L,k} \big),
\end{align} where $\langle \bm{c}_{l^\prime}\bm{c}_l\rangle \equiv \text{Tr}_\mathcal{E}(\bm{c}_{l^\prime}\bm{c}_l\rho_\mathcal{E}).$ Defining the function $\mathcal{C}_k(\tau)\equiv e^{-i\phi_k\tau}(\mathcal{M}_{1,k} + \mathcal{M}_{1+L,k})$ and using $\phi_{k+L} = -\phi_k$ and $\mathcal{M}_{l,k}^* = \mathcal{M}_{l+L,k+L},$ this becomes:
\begin{equation}\label{eq:general_KIC_correlations_appendix}
   g_{\tau,\tau^\prime}=  \bm{\mathcal{C}}^\dagger(\tau)\mathcal{M}^\dagger\Lambda \mathcal{M} \bm{\mathcal{C}}(\tau^\prime),
\end{equation} where $\bm{\mathcal{C}}(\tau)\equiv \big(\mathcal{C}^*_1(\tau),\hdots,\mathcal{C}^*_L(\tau),\mathcal{C}_1(\tau),\hdots,\mathcal{C}_L(\tau)\big)^T.$ The initial state is encoded in the correlation matrix $\Lambda =\text{Tr}_\mathcal{E}\big(\bm{c}\cdot\bm{c}^\dagger\rho_\mathcal{E}\big).$\\

From Eq.~(\ref{eq:general_KIC_correlations_appendix}), one can see that only the imaginary part of $g_{\tau,\tau^\prime}$ depends on both variables $\tau,\tau^\prime$ independently, while the real part is only a function of $\Delta\tau=\tau-\tau^\prime.$ For this, we note that the unitarily transformed correlation matrix has the general structure of a fermionic correlation matrix:
$$\mathcal{M}^\dagger \Lambda \mathcal{M}=\begin{pmatrix}
\mathcal{A} & \mathcal{B}\\
\mathcal{B}^\dagger & \mathds{1} - \mathcal{A}^T
\end{pmatrix},$$ where $\mathcal{A}$ is hermitian and $\mathcal{B}$ is antisymmetric.

Using these properties, we derive:
\begin{align*}
    g_{\tau,\tau^\prime} =& \sum_{k,l=1}^L\Bigg(\mathcal{C}_k(\tau) \mathcal{A}_{kl}\mathcal{C}^*_l(\tau^\prime) + \mathcal{C}_k(\tau) \mathcal{B}_{kl}\mathcal{C}_l(\tau^\prime)\\
    & -\mathcal{C}^*_k(\tau) \mathcal{B}^*_{kl}\mathcal{C}^*_l(\tau^\prime) + \mathcal{C}^*_k(\tau) (\delta_{k,l}-\mathcal{A}_{kl}^*)\mathcal{C}_l(\tau^\prime)\Bigg)\\
    =& 2i\,\text{Im}\Bigg[ \sum_{k,l=1}^L\Big(\mathcal{C}_k(\tau) \mathcal{A}_{kl}\mathcal{C}^*_l(\tau^\prime) + \mathcal{C}_k(\tau) \mathcal{B}_{kl}\mathcal{C}_l(\tau^\prime)\Bigg)\Bigg]\\
    &+ \sum_{k=1}^L \mathcal{C}_k(0)| \cos\big(\phi_k (\tau-\tau^\prime)\big).
\end{align*} Hence, the real part of $g_{\tau,\tau^\prime}$ depends only on the quasi-energy spectrum $\phi_k$ and the difference of times $\tau-\tau^\prime,$ while it is independent of the initial state.

For our purposes, let $\mathcal{N}$ be the matrix that diagonalizes the correlation matrix: $$\mathcal{N}^\dagger \Lambda \mathcal{N} = \text{diag}(\lambda_1,\lambda_2,..,\lambda_{L},\lambda_{L+1},..,\lambda_{2L}),$$ where we order the columns of $\mathcal{N}$ in such a way that the first $L$ eigenvalues lie in the interval $\{1/2,1\}$ and $\lambda_{L+i} = 1-\lambda_i.$ We parametrize the eigenvalues as $\lambda_i =1/(1+e^{-\kappa_i}),$ and $\lambda_{i+L} =1/(1+e^{\kappa_i})$ with $\kappa_i\geq 0.$ \\

If the initial state, encoded in $\Lambda$,  is diagonalized by the same unitary that diagonalizes the effective Hamiltonian, one has $\mathcal{N}^\dagger \mathcal{M}=\mathds{1}.$ In the thermodynamic limit, one obtains thus
\begin{align}\label{eq:G_lesser_eigenstate}
g_{\tau,\tau^\prime}=& \sum_{m=1}^{L} |\mathcal{C}_m(0)|^2 \Big(\frac{e^{-i\phi_m(\tau-\tau^\prime)}}{1+e^{-\kappa_m}} +\frac{e^{+i\phi_m(\tau-\tau^\prime)}}{1+e^{+\kappa_m}}\Big) \\
=&  \sum_{m=1}^{L} |\mathcal{C}_m(0)|^2 \Bigg(\cos\big(\phi_m(\tau-\tau^\prime)\big)\\
&-i\sin\big(\phi_m(\tau-\tau^\prime)\big) \tanh\Big(\frac{\kappa_m}{2}\Big)\Bigg) .
\end{align}
Further parametrizing the spectrum of $\Lambda$ in terms of effective temperatures, $\kappa_m = \beta_m\phi_m,$ and using $\mathcal{C}_m\equiv\mathcal{C}_m(0),$ we arrive at Eq.~(\ref{eq:G_lesser_finiteT}).

\section{Temporal finite-size effects}
\label{Sec:temp_finite_size}
In this Section, we will explain the subtle issue of finite size effects in the temporal domain that are present in the trotterized XY model. These are caused by the breaking of time-translation invariance due to effective non-unitary evolution of the environment, explained above as described in App.~\ref{app:generalized_keldysh_greens_functions}. 

In formulas, this can be understood by examplarily considering a correlation function of the type \begin{align} g_{\tau,\tau^\prime} &= \text{Tr}_\mathcal{E}\Big((\tilde{\mathcal{U}}_\mathcal{E})^{T-\tau}\hat{O}(\tilde{\mathcal{U}}_\mathcal{E})^{\tau-\tau^\prime}\hat{O}(\tilde{\mathcal{U}}_\mathcal{E})^{\tau^\prime}\rho_\mathcal{E} (\tilde{\mathcal{U}}_\mathcal{E}^{\dagger})^T\Big)\\
&=\text{Tr}_\mathcal{E}\Big(\hat{O}^\prime_{\tau-\tau^\prime}(\tilde{\mathcal{U}}_\mathcal{E})^{\tau^\prime-T}\rho_\mathcal{E}^{(T)} (\tilde{\mathcal{U}}_\mathcal{E})^{T-\tau}\Big),
\end{align} where we defined the translation invariant operator $\hat{O}_{\tau-\tau^\prime}^\prime \equiv  \hat{O}(\tilde{\mathcal{U}}_\mathcal{E})^{\tau - \tau^\prime} \hat{O}$ and $\rho_\mathcal{E}^{(T)}$ is the dressed density matrix defined in Eq.~(\ref{eq:dress_rho_def}). One sees that $g_{\tau,\tau^\prime}$ is only time-translation invariant only if $[\tilde{\mathcal{U}}_\mathcal{E},\rho_\mathcal{E}^{(T)}] = 0,$ such that $$(\tilde{\mathcal{U}}_\mathcal{E})^{\tau^\prime-T}\rho_\mathcal{E}^{(T)} (\tilde{\mathcal{U}}_\mathcal{E})^{T-\tau} = \rho_\mathcal{E}^{(T)} (\tilde{\mathcal{U}}_\mathcal{E})^{\tau^\prime - \tau}.$$ For the stationary initial states considered in this work, this is precisely the case when $\tilde{\mathcal{U}}_\mathcal{E}^{\dagger} = \big(\tilde{\mathcal{U}}_\mathcal{E}\big)^{-1},$ i.e. when time evolution is unitary.

\begin{figure}[H]
\centering
\includegraphics[width=0.5\textwidth]{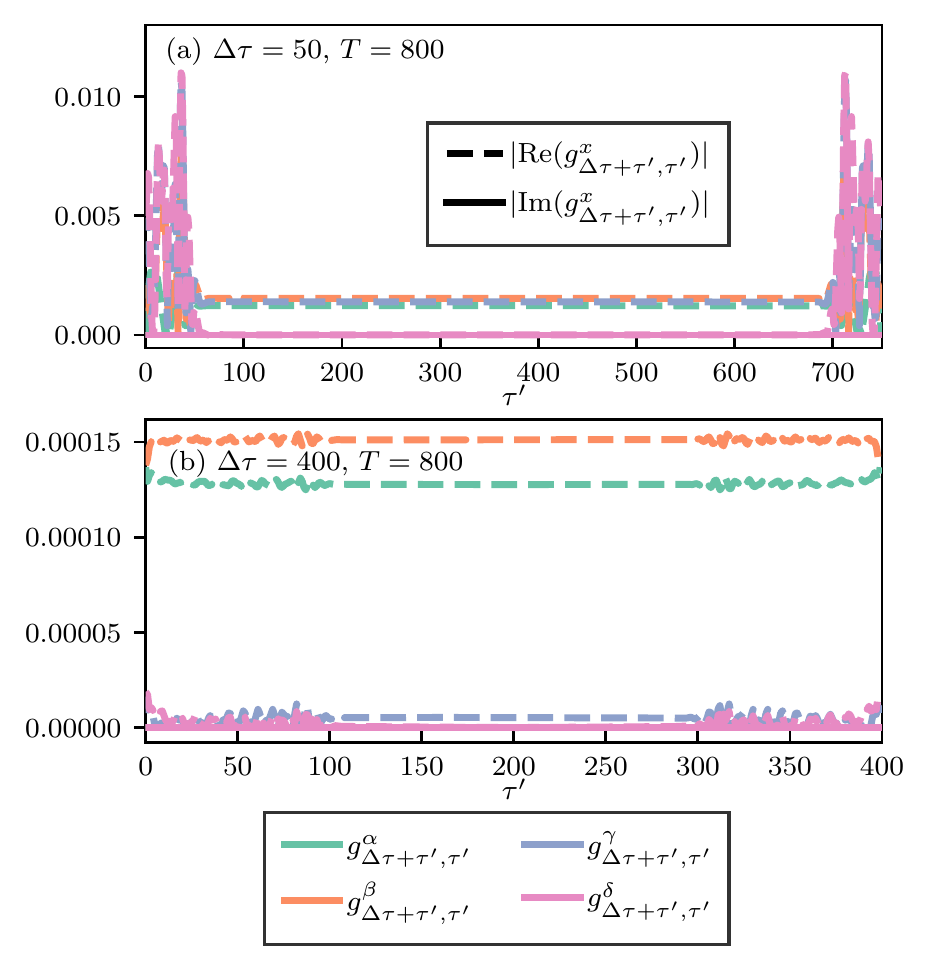}
\caption{Response functions for the trotterized XY model for the {\it infinite temperature initial state}, plotted from the same dataset as Fig.~{\ref{fig:scaling_xy_IT}} [data not rescaled here].}
\label{Fig:temporal_finite_size_XY_IT}
\end{figure}

\begin{figure}[H]
\centering
\includegraphics[width=0.5\textwidth]{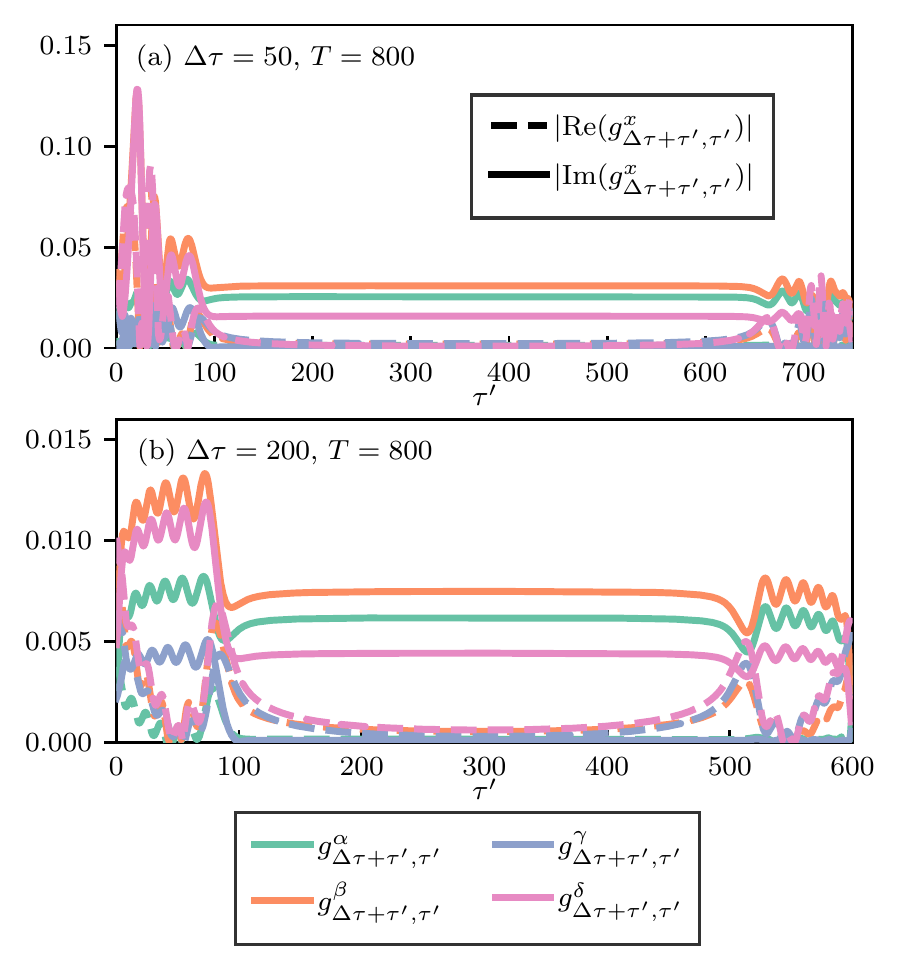}
\caption{Response functions for the trotterized XY model for the {\it critical initial state}, plotted from the same dataset as Fig.~{\ref{fig:scaling_xy_bog_imag}} [data not rescaled here]. }
\label{Fig:temporal_finite_size_XY_crit}
\end{figure}

In Figs.~(\ref{Fig:temporal_finite_size_XY_IT},\ref{Fig:temporal_finite_size_XY_crit}), we demonstrate that for $T\rightarrow \infty,$ however, there is approximate time translation invariance in the temporal bulk, $0 \ll \tau,\tau^\prime \ll T.$ Shown are the response functions $g^{x}_{\Delta\tau+\tau^\prime,\tau^\prime},\, x=\alpha,\beta,\gamma,\delta,$ of the trotterized XY model for the infinite temperature and the critical initial state [corresponding to Figs.~{\ref{fig:scaling_xy_IT},b}] for two different fixed values of $\Delta\tau.$ In both cases, for $\Delta\tau=50,$ we observe that there is a window around $\tau^\prime\approx (T-\Delta\tau)/2,$ where all response functions approximately reach a constant value, corresponding to the ``temporal bulk'', where finite size effects are minimal. For this reason, we constrain our study of response functions in the main text to the ``temporal bulk'' response functions $g^{x}_{[\Delta\tau]} \equiv g^{x}_{\frac{T+\Delta\tau}{2},\frac{T-\Delta\tau}{2}}$, see Fig.~\ref{fig:xy_scaling}. 
As the interval $\Delta\tau$ is increased [we choose here the intervals which are plotted in Figs.~{\ref{fig:scaling_xy_IT},b}], the window with minimal finite size effects decreases: In Fig.~\ref{Fig:temporal_finite_size_XY_crit}, the real part does not reach a truly saturated value, indicating temporal finite size effects which are responsible for the deviation from the power-law in Fig.~{\ref{fig:scaling_xy_bog_imag}}.
\section{Subsystem size scaling in the trotterized XY model}
\label{Sec:block_size_XY}
\begin{figure}[h]
\centering
\includegraphics[width=0.5\textwidth]{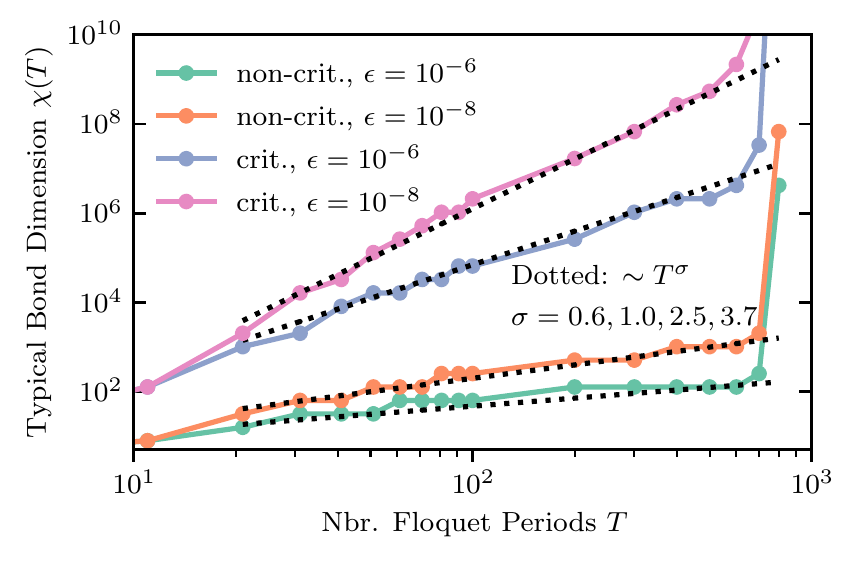}
\caption{Typical bond dimension associated with the
average subsystem block size
needed to fully diagonalize the correlation matrix $\Lambda$ of the IM
of the kicked Ising model over $T$ Floquet periods [corresponding to the data presented Fig.~{\ref{fig:xy_scaling}}].
 For $T\gtrsim L$, the results show consistent strong finite size effects: for $T \gg L=400$, long-range temporal correlations appear in the IM due to reflections at the edge of the reservoir~\cite{lerose2022overcoming}. }
\label{Fig:blocksize_XY}
\end{figure}

Here, we present results for the trotterized XY model analogous to those for the kicked Ising model in Fig.~\ref{fig:subsystem_scaling_Ising}. In Fig.~\ref{Fig:blocksize_XY}, we show the typical bond dimension $\chi$ associated with the average subsystem size $n_\text{av}$ needed to fully diagonalize the IM over $T$ Floquet periods with localization tolerance $\epsilon$. A crucial difference to the study in the kicked Ising model is the fact that we do not have an analogous formula to Eq.~({\ref{eq:G_lesser_finiteT}}) here, which would allow us to compute results in the thermodynamic limit. Therefore, while the qualitative discussion from Sec.~{\ref{sec:subsystem_scaling_ising}} applies here for $T\lesssim L$ [results are shown for a system size of $L=400$], the bond dimension shoots up as spatial finite size effects become relevant at $T=L.$ 

\end{document}